\documentclass[pdflatex,sn-mathphys-num]{sn-jnl}

\usepackage[version=4]{mhchem}
\usepackage{siunitx}
\usepackage[fontsize=10pt]{scrextend}
\usepackage{multirow}
\usepackage{color}
\usepackage{booktabs}
\usepackage{bookmark}

\theoremstyle{thmstyleone}%
\theoremstyle{thmstyletwo}%

\theoremstyle{thmstylethree}%

\begin{document}

\title[Article Title]{
Equilibrium Thermochemistry and Crystallographic Morphology of Manganese Sulfide Nanocrystals
}

\author[1]{\fnm{Junchi} \sur{Chen}}

\author[2,3]{\fnm{Tamilarasan} \sur{Subramani}}

\author[4]{\fnm{Deep} \sur{Mekan}}

\author[4]{\fnm{Danielle} \sur{Gendler}}

\author[1,5]{\fnm{Ray} \sur{Yang}}

\author[1]{\fnm{Manish} \sur{Kumar}}

\author[2,6]{\fnm{Megan} \sur{Householder}}

\author[2,3,7]{\fnm{Alexis} \sur{Rosado Ortiz}}

\author[4]{\fnm{Emil A.} \sur{Hernandez-Pagan}}

\author[2,3]{\fnm{Kristina} \sur{Lilova}}

\author*[1]{\fnm{Robert B.} \sur{Wexler}}
\email{wexler@wustl.edu}

\affil*[1]{\orgdiv{Department of Chemistry and Institute of Materials Science and Engineering}, \orgname{Washington University in St.\ Louis}, \orgaddress{\city{St.\ Louis}, \postcode{63130}, \state{MO}, \country{USA}}}

\affil[2]{\orgdiv{Center for Materials of the Universe}, \orgname{Arizona State University}, \orgaddress{\city{Tempe}, \postcode{85287}, \state{AZ}, \country{USA}}}

\affil[3]{\orgdiv{School of Molecular Sciences}, \orgname{Arizona State University}, \orgaddress{\city{Tempe}, \postcode{85287}, \state{AZ}, \country{USA}}}

\affil[4]{\orgdiv{Department of Chemistry and Biochemistry}, \orgname{University of Delaware}, \orgaddress{\city{Newark}, \postcode{19711}, \state{DE}, \country{USA}}}

\affil[5]{\orgdiv{Department of Computer Science and Engineering, McKelvey School of Engineering}, \orgname{Washington University in St.\ Louis}, \orgaddress{\city{St.\ Louis}, \postcode{63130}, \state{MO}, \country{USA}}}

\affil[6]{\orgdiv{School of Earth and Space Exploration}, \orgname{Arizona State University}, \orgaddress{\city{Tempe}, \postcode{85287}, \state{AZ}, \country{USA}}}

\affil[7]{\orgdiv{School of Pharmacy}, \orgname{Massachusetts College of Pharmacy and Health Sciences}, \orgaddress{\city{Boston}, \postcode{02115}, \state{MA}, \country{USA}}}

\abstract{
Manganese sulfide (MnS) is a p-type magnetic semiconductor whose physicochemical properties are sensitive to nanocrystal (NC) morphology, yet the thermodynamic driving forces governing morphology across MnS polymorphs remain poorly understood.
Here, we use density functional theory (DFT) to predict the equilibrium morphologies of rock salt (RS), zinc blende (ZB), and wurtzite (WZ) MnS NCs as a function of the relative chemical potential of sulfur, {\unboldmath $\Delta \mu_{\ce{S}}$}.
Benchmarking against Heyd--Scuseria--Ernzerhof (HSE06) hybrid functional calculations reveals that the {\unboldmath r$^2$SCAN} meta-generalized gradient approximation reproduces experimental lattice constants and thermochemical reaction energies but underestimates S-terminated polar surface energies by up to a factor of five; applying a Hubbard {\unboldmath $U$} correction ({\unboldmath r$^2$SCAN+$U$, $U = 2.7$}~eV) to the Mn 3d states brings the results into close agreement with HSE06.
Using the benchmarked {\unboldmath r$^2$SCAN+$U$} framework with the Gibbs--Wulff theorem, we predict that RS-MnS NCs favor nanocubes across nearly the entire stability window, ZB-MnS NCs transform from rhombic dodecahedra (Mn-rich) to polyhedra with 16 triangular faces (S-rich), and WZ-MnS NCs adopt rod-like morphologies with {\unboldmath $\Delta \mu_{\ce{S}}$}-sensitive base truncation.
Synthesized RS-MnS NCs confirm the predicted cubic morphology, and high-temperature oxidative solution calorimetry yields an apparent surface energy of {\unboldmath $1.15 \pm 0.38$~J$\cdot$m$^{-2}$}, higher than the theoretical equilibrium value (0.42--0.43~J{\unboldmath $\cdot$m$^{-2}$}), which we attribute to high-index facet exposure, surface area uncertainty, and non-ideal surface configurations in real samples.
This work establishes a framework for predicting the equilibrium morphologies of metal chalcogenide NCs and provides a foundation for future studies of solvent effects, ligand adsorption, and the kinetic mechanisms governing NC formation.
}


\maketitle

\section{Introduction}\label{sec1}

As a representative class of zero-dimensional nanomaterials, semiconductor nanocrystals (NCs) have been a central focus of nanoscience research since the 1980s\cite{ekimov_quantum_2023, rossetti_quantum_1983, murray_synthesis_1993} owing to their broad potential applications.\cite{alivisatos_semiconductor_1996, kovalenko_prospects_2015, ghosh_many_2018, mirkin_33_2025, ibanez_prospects_2025}
Since the physicochemical properties of NCs are highly sensitive to size and morphology,\cite{nirmal_luminescence_1999, owen_coordination_2015, boles_erratum_2016} precise synthetic control over these attributes is essential.\cite{chen_pure_2013}
Manganese sulfide (MnS), a p-type magnetic semiconductor with a wide band gap (2.7--3.7~eV)\cite{lokhande_process_1998, alanazi_structural_2021} and antiferromagnetic order,\cite{corliss_magnetic_1956} exemplifies this morphology--property relationship.
In particular, recent work has shown that \{100\}-faceted cubes and \{111\}-faceted octahedra of rock salt MnS exhibit distinct band gaps, with cubes showing higher electrical conductivity and stronger low-temperature magnetization than octahedra.\cite{chen_low-temperature_2023}
Such facet-dependent optical, electrical, and magnetic properties motivate shape control for applications ranging from magnetic resonance imaging\cite{chilton_use_1984, meng_phase_2016} to lithium-ion batteries.\cite{zhang_hydrothermal_2008, tang_morphology_2015}

\begin{figure*}[htb]
    \centering
    \includegraphics[width=0.85\linewidth]{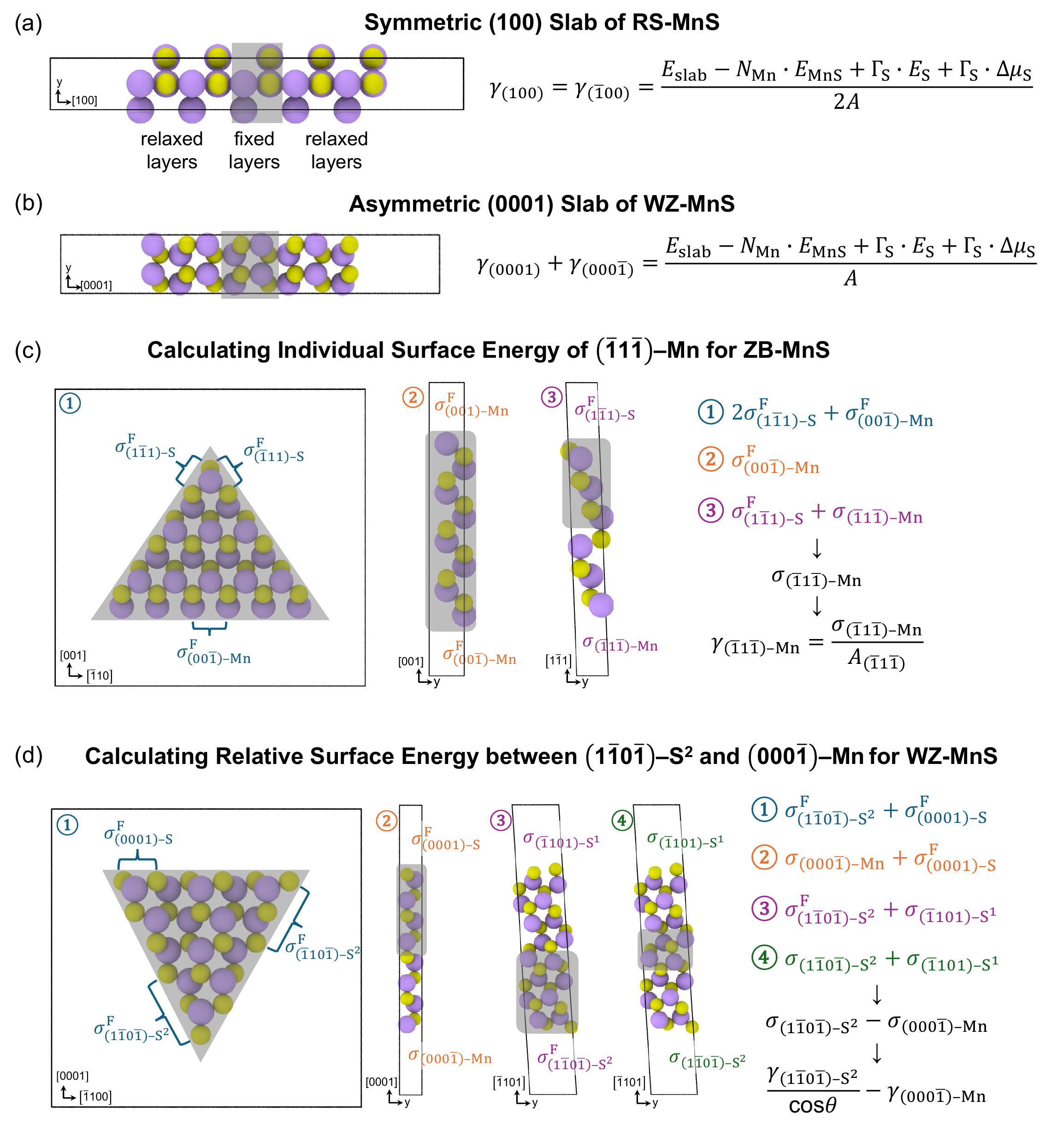}
    \caption{
Schematics illustrating the calculation of (a) the individual surface energy of the (100) facet in rock salt (RS) MnS from a symmetric slab, (b) the combined (0001) and (000$\overline{1}$) surface energies in wurtzite (WZ) MnS from an asymmetric slab, (c) the individual polar surface energy of the ($\overline{1}$1$\overline{1}$)--Mn facet in zinc blende (ZB) MnS from combined slab and wedge models, and (d) the relative surface energy between the (1$\overline{1}$0$\overline{1}$)--S$^2$ and (000$\overline{1}$)--Mn facets in WZ-MnS from combined slab and wedge models.
Gray-shaded regions denote atoms fixed (F) during geometry optimization.
$E_{\mathrm{slab}}$, $E_{\ce{MnS}}$, and $E_{\ce{S}}$ are the total energies of the slab, bulk MnS per formula unit, and elemental sulfur per atom, respectively.
$\sigma = A \cdot \gamma$, where $\gamma$ is the surface energy and $A$ is the surface area of one side of the unit slab.
$\Gamma_{\ce{S}} = N_{\ce{Mn}} - N_{\ce{S}}$, where $N_{\ce{Mn}}$ and $N_{\ce{S}}$ are the numbers of Mn and S atoms in the slab, respectively.
$\Delta \mu_{\mathrm{S}}$ is the chemical potential of sulfur relative to its standard state.
$\theta$ is the dihedral angle between two planes.
See the Ab Initio Thermodynamics subsection of Methods for details.
}
    \label{fig:fig1}
\end{figure*}

Despite significant advances in MnS NC synthesis,\cite{lu_metastable_2001, kan_synthesis_2001, jun_architectural_2002, joo_generalized_2003, michel_hydrothermal_2006, puglisi_monodisperse_2010, gui_hydrothermal_2011, yang_polymorphism_2012, gendler_halide-driven_2023} morphology control remains largely empirical.
Varying precursors,\cite{gendler_halide-driven_2023} temperatures,\cite{jun_architectural_2002} and reaction times\cite{chen_low-temperature_2023} can yield nanocubes, nanooctahedra, nanorods, and other morphologies across the three MnS polymorphs (rock salt [RS], wurtzite [WZ], and zinc blende [ZB]), yet the thermodynamic driving forces that underlie these morphological outcomes are poorly understood.
Which facets are intrinsically stable for each polymorph?
How do synthesis conditions (encoded in the relative chemical potential of sulfur, $\Delta \mu_{\ce{S}}$) shift the equilibrium morphology?
Answering these questions would enable predictive design of MnS NC syntheses.

The Gibbs--Wulff theorem provides a theoretical foundation for predicting equilibrium morphology from surface energies, and density functional theory (DFT) has been widely used to calculate such energies for elemental crystals\cite{vitos_surface_1998, xia_shape-controlled_2015, tran_surface_2016, fichthorn_theory_2023} and catalysts.\cite{xiao_high-index-facet-_2020, boukouvala_approaches_2021, sanspeur_wherewulff_2023}
However, applying this framework to MnS NCs faces two obstacles.
First, the noncentrosymmetric ZB and WZ structures lack key symmetry operations along certain crystallographic directions,\cite{sun_efficient_2013} preventing construction of symmetric slabs and direct evaluation of individual polar surface energies (Figures~\ref{fig:fig1}a~and~\ref{fig:fig1}b).
While approximate solutions exist,\cite{jp0445573, tian_dft_2018, yoo_efficient_2021, stuart_method_2023} general methods applicable to all MnS polymorphs have not been systematically developed.
Second, the accuracy of standard DFT approaches for transition-metal chalcogenide surfaces has not been validated.
The strongly constrained and appropriately normed (SCAN)\cite{sun_strongly_2015} meta-generalized gradient approximation (meta-GGA) functional and its regularized form r$^2$SCAN\cite{furness_accurate_2020} are now considered state-of-the-art for solid-state simulations\cite{peng_versatile_2016, ning_workhorse_2022, kothakonda_testing_2023} and have been incorporated into the Materials Project,\cite{kingsbury_performance_2022, horton_accelerated_2025} but their performance for MnS surfaces---particularly the polar, S-terminated facets that play critical roles in morphology determination---is unknown.

In this work, we address both obstacles to enable predictive morphology modeling for MnS NCs.
We employ wedge models\cite{zhang_surface_2004, dreyer_absolute_2014, akiyama_modified_2019, jin_absolute_2021} to decouple individual polar surface energies in ZB-MnS (Figure~\ref{fig:fig1}c) and relative surface energy methods\cite{li_computing_2015} for WZ-MnS (Figure~\ref{fig:fig1}d), providing a complete framework for Wulff construction across all three polymorphs.
We benchmark DFT methods against hybrid functional calculations and find that r$^2$SCAN alone underestimates S-terminated polar surface energies by up to a factor of five.
This deficiency, corrected by adding a Hubbard $U$ term ($U = 2.7$~eV), has implications for computational studies of other transition-metal chalcogenide NCs.

Using the benchmarked r$^2$SCAN+$U$ framework, we calculate surface energies for RS-, ZB-, and WZ-MnS facets and construct equilibrium morphologies as a function of $\Delta \mu_{\ce{S}}$.
These thermodynamic Wulff constructions describe facet preferences within a given crystal structure rather than polymorph selection among competing phases, which can depend on nucleation barriers, precursor chemistry, temperature, and growth kinetics.
Our predictions for RS-MnS (nanocubes across nearly the entire stability window) are consistent with experimental observations and with the morphology of the RS-MnS NCs synthesized here, whose apparent surface energy we measure by high-temperature oxidative solution calorimetry.
For the less-studied ZB-MnS, we predict a morphological transition from rhombic dodecahedra (Mn-rich) to 16-faced polyhedra (S-rich), providing testable guidance for future synthesis efforts.
WZ-MnS NCs are predicted to adopt rod-like morphologies with $\Delta \mu_{\ce{S}}$-sensitive base truncation.
Together, these results establish a quantitative foundation for morphology design in MnS NCs and provide a framework extensible to other metal chalcogenide systems.

\section{Results}\label{sec2}

\subsection{Computational framework validation}

\subsubsection{Bulk properties benchmarking}

Before investigating surface energies, we first identified an appropriate DFT framework by benchmarking exchange-correlation (XC) functionals against experimental lattice constants and reaction energies.
The functionals tested include the Perdew--Burke--Ernzerhof (PBE) generalized gradient approximation (GGA)\cite{Perdew:1996pki} and its D3 dispersion-corrected variant with Becke--Johnson damping, PBE-D3+BJ;\cite{grimme_consistent_2010, grimme_effect_2011} the SCAN\cite{sun_strongly_2015} and r$^2$SCAN\cite{furness_accurate_2020} meta-GGAs; and their corresponding van der Waals (vdW) dispersion-corrected variants, SCAN-rVV10\cite{peng_versatile_2016} and r$^2$SCAN-rVV10.\cite{ning_workhorse_2022}

\begin{figure}[htb]
    \centering
    \includegraphics[width=0.80\linewidth]{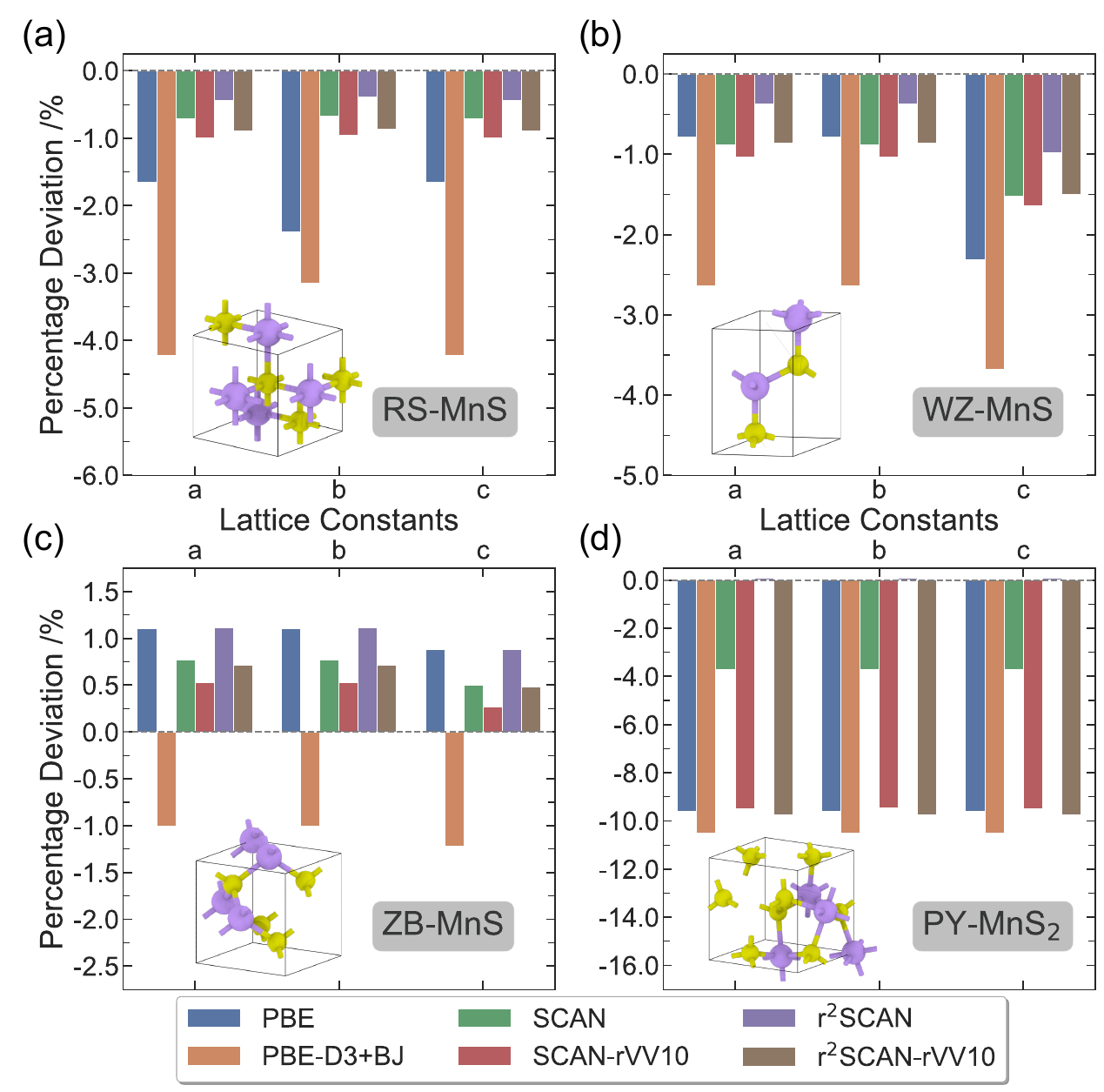}
    \caption{
Percentage deviations of DFT-predicted lattice constants from experimental values for (a) rock salt (RS) MnS, (b) wurtzite (WZ) MnS, (c) zinc blende (ZB) MnS, and (d) pyrite (PY) \ce{MnS2}.
Six exchange-correlation functionals are compared: PBE, PBE-D3+BJ, SCAN, SCAN-rVV10, r$^2$SCAN, and r$^2$SCAN-rVV10.
Numerical values are listed in Table~S1.
}
    \label{fig:fig2}
\end{figure}

Figure~\ref{fig:fig2} compares predicted lattice constants with experimental values\cite{Zagorac:in5024} for the three MnS polymorphs (RS, WZ, and ZB) as well as pyrite (PY) \ce{MnS2} (data in Table~S1).
For RS- and WZ-MnS, all functionals except PBE-D3+BJ slightly underestimate the lattice constants; r$^2$SCAN achieves the closest agreement, with deviations below 1\%.
For ZB-MnS, deviations are comparably small across all functionals ($\approx$1\%), with SCAN performing marginally better.
The greatest differentiation among functionals occurs for PY-\ce{MnS2}, where pure meta-GGA functionals substantially outperform GGA-PBE; r$^2$SCAN nearly perfectly reproduces the lattice constants, with the \ce{Mn^2+} magnetic moment correctly converging to the high-spin value.
Overall, these results establish the reliability of meta-GGA functionals, particularly r$^2$SCAN, for predicting the solid-state structures of MnS.

\begin{figure*}[htb]
    \centering
    \includegraphics[width=1.0\linewidth]{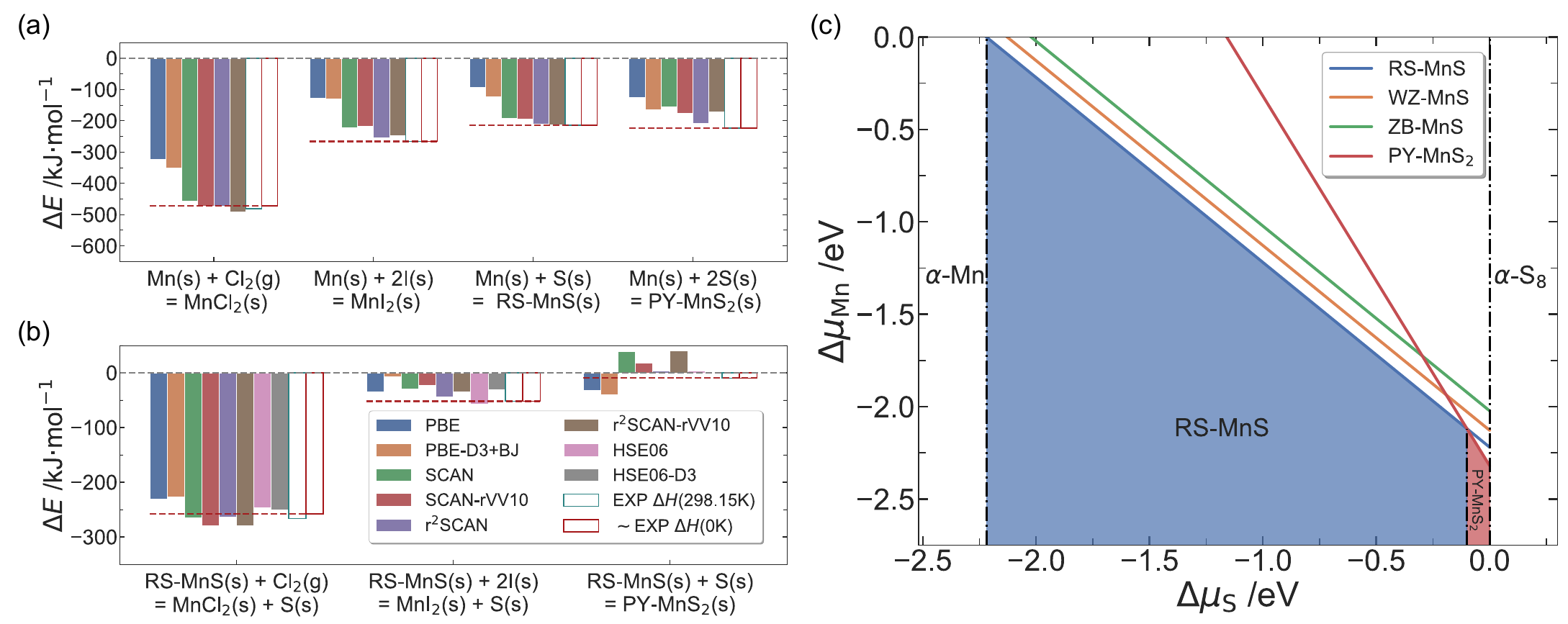}
    \caption{
Comparison of calculated and experimental (a) formation energies for \ce{MnCl2}, \ce{MnI2}, rock salt (RS) MnS, and pyrite (PY) \ce{MnS2}, and (b) reaction energies of RS-MnS with \ce{Cl2}, iodine (I), and S.
In both panels, filled bars correspond to six exchange-correlation functionals (PBE, PBE-D3+BJ, SCAN, SCAN-rVV10, r$^2$SCAN, and r$^2$SCAN-rVV10) and two hybrid-functional benchmarks (HSE06 and HSE06-D3); unfilled green bars denote experimental reaction enthalpies at 298.15~K, and unfilled red bars with red dashed lines extending from their bases indicate experimental values extrapolated to 0~K.
(c) Phase diagram of bulk MnS polymorphs as a function of the relative chemical potentials $\Delta \mu_{\ce{Mn}}$ and $\Delta \mu_{\ce{S}}$.
Solid lines are derived from formation energies; vertical dash-dotted lines denote polymorph phase boundaries.
}
    \label{fig:fig3}
\end{figure*}

We next benchmarked thermochemical accuracy by computing formation energies for \ce{MnCl2}, \ce{MnI2}, RS-MnS, and PY-\ce{MnS2}, as well as reaction energies of RS-MnS with \ce{Cl2}, iodine (I), and S.
As illustrated in Figures~\ref{fig:fig3}a~and~\ref{fig:fig3}b, experimental reaction enthalpies measured at room temperature (unfilled green bars) were extrapolated to 0~K (unfilled red bars) to serve as references, with only gaseous-species contributions to the heat capacity taken into account.\cite{thomas_c_allison_nist-janaf_2013}
Reaction energies from single-point self-consistent field (SCF) calculations using the Heyd--Scuseria--Ernzerhof (HSE06) hybrid functional\cite{krukau_influence_2006} (filled pink bars) and its vdW variant HSE06-D3\cite{moellmann_dft-d3_2014} (filled gray bars), both on r$^2$SCAN-optimized structures, serve as theoretical benchmarks.
For the three reactions involving Mn, \ce{Cl2}, and S that yield \ce{MnCl2} or RS-MnS (see Figures~\ref{fig:fig3}a~and~\ref{fig:fig3}b), r$^2$SCAN (filled purple bars) is nearly identical to the 0~K experimental values.
For the remaining four reactions, r$^2$SCAN predictions are slightly higher (by up to 16.80 kJ$\cdot$mol$^{-1}$) but remain the most accurate among all candidate functionals.
Notably, r$^2$SCAN-rVV10 yields slightly less accurate reaction energies than r$^2$SCAN, consistent with its reported performance for solid-state thermodynamics.\cite{kothakonda_testing_2023}

Using the benchmarked energies, we constructed a phase diagram for bulk MnS polymorphs as a function of the relative chemical potentials $\Delta \mu_{\ce{S}}$ and $\Delta \mu_{\ce{Mn}}$ (defined in Equation~\ref{3rd_eq}; see Methods).
From the 298.15~K experimental formation enthalpies of RS-MnS and PY-\ce{MnS2},\cite{thermochemical_data_barin} we derived the linear relationships between $\Delta \mu_{\ce{S}}$ and $\Delta \mu_{\ce{Mn}}$ shown as the blue and red lines in Figure~\ref{fig:fig3}c.
Their intersections with the elemental stability boundaries define the thermodynamically stable region for bulk RS-MnS: $-2.22$~eV~$\leq \Delta \mu_{\ce{S}} \leq -0.10$~eV (blue shaded region).
The lower $\Delta \mu_{\ce{S}}$ limit corresponds to Mn-rich/S-poor conditions, where formation of elemental Mn would become thermodynamically allowed, whereas the upper limit corresponds to S-rich conditions, where formation of the sulfur-rich PY-\ce{MnS2} phase would be favored.
We therefore restrict the morphology predictions of RS-MnS NCs to this stability window.
Since experimental formation enthalpies for WZ- and ZB-MnS are unavailable, we estimated them by combining the formation enthalpy of RS-MnS with polymorph energy differences from single-point HSE06-D3 calculations on r$^2$SCAN-optimized structures.
The resulting stability ranges are $-2.13$~eV~$\leq \Delta \mu_{\ce{S}} \leq -0.19$~eV for WZ-MnS and $-2.03$~eV~$\leq \Delta \mu_{\ce{S}} \leq -0.30$~eV for ZB-MnS (orange and green lines in Figure~\ref{fig:fig3}c), and the corresponding morphology predictions are likewise restricted to these polymorph-specific windows.

In summary, r$^2$SCAN provides the best overall agreement with experimental lattice constants and reaction energies among the functionals tested and was therefore adopted for subsequent surface energy calculations.

\subsubsection{Surface energy corrections with \texorpdfstring{r$^{\text{2}}$SCAN+$U$}{r²SCAN+U}}\label{surf-bench}

Surface energy calculations (Equations~\ref{4th_eq}~and~\ref{5th_eq}) involve slab optimizations, which are inherently more demanding for DFT than bulk calculations: dangling bonds on coordinatively unsaturated surface atoms give rise to localized electronic states that challenge semilocal XC functionals.
To assess the reliability of r$^2$SCAN for MnS surfaces, we optimized a series of RS-MnS slabs---including the nonpolar (010) and (100), the polar low-index (111), and the polar high-index (131) and (311) facets---and evaluated their surface energies.
HSE06 single-point SCF calculations on these r$^2$SCAN-optimized structures provide theoretical benchmarks for the r$^2$SCAN results.

\begin{figure*}[htb]
    \centering
    \includegraphics[width=1.0\linewidth]{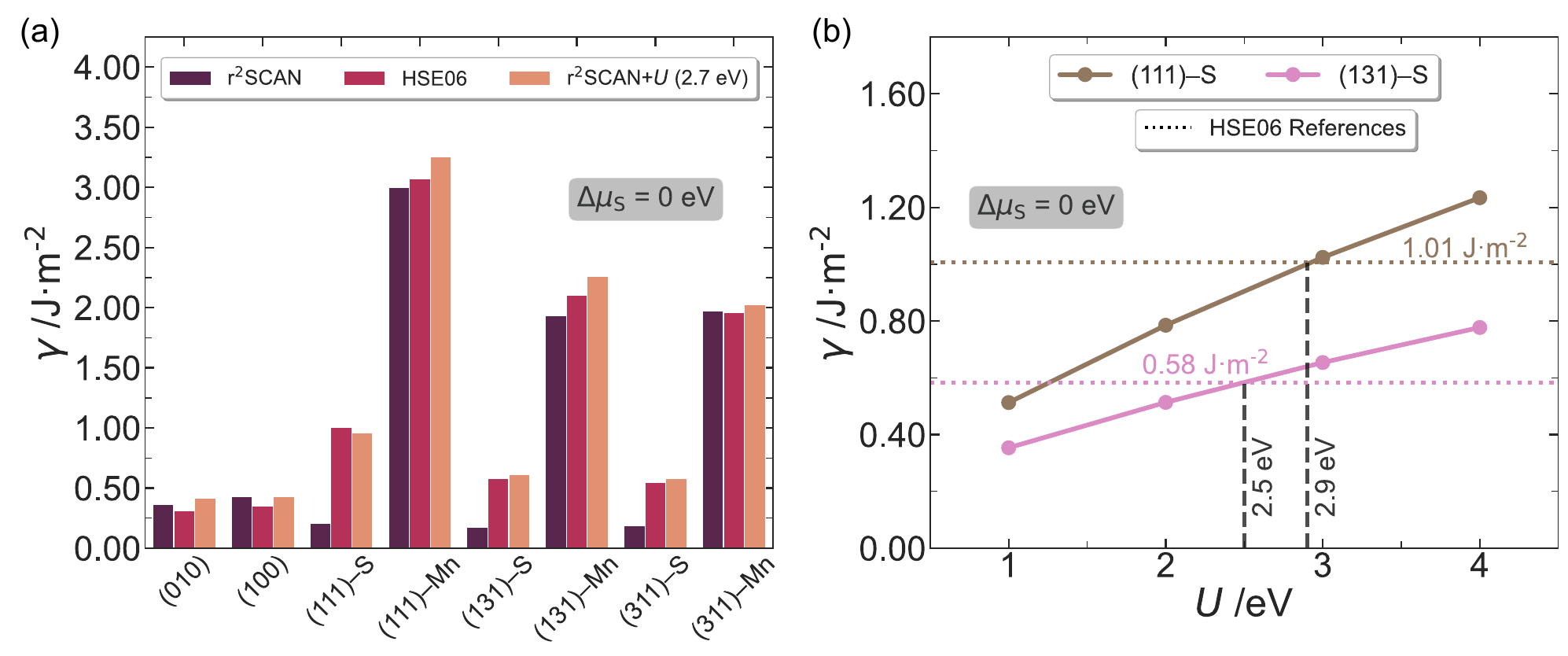}
    \caption{
(a) Surface energies of two nonpolar and six polar rock salt (RS) MnS facets calculated using r$^2$SCAN, HSE06, and r$^2$SCAN+$U$ ($U = 2.7$~eV).
S and Mn labels following the Miller indices denote the surface termination.
(b) Dependence of S-terminated (111) and (131) RS-MnS surface energies on the Hubbard $U$ value in r$^2$SCAN+$U$ calculations.
Horizontal dotted lines indicate the corresponding HSE06 surface energies, with values annotated; vertical dashed lines mark the optimal $U$ for each facet (2.9~eV for (111)--S and 2.5~eV for (131)--S) that minimizes the deviation from HSE06.
In both panels, polar surface energies are evaluated at $\Delta \mu_{\ce{S}} = 0$~eV.
}
    \label{fig:fig4}
\end{figure*}

The bulk energy is evaluated by fitting slab energies as a linear function of the number of atoms (Figure~S1 and Table~S2),\cite{fiorentini_extracting_1996} while the converged surface energies from r$^2$SCAN and HSE06 are compared in Figure~\ref{fig:fig4}a.
For the nonpolar (010) and (100) facets, the two methods agree to within 0.08~J$\cdot$m$^{-2}$.
r$^2$SCAN also slightly underestimates the surface energies of Mn-terminated polar facets, though the discrepancies are small relative to their large magnitudes.
In contrast, r$^2$SCAN dramatically underestimates the surface energies of S-terminated polar facets, by nearly a factor of five for the (111)--S facet.
S-terminated polar surfaces therefore present a greater challenge for semilocal r$^2$SCAN than nonpolar or Mn-terminated surfaces (Subsection S2.7).
We therefore adopted the r$^2$SCAN+$U$ approach to correct this deficiency.

To determine the appropriate Hubbard $U$ for Mn 3d states, we varied $U$ from 1~to~4~eV and performed r$^2$SCAN+$U$ single-point SCF calculations on the r$^2$SCAN-optimized (111)--S and (131)--S slabs.
As shown in Figure~\ref{fig:fig4}b, matching the r$^2$SCAN+$U$ surface energies to the HSE06 references yields optimal $U$ values of 2.9~eV and 2.5~eV for the (111)--S and (131)--S facets, respectively.
We adopted $U=2.7$~eV, the average of these two values, as a balanced correction.
This value coincides with the $U = 2.7$~eV determined independently by Gautam and Carter for Mn oxides within the SCAN+$U$ framework, where $U$ was fitted to experimental oxidation enthalpies rather than hybrid DFT surface energies.\cite{sai_gautam_evaluating_2018}
The convergence of these two approaches, which used different fitting targets and different materials classes, suggests that $U = 2.7$~eV is an appropriate correction for the thermochemistry of Mn compounds, rather than reflecting an artifact of our fitting procedure.
However, as an effective correction, the optimal $U$ value may depend on the calibration target, such as bulk electronic structure properties (Figure~S2).
We further tested its transferability across facets by calculating corrected surface energies for all considered RS-MnS facets.

As shown in Figure~\ref{fig:fig4}a, the $U=2.7$~eV correction exerts only a minor influence on the nonpolar surface energies and slightly overestimates those of Mn-terminated facets, while substantially correcting the S-terminated polar surface energies.
The slight overestimation of Mn-terminated facet energies has negligible impact on the predicted morphology because these facets already possess high surface energies and contribute minimally to the Wulff shape.
The corrections to S-terminated facets, by contrast, are consequential: without them, qualitatively different morphologies are predicted (compare Figure~\ref{fig:fig5}, Figure~S3, and Figure~S4).
Furthermore, for representative polar ZB- and WZ-MnS slabs, r$^2$SCAN+$U$ reduces the deviation of the combined surface energy from HSE06 from $-0.49$ to +0.12 J$\cdot$m$^{-2}$ (ZB) and from $-0.50$ to +0.22 J$\cdot$m$^{-2}$ (WZ) (Table S3), which provides preliminary support for applying the same correction across the three polymorphs.
We therefore transfer the RS-derived $U = 2.7$~eV value to the ZB- and WZ-MnS surface calculations.

In summary, r$^2$SCAN alone markedly underestimates S-terminated polar surface energies, but applying a Hubbard $U=2.7$~eV correction to the Mn 3d states brings the RS-MnS surface energies into close agreement with HSE06.
Having established r$^2$SCAN optimization followed by r$^2$SCAN+$U$ correction as a reliable framework for MnS surface calculations, we now apply it systematically to predict equilibrium morphologies across all three polymorphs as a function of $\Delta\mu_{\ce{S}}$.
Thermodynamic Wulff constructions are performed for each specified polymorph and do not account for transformations between polymorphs, which lie beyond the scope of the approach used here.\cite{barnard_model_2004, boukouvala_approaches_2021}

\subsection{Predicted equilibrium morphologies}

\subsubsection{RS-MnS}

\begin{figure*}[htb]
    \centering
    \includegraphics[width=1.0\linewidth]{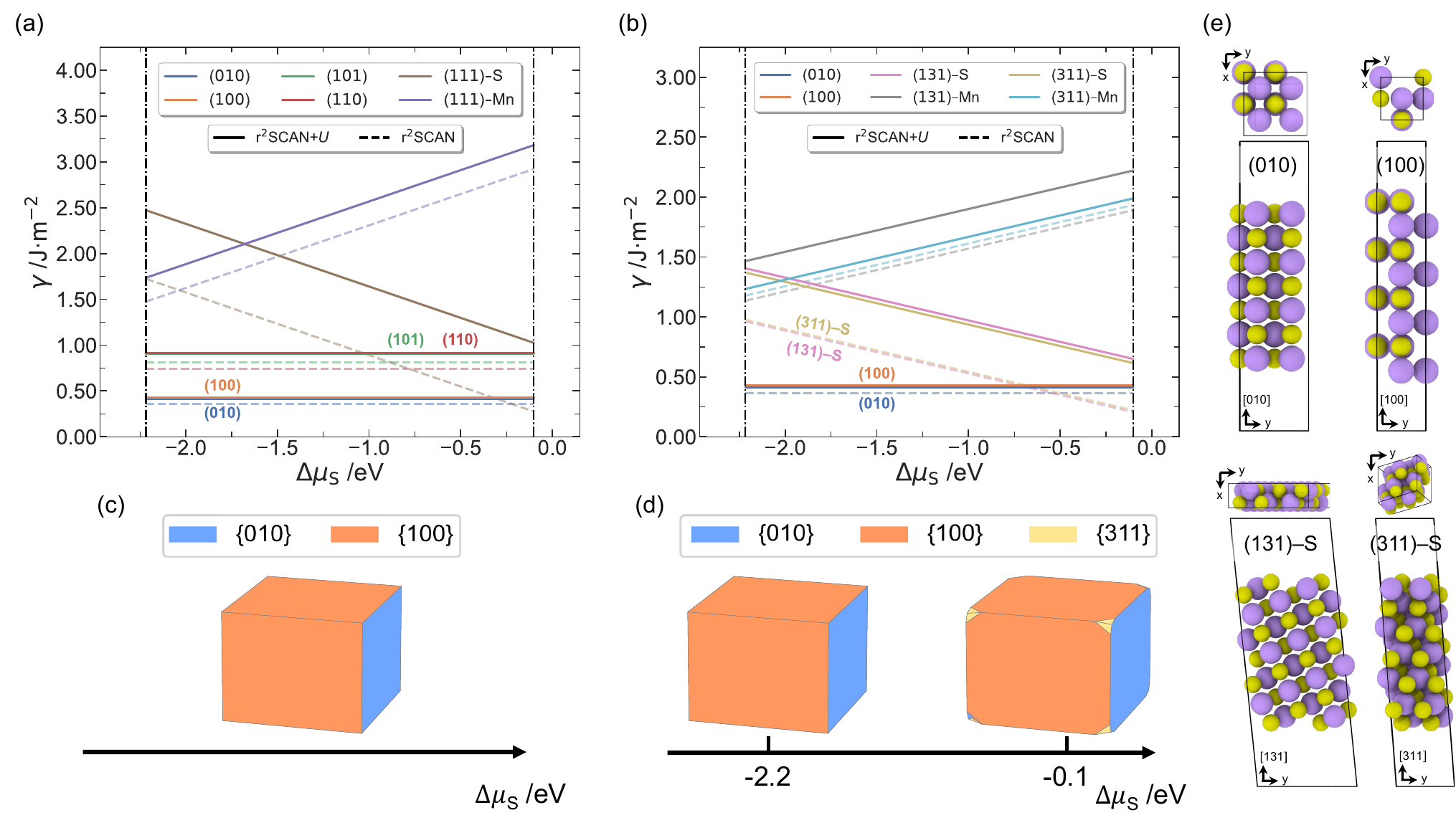}
    \caption{
Dependence of rock salt (RS) MnS surface energies on the relative chemical potential of sulfur $\Delta \mu_{\ce{S}}$ for (a) low-index facets and (b) the six lowest-energy facets among all $\{h,k,l\} \leq 3$ facets.
Solid and dashed lines represent r$^2$SCAN+$U$ and r$^2$SCAN results, respectively.
The $\Delta \mu_{\ce{S}}$ range between the two vertical dash-dotted lines ($-2.22$~eV $\leq \Delta \mu_{\ce{S}} \leq -0.10$~eV) indicates the thermodynamically stable region of bulk RS-MnS.
Wulff constructions of RS-MnS nanocrystals considering (c) only low-index facets and (d) all facets with Miller indices up to 3.
Each Wulff shape corresponds to the $\Delta \mu_{\ce{S}}$ value indicated beneath it.
If no specific $\Delta \mu_{\ce{S}}$ value is given, the shape is invariant across the entire stability window.
(e) Four representative RS-MnS slab models showing the (010), (100), (131)--S, and (311)--S facets.
}
    \label{fig:fig5}
\end{figure*}

We now examine the surface energies and equilibrium morphologies of RS-MnS as a function of $\Delta \mu_{\ce{S}}$, considering facets with Miller indices up to 3 (Table~S4).
Here, ($hkl$) denotes an individual crystallographic facet, [$hkl$] represents a crystallographic direction (normal to ($hkl$) in cubic systems), and $\{hkl\}$ refers to the family of symmetry-equivalent facets.
The high symmetry of RS-MnS (space group $Fm\overline{3}m$), which includes an inversion center, permits symmetric slabs with identical terminations along any cleavage direction, so individual surface energies can be calculated directly from Equation~\ref{fig:fig5}.
Low-index facets ($\{h,k,l\} \leq 1$) generally possess fewer dangling bonds and correspondingly lower surface energies, so they tend to dominate the equilibrium morphology.
To assess consistency, we compare Wulff constructions derived from low-index facets alone with those including all facets considered.
Counterpart r$^2$SCAN results are provided in Supplementary Information (Figure~S3 and Table~S5 in Subsection~S2.1).

Figure~\ref{fig:fig5}a shows the low-index surface energies as a function of $\Delta \mu_{\ce{S}}$ within the RS-MnS stability window.
Nonpolar facets (such as the (010) slab in Figure~\ref{fig:fig5}e) are stoichiometric and therefore exhibit $\Delta \mu_{\ce{S}}$-independent surface energies.
S-terminated polar facets, by contrast, decrease in surface energy with increasing $\Delta \mu_{\ce{S}}$, while Mn-terminated polar facets increase.
Throughout the stability window, r$^2$SCAN+$U$ consistently predicts the (100) and (010) facets, which are slightly inequivalent due to the antiferromagnetic ordering of RS-MnS, to have the lowest surface energies, and the corresponding Wulff constructions confirm that nanocubes are the only equilibrium morphology (Figure~\ref{fig:fig5}c).
Including high-index facets does not alter this conclusion: (100) and (010) remain the lowest-energy surfaces (Figure~\ref{fig:fig5}b), and nanocubic RS-MnS dominates the resulting Wulff constructions across nearly the entire $\Delta \mu_{\ce{S}}$ range (Figure~\ref{fig:fig5}d and Table~S6).
Only at the S-rich limit, approaching the RS-MnS $\rightarrow$ PY-\ce{MnS2} transition, do the nanocubes become slightly truncated by S-terminated \{311\} facets.
Notably, the r$^2$SCAN+$U$ Wulff constructions are consistent regardless of whether high-index facets are included, unlike the uncorrected r$^2$SCAN results, which yield qualitatively different morphologies when high-index facets are considered (Figures~S3c~and~S3d).

The r$^2$SCAN+$U$ results thus predict that RS-MnS NCs favor cubic morphologies across nearly the entire stability window.
Nanocubes are indeed commonly observed experimentally across multiple synthesis routes, including aqueous solution,\cite{chen_low-temperature_2023} hydrothermal,\cite{michel_hydrothermal_2006, zhang_hydrothermal_2008} solvothermal,\cite{yang_size-controlled_2012} and colloidal\cite{jun_architectural_2002, gendler_halide-driven_2023} syntheses.
However, octahedral RS-MnS NCs have also been reported in hydrothermal studies\cite{zhang_hydrothermal_2008, michel_hydrothermal_2006, puglisi_monodisperse_2010, gui_hydrothermal_2011} and in aqueous solution synthesis,\cite{chen_low-temperature_2023} a morphology not captured by our vacuum Wulff shapes.
The roles of solvent polarity and ligand adsorption in stabilizing non-cubic morphologies are addressed in the Discussion.

\begin{figure*}[htb]
    \centering
    \includegraphics[width=1.0\linewidth]{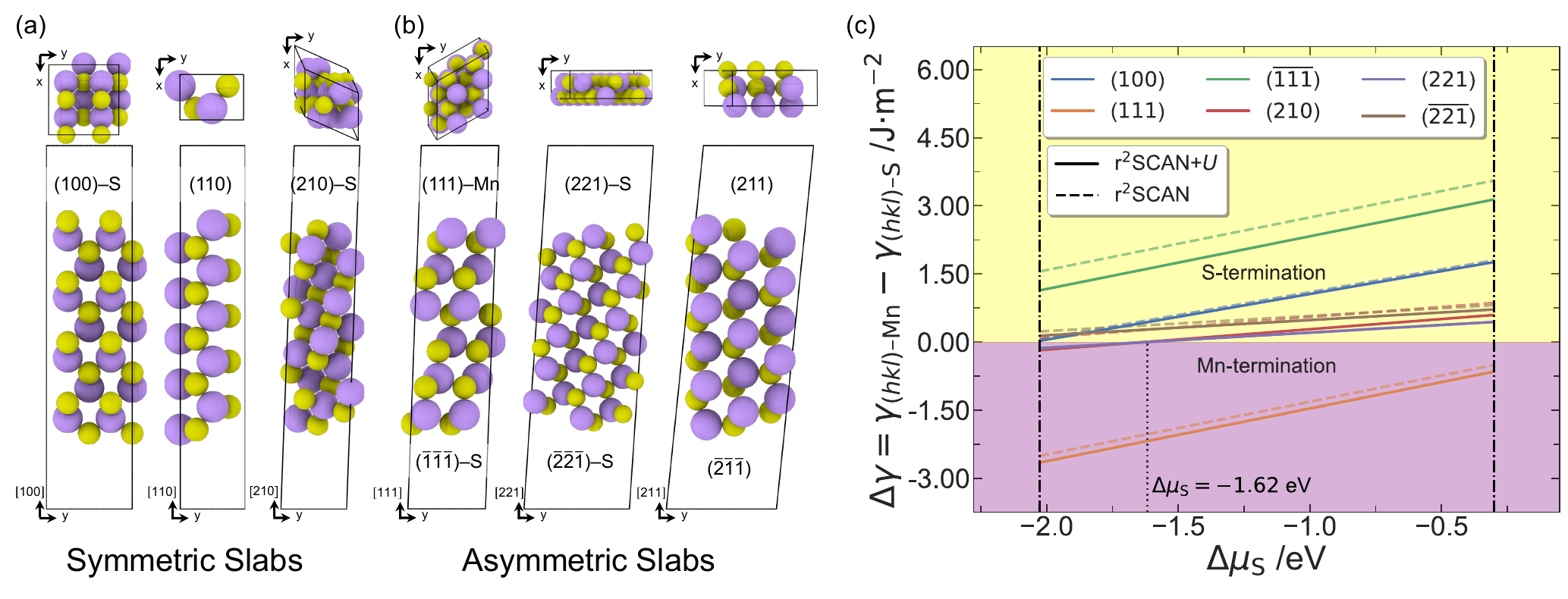}
    \caption{
Representative slab models for (a) symmetric and (b) asymmetric zinc blende (ZB) MnS surfaces.
(c) Termination stability comparison for six ZB-MnS facets: (100), (111), ($\overline{111}$), (210), (221), and ($\overline{221}$).
The yellow region ($\Delta \gamma > 0$) indicates that the S-terminated configuration is more stable, and the purple region ($\Delta \gamma < 0$) indicates that the Mn-terminated configuration is more stable.
Solid and dashed lines represent r$^2$SCAN+$U$ and r$^2$SCAN results, respectively.
The $\Delta \mu_{\ce{S}}$ interval between the two vertical dash-dotted lines ($-2.03$~eV $\leq \Delta \mu_{\ce{S}} \leq -0.30$~eV) indicates the thermodynamically stable region of bulk ZB-MnS; the vertical dotted line marks the $\Delta \mu_{\ce{S}}$ value ($-1.62$~eV) at which the (221) and (210) facets undergo a termination transition from Mn to S.
}
    \label{fig:fig6}
\end{figure*}

\subsubsection{ZB-MnS}

ZB-MnS crystallizes in the $F\overline{4}3m$ space group, which lacks centrosymmetry and produces a net dipole along certain stacking directions.
Symmetric slabs therefore cannot be constructed for all facets (Figures~\ref{fig:fig6}a~and~\ref{fig:fig6}b), precluding direct application of Equation~\ref{5th_eq}.
We instead adopt a combined slab-and-wedge approach (see Methods and Figure~\ref{fig:fig1}c) to obtain individual polar surface energies.
Facets with Miller indices up to 2 are considered (Table~S7), with all surface energies calculated using r$^2$SCAN+$U$ ($U=2.7$~eV).
As for RS-MnS, we compare Wulff constructions from low-index facets alone with those including all considered facets.
Counterpart r$^2$SCAN results are provided in Supplementary Information (Figure~S5 and Table~S8 in Subsection~S2.3).

Nonstoichiometric facets in ZB-MnS can adopt multiple surface terminations; for example, the (100) and (111) facets (Figures~\ref{fig:fig6}a~and~\ref{fig:fig6}b) may be either S- or Mn-terminated.
We therefore first determine the preferred termination for each polar facet.
Within the ZB-MnS stability window (Figure~\ref{fig:fig6}c), the (111) facet remains Mn-terminated, while the (100), ($\overline{111}$), and ($\overline{221}$) facets prefer S-termination throughout.
The (210) and (221) facets undergo a transition from Mn- to S-termination at $\Delta \mu_{\ce{S}} = -1.62$~eV, which may affect the Wulff shape if these facets have sufficiently low surface energies to be exposed.

\begin{figure*}[htb]
    \centering
    \includegraphics[width=1.0\linewidth]{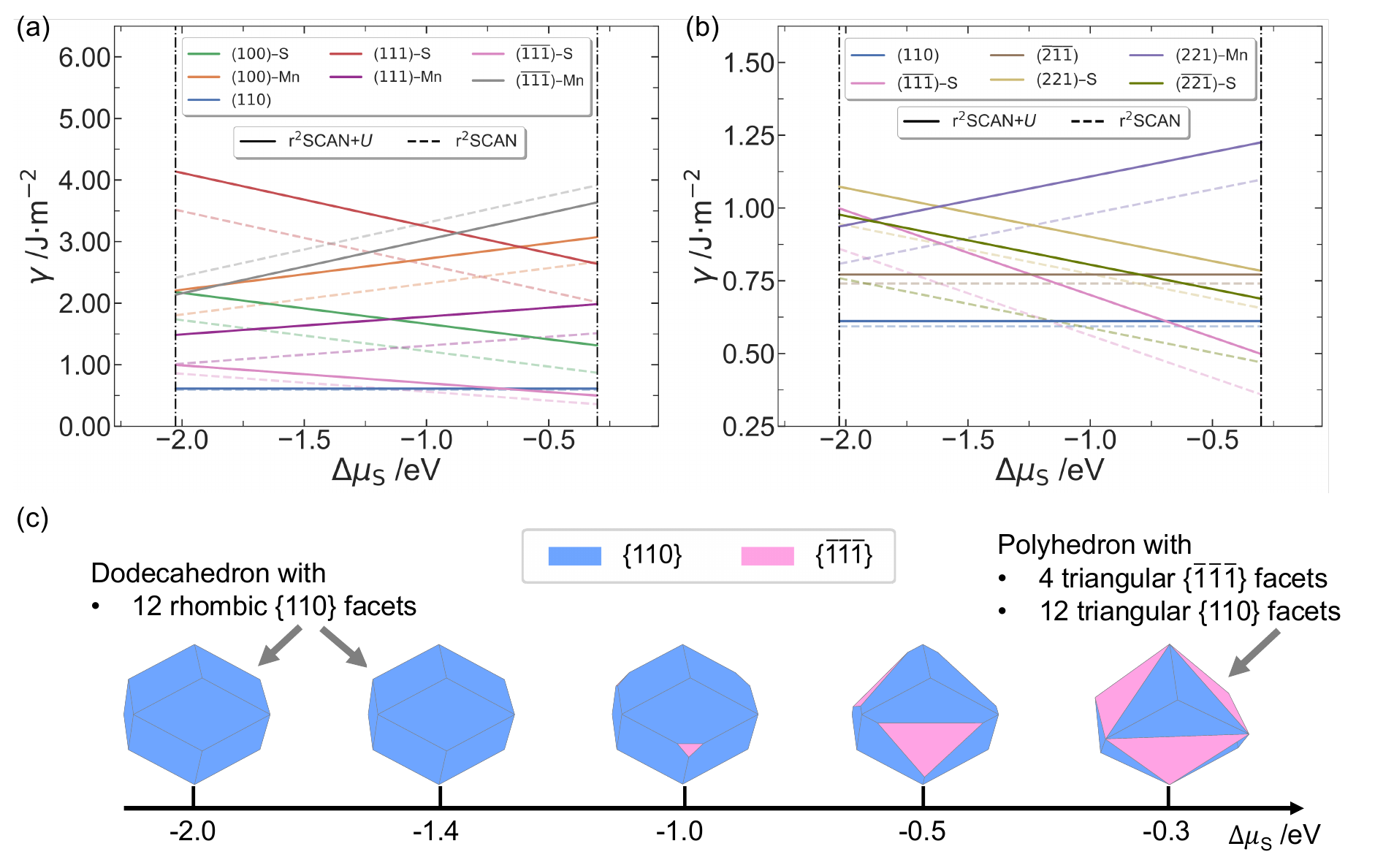}
    \caption{
Dependence of zinc blende (ZB) MnS surface energies on the relative chemical potential of sulfur $\Delta \mu_{\ce{S}}$ for (a) low-index facets and (b) the six lowest-energy facets among all $\{h,k,l\} \leq 2$ facets.
Solid and dashed lines represent r$^2$SCAN+$U$ and r$^2$SCAN results, respectively.
The $\Delta \mu_{\ce{S}}$ interval between the two vertical dash-dotted lines ($-2.03$~eV $\leq \Delta \mu_{\ce{S}} \leq -0.30$~eV) indicates the thermodynamically stable region of bulk ZB-MnS.
(c) Identical Wulff constructions of ZB-MnS nanocrystals obtained by considering only low-index facets or all facets with Miller indices up to 2.
Each Wulff shape corresponds to the $\Delta \mu_{\ce{S}}$ value indicated beneath it.
}
    \label{fig:fig7}
\end{figure*}

Figure~\ref{fig:fig7}a shows the low-index surface energies across the ZB-MnS stability window, with the (110) and ($\overline{111}$)--S facets being the two most stable facets.
The resulting morphological evolution (Figure~\ref{fig:fig7}c) proceeds from rhombic dodecahedra enclosed by \{110\} facets under Mn-rich conditions, to 16-faced polyhedra bounded by four triangular \{$\overline{111}$\} facets and twelve triangular \{110\} facets under S-rich conditions.
When high-index facets are included (Figure~\ref{fig:fig7}b), the (110) and ($\overline{111}$)--S facets remain the lowest-energy surfaces under Mn-rich and S-rich conditions, respectively, with the crossover at $\Delta \mu_{\ce{S}} = -0.69$~eV.
The full Wulff constructions (Figure~\ref{fig:fig7}c and Table~S9) confirm that the overall morphological evolution is fully consistent with the results of the low-index analysis.
Notably, in agreement with the observations for RS-MnS, r$^2$SCAN+$U$ substantially corrects the description of S-terminated ZB-MnS facets. 
Consequently, the inclusion of high-index facets does not alter the predicted Wulff shapes, whereas the uncorrected r$^2$SCAN yields inconsistent morphologies featuring spurious \{221\}--S and \{$\overline{221}$\}--S facets (Figures~S5d~and~S5e).

In summary, the nonpolar \{110\} and polar \{$\overline{111}$\}--S facets possess the lowest surface energies under Mn-rich and S-rich conditions, respectively, driving a morphological transition from rhombic dodecahedra (Mn-rich) to 16-faced polyhedra (S-rich).
Although ZB-MnS NCs have been synthesized by colloidal cation exchange\cite{anie.201701087} and solvothermal\cite{yang_polymorphism_2012, HAO20179} routes, well-faceted ZB-MnS NCs with identifiable equilibrium morphologies have, to our knowledge, not yet been reported.

\begin{figure*}[htb]
    \centering
    \includegraphics[width=1.0\linewidth]{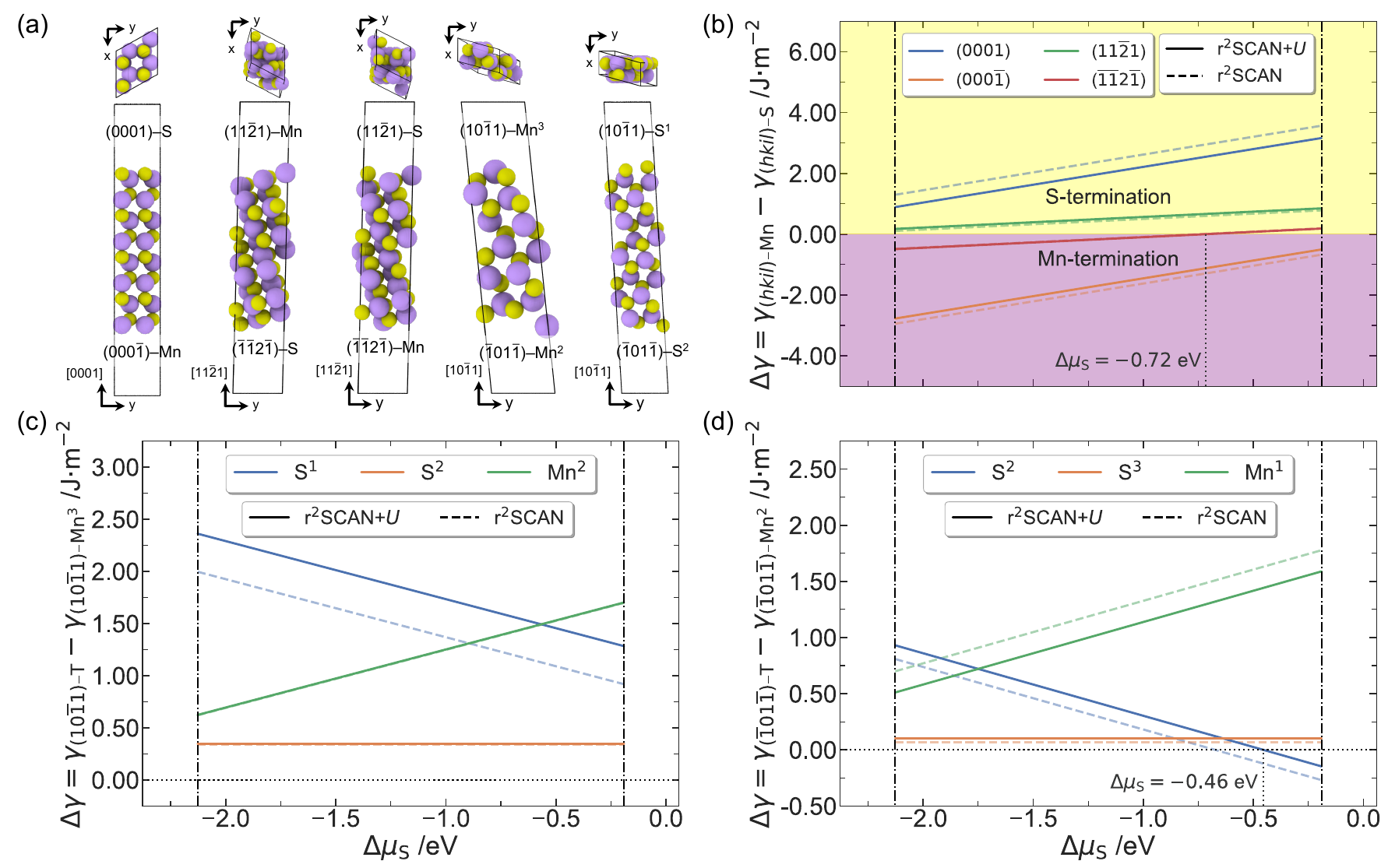}
    \caption{
(a) Representative asymmetric slab models for wurtzite (WZ) MnS polar surfaces.
(b) Termination stability comparison between S-terminated and Mn-terminated configurations for the (0001), ($000\overline{1}$), ($11\overline{2}1$), and ($\overline{11}2\overline{1}$) WZ-MnS polar facets.
The yellow region ($\Delta \gamma > 0$) indicates that the S-terminated configuration is more stable, and the purple region ($\Delta \gamma < 0$) indicates that the Mn-terminated configuration is more stable.
The $\Delta \mu_{\ce{S}}$ interval between the two vertical dash-dotted lines ($-2.13$~eV $\leq \Delta \mu_{\ce{S}} \leq -0.19$~eV) indicates the thermodynamically stable region of bulk WZ-MnS.
Termination stability comparison of four configurations for the (c) ($10\overline{1}1$) and (d) ($\overline{1}01\overline{1}$) facets; for each facet, terminations are labeled by the species and coordination number of the outermost surface atom: S$^1$, S$^2$, Mn$^2$, and Mn$^3$ for ($10\overline{1}1$); S$^2$, S$^3$, Mn$^1$, and Mn$^2$ for ($\overline{1}01\overline{1}$).
Solid and dashed lines represent r$^2$SCAN+$U$ and r$^2$SCAN results, respectively.
The vertical dotted lines mark the $\Delta \mu_{\ce{S}}$ values ($-0.72$~eV and $-0.46$~eV) at which the ($\overline{11}2\overline{1}$) and ($\overline{1}01\overline{1}$) facets undergo termination transitions from Mn to S, respectively.
}
    \label{fig:fig8}
\end{figure*}

\subsubsection{WZ-MnS}

WZ-MnS (space group $P6_3mc$) lacks both inversion and mirror symmetry along [0001], so the wedge-based strategy used for ZB-MnS cannot yield individual polar surface energies.
We instead compute relative surface energies using combined slab and wedge models (see Methods and Figure~\ref{fig:fig1}d).
Stacking along any [$hkil$] direction with $l \neq 0$ produces a net dipole and multiple termination configurations (Figure~\ref{fig:fig8}a), substantially increasing computational cost.
We therefore restrict the analysis to facets with Miller indices up to 1 (Table~S10).
All surface energies are calculated using r$^2$SCAN+$U$ ($U=2.7$~eV); counterpart r$^2$SCAN results are provided in Supplementary Information (Figure~S6 and Table~S11 in Subsection~S2.5).

\begin{figure*}[htb]
    \centering
    \includegraphics[width=1.0\linewidth]{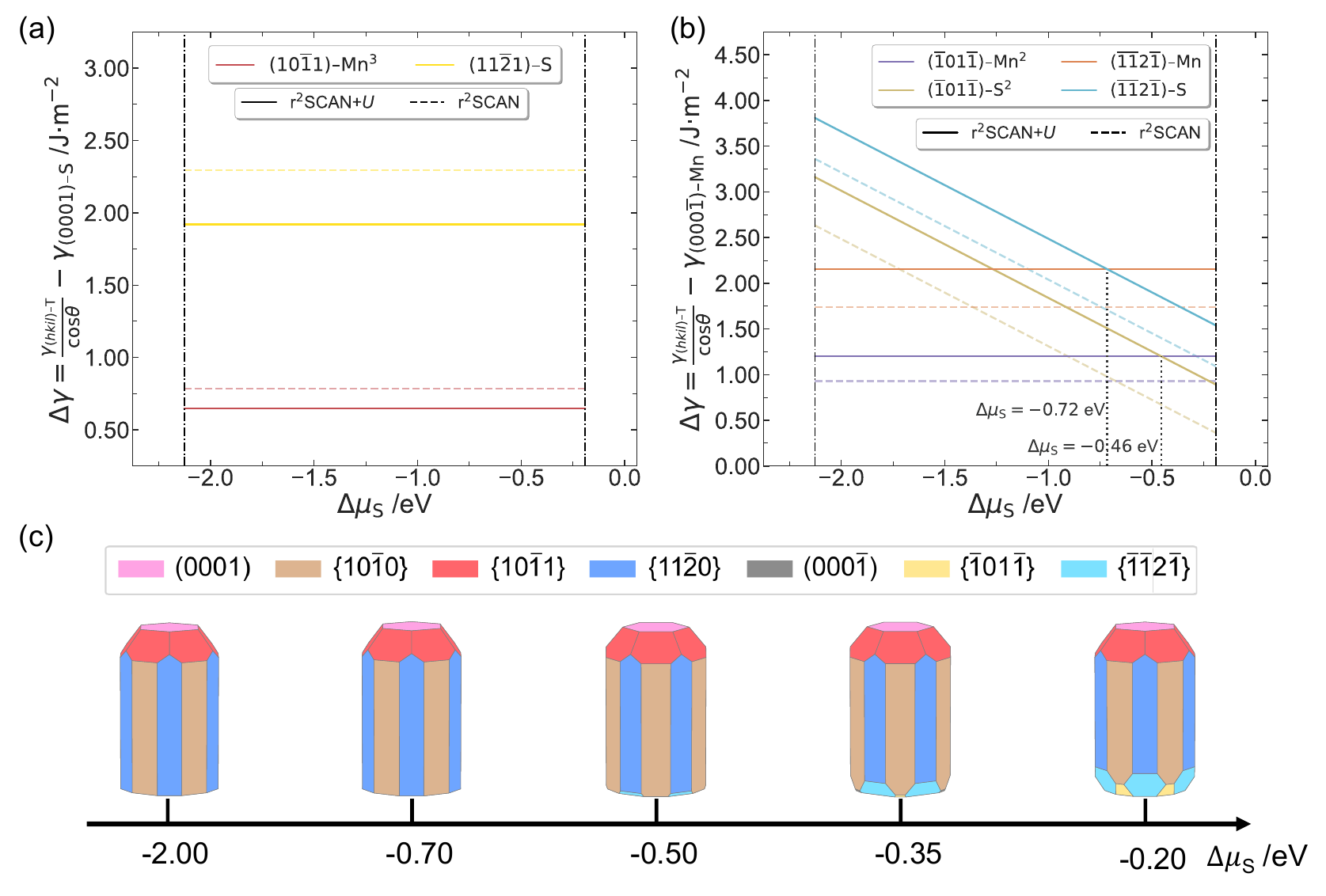}
    \caption{
Dependence of wurtzite (WZ) MnS relative surface energies $\Delta \gamma$ on the relative chemical potential of sulfur $\Delta \mu_{\ce{S}}$ along (a) the [0001] direction and (b) the [$000\overline{1}$] direction.
Solid and dashed lines represent r$^2$SCAN+$U$ and r$^2$SCAN results, respectively.
The $\Delta \mu_{\ce{S}}$ interval between the two vertical dash-dotted lines ($-2.13$~eV $\leq \Delta \mu_{\ce{S}} \leq -0.19$~eV) indicates the thermodynamically stable region of bulk WZ-MnS.
The vertical dotted lines in (b) mark the $\Delta \mu_{\ce{S}}$ values ($-0.72$~eV and $-0.46$~eV) at which the ($\overline{11}2\overline{1}$) and ($\overline{1}01\overline{1}$) facets undergo termination transitions from Mn to S, respectively.
(c) Wulff constructions of WZ-MnS nanocrystals considering only low-index facets.
Each Wulff shape corresponds to the $\Delta \mu_{\ce{S}}$ value indicated beneath it.
}
    \label{fig:fig9}
\end{figure*}

We next determine the preferred termination for each nonstoichiometric low-index facet.
Along [0001] and [11$\overline{2}$1] and their opposites, only one S- and one Mn-terminated surface exist, whereas four distinct terminations arise along [10$\overline{1}$1] and [$\overline{1}$01$\overline{1}$], labeled S$^1$, S$^2$, Mn$^2$, and Mn$^3$ for (10$\overline{1}$1) and S$^2$, S$^3$, Mn$^1$, and Mn$^2$ for ($\overline{1}$01$\overline{1}$) by the species and coordination number of the outermost surface atoms.
Among the facets along [0001] and [11$\overline{2}$1] (Figure~\ref{fig:fig8}b), Mn-termination is consistently favored on (000$\overline{1}$), while (0001) and (11$\overline{2}$1) are always S-terminated.
The ($\overline{11}$2$\overline{1}$) facet transitions from Mn- to S-termination at $\Delta \mu_{\ce{S}} = -0.72$~eV.

For (10$\overline{1}$1), which admits four distinct terminations, the Mn$^3$ configuration is most stable throughout the stability window (Figure~\ref{fig:fig8}c), while ($\overline{1}$01$\overline{1}$) transitions from Mn$^2$ to S$^2$ termination at $\Delta \mu_{\ce{S}} = -0.46$~eV (Figure~\ref{fig:fig8}d).
In summary, four polar facets---(0001)--S, (000$\overline{1}$)--Mn, (11$\overline{2}$1)--S, and (10$\overline{1}$1)--Mn$^3$---maintain a single preferred termination across the entire stability window, while ($\overline{11}$2$\overline{1}$) and ($\overline{1}$01$\overline{1}$) undergo Mn-to-S termination transitions at $\Delta \mu_{\ce{S}} = -0.72$~eV and $-0.46$~eV, respectively.

Using these preferred terminations, we calculated relative surface energies $\Delta \gamma$ along both polar directions.
Within the Wulff construction framework, a larger $\Delta \gamma$ corresponds to a greater distance from the crystal center along the reference crystallographic direction (i.e., [0001] or [000$\overline{1}$]), so facets with smaller $\Delta \gamma$ are preferentially exposed in the equilibrium morphology.
Along [0001] (Figure~\ref{fig:fig9}a), the $\Delta \gamma$ values for (10$\overline{1}$1)--Mn$^3$ and (11$\overline{2}$1)--S relative to (0001)--S are independent of $\Delta \mu_{\ce{S}}$, indicating that the top of the WZ-MnS NC is morphologically invariant.
The (10$\overline{1}$1)--Mn$^3$ facet has a substantially smaller $\Delta \gamma$ (0.65~J$\cdot$m$^{-2}$) than (11$\overline{2}$1)--S (1.92~J$\cdot$m$^{-2}$), so it is the facet exposed adjacent to (0001)--S in the Wulff shape.
Along [000$\overline{1}$] (Figure~\ref{fig:fig9}b), the morphology is $\Delta \mu_{\ce{S}}$-sensitive.
Under Mn-rich conditions, both ($\overline{1}$01$\overline{1}$)--Mn$^2$ and ($\overline{11}$2$\overline{1}$)--Mn exhibit constant $\Delta \gamma$ values.
As $\Delta \mu_{\ce{S}}$ increases, S-terminated configurations become preferred, specifically ($\overline{11}$2$\overline{1}$)--S at $\Delta \mu_{\ce{S}} = -0.72$~eV and ($\overline{1}$01$\overline{1}$)--S$^2$ at $\Delta \mu_{\ce{S}} = -0.46$~eV, progressively lowering $\Delta \gamma$ and stabilizing these facets.

The resulting Wulff constructions (Figure~\ref{fig:fig9}c and Table~S12) show that WZ-MnS NCs adopt rod-like morphologies.
The top of the rod, enclosed by (0001)--S and \{10$\overline{1}$1\}--Mn$^3$ facets, is invariant with $\Delta \mu_{\ce{S}}$, consistent with the $\Delta \mu_{\ce{S}}$-independent $\Delta \gamma$ values along [0001].
The lateral surfaces are dominated by the nonpolar \{$10\overline{1}0$\} and \{$11\overline{2}0$\} facets.
Only the base evolves: under Mn-rich conditions it is a flat (000$\overline{1}$)--Mn facet, but as $\Delta \mu_{\ce{S}}$ increases beyond $-0.50$~eV, progressive truncation by S-terminated \{$\overline{11}2\overline{1}$\} and \{$\overline{1}01\overline{1}$\} facets produces a polyhedral base.
Although \{$\overline{1}01\overline{1}$\} facets have lower $\Delta \gamma$ values than \{$\overline{11}2\overline{1}$\} at the S-rich limit, they occupy a smaller fraction of the base due to the smaller dihedral angle between \{$\overline{1}01\overline{1}$\} and the ($000\overline{1}$) facet.

In summary, WZ-MnS NCs are predicted to adopt rod-like morphologies whose tops remain invariant with $\Delta \mu_{\ce{S}}$ while the base progressively develops polyhedral truncation under S-rich conditions.
These predictions, encompassing both rod-like and bullet-like morphologies, are consistent with experimental observations for WZ-MnS NCs using hydrothermal,\cite{michel_hydrothermal_2006, gui_hydrothermal_2011, tang_morphology_2015} solvothermal,\cite{lu_metastable_2001} and colloidal syntheses.\cite{jun_architectural_2002, joo_generalized_2003, gendler_halide-driven_2023}
The roles of solvent polarity and ligand adsorption in modifying these morphologies are discussed further in the Discussion.

\subsection{Comparative morphological behavior across MnS polymorphs}

The preceding analysis reveals both universal principles and polymorph-specific behaviors governing MnS NC morphologies (Table~\ref{tab:polymorph-summary} and Subsection S2.8).

\begin{table*}[tb]
\renewcommand{\arraystretch}{1.5}
\caption{
Summary of predicted equilibrium morphological properties for rock salt (RS), zinc blende (ZB), and wurtzite (WZ) MnS nanocrystals.
For each polymorph, the space group, coordination geometry, principal exposed facets across the stability window, and equilibrium morphologies under Mn-rich and S-rich conditions are listed, along with the sensitivity of the morphology to the relative chemical potential of sulfur $\Delta \mu_{\ce{S}}$.
Principal facets and morphologies are derived from the r$^2$SCAN+$U$ Wulff constructions presented in Figures~\ref{fig:fig5},~\ref{fig:fig7},~and~\ref{fig:fig9}.
Experimental support indicates whether the predicted morphologies are consistent with synthesis data from this work or from the literature.
}
\label{tab:polymorph-summary}
\centering
\resizebox{\linewidth}{!}{
    \begin{tabular}{@{} l ccc @{}}
    \toprule
     & RS-MnS & ZB-MnS & WZ-MnS \\
    \midrule
    Space group             & $Fm\overline{3}m$ & $F\overline{4}3m$ & $P6_3mc$ \\
    Coordination            & Octahedral        & Tetrahedral        & Tetrahedral \\
    Principal facets        & \{100\}, \{010\}  & \{110\}, \{$\overline{111}$\}--S & \{10$\overline{1}$0\}, \{11$\overline{2}$0\} \\
    Morphology (Mn-rich)    & Nanocube          & Rhombic dodecahedron   & Nanorod (flat base) \\
    Morphology (S-rich)     & Nanocube          & 16-faced polyhedron & Nanorod (faceted base) \\
    $\Delta\mu_{\ce{S}}$ sensitivity & Negligible & High            & Moderate (base only) \\
    Experimental support & Yes (this work)   & Limited            & Yes (Refs.~\citenum{joo_generalized_2003}, \citenum{michel_hydrothermal_2006}, \citenum{gui_hydrothermal_2011}, \citenum{gendler_halide-driven_2023}) \\
    \bottomrule
    \end{tabular}
}
\end{table*}

\subsubsection{Intrinsic morphology trends}
The predicted equilibrium morphologies differ markedly among the three MnS polymorphs, reflecting their distinct coordination geometries and surface-energy landscapes.
In octahedrally coordinated RS-MnS (space group $Fm\overline{3}m$), the morphology is nearly independent of $\Delta\mu_{\ce{S}}$: nonpolar $\{100\}$ and $\{010\}$ families contribute $\sim$100\% of the surface area across nearly the entire $\Delta\mu_{\ce{S}}$ stability window (Table S6), giving nanocubes (Figure~\ref{fig:fig5}).
Polar solvents may further reduce the surface energy of polar facets, thereby modifying the equilibrium morphology of RS-MnS NCs beyond the nanocube predicted in our vacuum calculations.\cite{zhang_hydrothermal_2008, michel_hydrothermal_2006, puglisi_monodisperse_2010, gui_hydrothermal_2011, chen_low-temperature_2023}
By contrast, tetrahedrally coordinated ZB- and WZ-MnS are more sensitive to $\Delta\mu_{\ce{S}}$.
ZB-MnS shifts from 100\% $\{110\}$-faceted rhombic dodecahedra under Mn-rich conditions to 16-faced polyhedra with increasing \{$\overline{111}$\}--S exposure under S-rich conditions (Figure~\ref{fig:fig7} and Table~S9), whereas WZ-MnS remains rod-like with $\{10\bar{1}0\}$ and $\{11\bar{2}0\}$ prism facets comprising 65--73\% of total surface area and $\Delta\mu_{\ce{S}}$-dependent changes localized mainly at the rod termini (Figure~\ref{fig:fig9} and Table~S12).

\subsubsection{Implications for synthesis design}

These comparisons suggest that the equilibrium morphology of RS-MnS remains cubic across nearly the entire $\Delta\mu_{\ce{S}}$ stability window (though polar solvents may alter this), while ZB- and WZ-MnS morphologies are tunable via $\Delta\mu_{\ce{S}}$.
For ZB-MnS, achieving the 16-faced polyhedral morphology with 12 \{110\} and 4 \{$\overline{111}$\}--S facets requires S-rich conditions approaching the stability boundary; for WZ-MnS, the rod aspect ratio should be relatively insensitive to $\Delta\mu_{\ce{S}}$, but the base truncation geometry is tunable.

In experiments, \(\Delta\mu_{\ce{S}}\) represents the effective sulfur richness of the reaction environment and is not set by any single synthesis parameter.
Precursor identity and ratio, temperature, and pressure can collectively shift the system toward Mn- or S-rich conditions.
For example, increasing the relative amount of sulfur precursor can promote a more S-rich chemical environment.
Providing sufficient thermal energy and reaction time is expected to make the experimental morphology more likely to approach the thermodynamic Wulff shape by enabling shape equilibration.\cite{michel_hydrothermal_2006}
While our \(\Delta\mu_{\mathrm{S}}\)-dependent predictions capture thermodynamic morphology trends, specific experimental outcomes may also reflect solvent polarities, ligand adsorption, and kinetic factors that are not explicitly included in the present framework.
To assess the validity of this computational framework and quantify deviations in real nanocrystal systems, we next turn to experimental measurements on RS-MnS NCs.

\begin{figure*}[htb]
    \centering
    \includegraphics[width=1.0\linewidth]{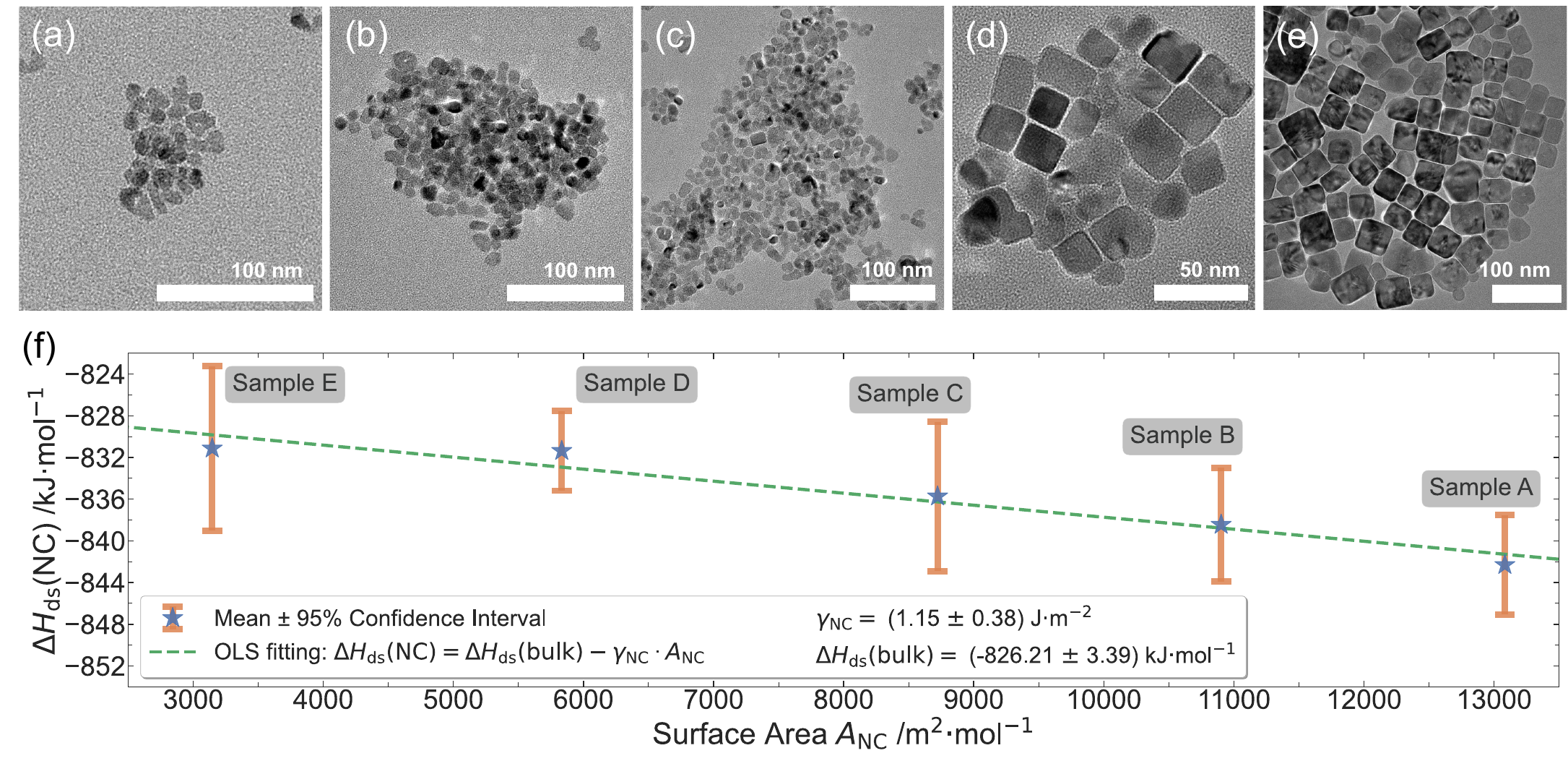}
    \caption{
(a--e) Transmission electron microscopy (TEM) images of rock salt (RS) MnS nanocrystals of increasing size: (a) sample A, $\sim$10~nm; (b) sample B, $\sim$12~nm; (c) sample C, $\sim$15~nm; (d) sample D, $\sim$23~nm; (e) sample E, $\sim$42~nm.
(f) Corrected drop solution enthalpies $\Delta H_{\mathrm{ds}}(\mathrm{NC})$ plotted against molar surface area $A_{\mathrm{NC}}$ for the five RS-MnS nanocrystal samples.
Data points are shown as mean values $\pm 95\%$ confidence intervals, estimated as $\pm 2$~times the standard error of the mean.
The ordinary least squares fit yields an apparent surface energy $\gamma_{\mathrm{NC}} = 1.15 \pm 0.38$~J$\cdot$m$^{-2}$ (slope) and an intercept $\Delta H_{\mathrm{ds}}(\mathrm{bulk}) = -826.21 \pm 3.39$~kJ$\cdot$mol$^{-1}$, consistent with the independently measured bulk value ($-827.20 \pm 5.13$~kJ$\cdot$mol$^{-1}$; Table~\ref{tab:calorimetry}).
}
    \label{fig:fig10}
\end{figure*}

\subsection{Synthesis and calorimetry of RS-MnS NCs}

To complement the theoretical predictions, we synthesized RS-MnS NCs and measured their apparent surface energies ($\gamma_{\text{NC}}$) by high-temperature oxidative solution calorimetry.
Because halide ligands introduce energy contributions that are difficult to correct for quantitatively, all NCs were synthesized using manganese(II) nitrate tetrahydrate and sodium diethyldithiocarbamate trihydrate in oleylamine (OLAM) solvent, thereby avoiding halide-containing precursors (see Methods).
Five samples of varying size are summarized in Table~S13, with additional experimental details in Supplementary Information (Subsection~S2.9).

Figures~\ref{fig:fig10}a--e show transmission electron microscopy (TEM) images of the five RS-MnS NC samples, ranging from $\sim$10~nm (sample A) to $\sim$42~nm (sample E).
Powder X-ray diffraction (PXRD) patterns (Figure~S7) confirm that all samples crystallize in the RS structure.
At shorter reaction times the NCs are smaller and morphologically irregular (samples A--C), whereas prolonged reaction yields larger, well-defined nanocubes (samples D--E), consistent with the theoretical prediction that the cubic morphology is thermodynamically favored.

\begin{table}[tb]
\renewcommand{\arraystretch}{1.5}
\caption{
Corrected average drop solution enthalpies ($\Delta H_{\mathrm{ds}}$) and formation enthalpies ($\Delta H_{\mathrm{f}}$) for each rock salt (RS) MnS nanocrystal sample and for bulk MnS.
The 95\% confidence interval is estimated as $\pm 2$~times the standard error of the mean; the number of replicate drops is shown in parentheses.
Elemental reference values for Mn and S are from References~\citenum{hayun_enthalpies_2020}~and~\citenum{lilova_synergistic_2020}, respectively.
Corrected drop solution enthalpies and formation enthalpies were calculated using the thermochemical cycles in Tables~S14~and~S15, respectively.
}
\label{tab:calorimetry}
\centering
\begin{tabular}{@{} l S[table-format=-3.2] @{\,$\pm$\,} S[table-format=1.2] c S[table-format=-3.2] @{\,$\pm$\,} S[table-format=1.2] @{}}
\toprule
{Species} & \multicolumn{3}{c}{$\Delta H_{\mathrm{ds}}$ /kJ$\cdot$mol$^{-1}$} & \multicolumn{2}{c}{$\Delta H_{\mathrm{f}}$ /kJ$\cdot$mol$^{-1}$} \\
\midrule
NC sample A & -842.30 & 4.77 & {(7)} & -197.19 & 5.72 \\
NC sample B & -838.45 & 5.46 & {(7)} & -201.05 & 6.30 \\
NC sample C & -835.73 & 7.16 & {(8)} & -203.77 & 7.82 \\
NC sample D & -831.36 & 3.83 & {(7)} & -208.14 & 4.74 \\
NC sample E & -831.13 & 7.89 & {(8)} & -208.37 & 8.49 \\
Bulk MnS    & -827.20 & 5.13 & {(6)} & -212.30 & 6.01 \\
\midrule
Mn          & -410.32 & 1.55 & {\cite{hayun_enthalpies_2020}} & \multicolumn{2}{c}{---} \\
S           & -629.18 & 2.73 & {\cite{lilova_synergistic_2020}} & \multicolumn{2}{c}{---} \\
\bottomrule
\end{tabular}
\end{table}

Drop solution enthalpies ($\Delta H_{\text{ds}}$) were measured by dropping samples into the calorimeter at 1073~K; at this temperature, adsorbed OLAM was fully combusted to \ce{CO2}, \ce{H2O}, and \ce{N2}.
The measured $\Delta H_{\text{ds}}$ therefore includes the heat content from 298~to~1073~K, the enthalpy of solution, and the enthalpy of oxidation.
The OLAM content of each sample was determined by furnace thermogravimetry (see Methods) and its contribution to $\Delta H_{\text{ds}}$ was subtracted using the heat of combustion at 1073~K (Subsection~S2.9).
Tables~S14~and~S15 present the thermochemical cycles for correcting $\Delta H_{\text{ds}}$ and calculating the formation enthalpies ($\Delta H_{\text{f}}$), respectively.
Average values are listed in Table~\ref{tab:calorimetry}, with individual measurements in Table~S16.
Formation enthalpies of all NC samples are less negative than that of bulk MnS, and become progressively less negative with decreasing particle size, reflecting the increasing surface energy penalty at higher surface-to-volume ratios.

The apparent surface energy $\gamma_{\text{NC}}$ is obtained from the slope of a linear fit of $\Delta H_{\text{ds}}(\text{NC})$ against molar surface area $A_{\text{NC}}$: $\Delta H_{\text{ds}}(\text{NC})=\Delta H_{\text{ds}}(\text{bulk})-\gamma_{\text{NC}}\cdot A_{\text{NC}}$.
Figure~\ref{fig:fig10}f shows the ordinary least squares fit across all individual measurements, yielding $\gamma_{\text{NC}} = 1.15 \pm 0.38$~J$\cdot$m$^{-2}$.
The fitted intercept $\Delta H_{\text{ds}}(\text{bulk}) = -826.21 \pm 3.39$~kJ$\cdot$mol$^{-1}$ agrees well with the independently measured bulk value ($-827.20 \pm 5.13$~kJ$\cdot$mol$^{-1}$ in Table~\ref{tab:calorimetry}), validating the extrapolation.

The experimentally determined $\gamma_{\text{NC}}$ ($1.15 \pm 0.38$~J$\cdot$m$^{-2}$) exceeds the r$^2$SCAN+$U$ prediction for equilibrium \{100\}-dominated nanocubes (0.42--0.43~J$\cdot$m$^{-2}$, Table~S6) by approximately a factor of three.
Several factors contribute to this discrepancy, including the exposure of high-energy facets in small, quasi-spherical NCs, uncertainty in TEM-based surface area estimates, and non-ideal surface configurations such as defects and residual adsorbates.
These factors are analyzed in the Discussion.

\section{Discussion}

\subsection{Theoretical morphologies and surface properties of MnS NCs}

The prediction that nanocubes are the equilibrium RS-MnS morphology across nearly the entire $\Delta\mu_{\ce{S}}$ window is consistent with temperature-dependent synthesis studies\cite{zhang_hydrothermal_2008, yang_size-controlled_2012, chen_low-temperature_2023, michel_hydrothermal_2006, gendler_halide-driven_2023}: Jun et al.\cite{jun_architectural_2002} reported that elevated temperatures favor nanocubes over nanospheres under otherwise identical conditions.
Additionally, our calculations indicate that the \{010\} facets exhibit ferromagnetism, consistent with the ferromagnetic surface properties observed in cubic NCs by Chen et al.\cite{chen_low-temperature_2023} and Yang et al.\cite{yang_size-controlled_2012}

Octahedral RS-MnS morphologies, predominantly synthesized by hydrothermal routes\cite{zhang_hydrothermal_2008, michel_hydrothermal_2006, puglisi_monodisperse_2010, gui_hydrothermal_2011} or at low temperatures in aqueous solution,\cite{chen_low-temperature_2023} were not captured in our theoretical predictions or observed in our experimental synthesis.
Under hydrothermal conditions, the formation of octahedral RS-MnS NCs competes kinetically with the metastable WZ-MnS phase\cite{michel_hydrothermal_2006, gui_hydrothermal_2011} and requires high concentrations of sulfur ions in solution.
Although solvent polarity and ligand adsorption can stabilize \{111\} facets, Michel et al.\cite{michel_hydrothermal_2006} reported that octahedral RS-MnS NCs progressively transform into nanocubes upon aging above 200~$^{\circ}$C, suggesting that the octahedral morphology represents a kinetically trapped state rather than the thermodynamic minimum.

Experimental morphological data for monodisperse ZB-MnS NCs remain scarce, largely because this metastable polymorph readily transforms to WZ- or RS-MnS at elevated temperatures (Figure~\ref{fig:fig3}c).
Yang et al.\cite{yang_polymorphism_2012} synthesized single-crystalline ZB-MnS nanoparticles by reacting \ce{MnCl2} with thioacetamide at 200~$^{\circ}$C, with high-resolution TEM confirming \{111\} facets, consistent with our prediction that \{$\overline{111}$\}--S and \{110\} facets dominate under S-rich conditions.
Incorporating solvation and ligand effects into the surface energy calculations would further stabilize \{$\overline{111}$\}--S, potentially favoring fully tetrahedral morphologies.
The structural similarity between ZB-MnS (111) and WZ-MnS (0001) facets also rationalizes the commonly observed branched wires with ZB tetrahedral seeds and WZ arms.\cite{tang_morphology_2015, lu_metastable_2001, jun_architectural_2002, yang_polymorphism_2012}

Our WZ-MnS predictions capture the bullet-like\cite{joo_generalized_2003, michel_hydrothermal_2006, gendler_halide-driven_2023} and spindle-like\cite{tang_morphology_2015, jun_architectural_2002, gui_hydrothermal_2011} nanorods observed experimentally.
Bullet-like morphologies, with an asymmetric rod terminus, correspond to the predicted base truncation under moderate-to-high $\Delta\mu_{\ce{S}}$, whereas spindle-like morphologies imply symmetric truncation at both termini, a feature not captured by our vacuum calculations, which yield distinct surface energies for the [0001] and [000$\overline{1}$] termini.
Because all facets with $l\neq0$ are sensitive to reaction conditions such as solvent polarity, ligand adsorption, and temperature, incorporating solvent models and surface--ligand interactions represents an important next step toward capturing the full range of observed WZ-MnS morphologies.

\subsection{Reconciling measured and predicted surface energies of RS-MnS NCs}

The OLAM coating introduces significant heat effects that increase measurement uncertainty, and renders Brunauer--Emmett--Teller surface area determination inapplicable: degassing at 333~K for 24~hours retains $\sim$80\% of the OLAM, while higher temperatures risk oxidation, agglomeration, or sulfur loss.
Despite these challenges, the measured apparent surface energy ($1.15 \pm 0.38$~J$\cdot$m$^{-2}$) is comparable to values reported for other nanoscale sulfides and oxides (Table~S17).

When samples are grouped by morphology (quasi-spherical [A--C], transitional [C--E], and cubic-like [D--E]), the apparent surface energy decreases progressively with increasing cubic character (Figure~S8).
Although the individual slopes are not statistically significant given the limited data per group, the trend supports the interpretation that the theory--experiment discrepancy arises primarily from high-index facet exposure and surface area uncertainty in smaller, less-faceted NCs.
As morphology approaches the equilibrium nanocube, the measured surface energy should converge toward the theoretical \{100\} value.
A similar morphology dependence has been reported for ZnO NCs by Zhang et al.\cite{zhang_surface_2007}

Since the smallest, quasi-spherical NCs (samples A--C) exert the greatest leverage on the fitted apparent surface energy, we evaluated the theoretical orientation-averaged surface energy in the near-spherical limit of RS-MnS using two independent approaches: a continuous sixth-order cubic-symmetric polynomial expansion (Table~S18 and Figure~S9) and a discrete spherical Voronoi tessellation (see Methods and Subsection~S2.10).
While the Voronoi tessellation yields systematically slightly higher values than the cubic expansion, both approaches produce the same decreasing trend with $\Delta \mu_{\ce{S}}$ (Figure~S10).
The resulting spherical average surface energy, 0.84--1.00~J$\cdot$m$^{-2}$, is much closer to the experimentally measured apparent surface energy ($1.15 \pm 0.38$~J$\cdot$m$^{-2}$) than the equilibrium Wulff value (0.42--0.43~J$\cdot$m$^{-2}$) and falls well within the range of surface energies obtained from fits to different sample subsets (Figure~S8).

\subsection{Perspectives on theoretical modeling of NC morphology}

We close by identifying three directions for extending this work.

First, the relative surface energy approach used here for WZ-MnS relies on the 6$_3$ screw axis and therefore extends to other $P6_3$-group crystals, but predicting equilibrium morphologies for crystals of still lower symmetry will require methods tailored to their specific symmetry elements.

Second, machine learning interatomic potentials (MLIPs) could substantially accelerate surface modeling, particularly for the computationally demanding wedge calculations.
Current foundational/universal MLIPs\cite{barroso-luque_open_2024, kuner_mp-aloe_2025, kaplan_foundational_2025, 10.1063/5.0297006, wood_uma_2025} such as MACE\cite{10.1063/5.0297006} and UMA\cite{wood_uma_2025} are trained predominantly on bulk structures, limiting their accuracy for surface properties.
Incorporating meta-GGA-level surface energies into the training data is a promising route toward MLIPs capable of reliable Wulff construction predictions.

Third, kinetic mechanisms, rather than thermodynamic equilibrium, often govern the polymorphic outcome of NC synthesis.
For example, Gendler et al.\cite{gendler_halide-driven_2023} showed that varying the halide precursor selects different polymorphic MnS NCs.
Consequently, elucidating the configurations of the pre-nucleation complexes induced by different ligands and uncovering their associated nucleation mechanisms are of considerable significance for achieving true predictive control over MnS NC synthesis.

\section{Conclusions}

In this work, we employed DFT with Wulff construction methods to investigate the surface thermodynamics and equilibrium morphologies of polymorphic MnS NCs.
Using r$^2$SCAN+$U$ ($U = 2.7$~eV), we evaluated surface energies for RS, ZB, and WZ MnS and predicted their equilibrium morphologies as a function of the relative chemical potential of sulfur, $\Delta \mu_{\ce{S}}$.
The predicted RS-MnS nanocubes and WZ-MnS nanorods are consistent with experimental observations.
For the less commonly synthesized ZB-MnS, our results reveal a morphological transition from rhombic dodecahedra to 16-faced polyhedra with increasing $\Delta \mu_{\ce{S}}$.
Experimentally, our synthesized RS-MnS NCs increasingly exhibit cubic morphology with increasing reaction time, supporting the predicted thermodynamic stability of nanocubes.
The experimentally measured apparent surface energy ($1.15 \pm 0.38$~J$\cdot$m$^{-2}$) exceeds the theoretical prediction (0.42--0.43~J$\cdot$m$^{-2}$), a discrepancy attributable to uncertainties in transmission-electron-microscopy-based surface area estimation, non-ideal surface configurations, and the exposure of high-energy facets in small, quasi-spherical NCs.
Overall, this study establishes a framework for evaluating the surface energetics and crystallographic morphologies of MnS NCs that is extensible to other metal chalcogenide systems.
The results provide a quantitative foundation for understanding the thermodynamic driving forces underlying MnS NC formation and offer guidance for future investigations of solvent and ligand effects, as well as the kinetic mechanisms governing nucleation and growth.

\section{Methods}

\subsection{Computational methods}

All DFT calculations were performed using the projector augmented wave (PAW) method\cite{kresse_ultrasoft_1999} as implemented in the Vienna Ab initio Simulation Package (VASP) version 6.4.2.\cite{kresse_ab_1993, kresse_efficient_1996, kresse_efficiency_1996}
XC functionals were evaluated by comparing predicted bulk reaction energies and lattice constants of MnS polymorphs (Table~S19) against experimental references.\cite{Zagorac:in5024, thomas_c_allison_nist-janaf_2013}
The functionals considered include the PBE GGA;\cite{Perdew:1996pki} the SCAN\cite{sun_strongly_2015} and r$^2$SCAN\cite{furness_accurate_2020} meta-GGAs; and the HSE06 range-separated hybrid functional,\cite{krukau_influence_2006} along with their van der Waals dispersion-corrected variants: DFT-D3 with Becke--Johnson (BJ) damping\cite{grimme_consistent_2010, grimme_effect_2011} for PBE and HSE06,\cite{moellmann_dft-d3_2014} and rVV10\cite{peng_versatile_2016, ning_workhorse_2022} for SCAN and r$^2$SCAN.
On the basis of these benchmarks, r$^2$SCAN was selected for structural relaxations of MnS surface slabs, with electrostatic corrections\cite{neugebauer_adsorbate-substrate_1992, makov_periodic_1995} applied to asymmetric slabs possessing net surface dipole moments.
Surface energies were computed via single-point calculations using the rotationally invariant Hubbard $U$\cite{dudarev_electron-energy-loss_1998} augmented r$^2$SCAN functional (r$^2$SCAN+$U$), with $U = 2.7$~eV determined by fitting to HSE06 surface energy references.
As a result, the corrected surface energies do not include changes in surface relaxation induced by the $U$ term.
Since meta-GGA pseudopotentials are not available in VASP, pseudopotentials were taken from the PBE PAW datasets (version 54) recommended by the Materials Project (Table~S20).\cite{kingsbury_performance_2022, horton_accelerated_2025}
A plane-wave kinetic energy cutoff of 520 eV was used, converging total energies of bulk structures to within 1 meV/atom.
Brillouin-zone integrations employed a $\Gamma$-centered Monkhorst--Pack $k$-mesh with reciprocal spacing $\leq 0.25$~\AA$^{-1}$ (including the $2\pi$ factor) and Gaussian smearing with a width of 0.05~eV.
Electronic minimization used the conjugate gradient algorithm, with the blocked Davidson iteration scheme adopted when convergence difficulties arose.
The SCF convergence criterion was $1 \times 10^{-6}$~eV/cell for total energy, and geometry optimizations were terminated when all atomic forces fell below 0.02~eV/\AA.
The VASP Accurate precision mode was used throughout, with projection operators evaluated in reciprocal space.
For $\alpha$-Mn, a low-spin ferromagnetic collinear configuration was adopted to reduce computational cost; although different magnetic configurations yield energy differences of up to $\sim 300$~meV$\cdot$atom$^{-1}$ with coarse $k$-point grids (Table~S21), the reaction energy evaluation requires only the per-atom energy of the ground-state configuration, so this variability does not affect our results.

\subsection{Ab initio thermodynamics}

Reaction energies relevant to MnS synthesis were approximated as total energy differences from DFT:
\begin{equation} \label{1st_eq}
    \Delta E_{\mathrm{rxn}} =\sum_p N_p \cdot E_p - \sum_r N_r \cdot E_r
\end{equation}
where $N_p$ and $E_p$ are the stoichiometric coefficients and total energies of the products, and $N_r$ and $E_r$ are those of the reactants.

The Gibbs formation energy of bulk Mn$_x$S$_y$ is:
\begin{equation} \label{2nd_eq}
    \Delta G_{\mathrm{f}}(\mathrm{Mn}_x\mathrm{S}_y) = x \cdot \Delta \mu_{\mathrm{Mn}} + y \cdot \Delta \mu_{\mathrm{S}}
\end{equation}
where $x$ and $y$ are the stoichiometric coefficients, $\Delta \mu_{\mathrm{S}} = \mu_{\mathrm{S}} - \mu^{\circ}_{\mathrm{S}}$ is the chemical potential of sulfur relative to its standard state ($\alpha$-\ce{S8}), and $\Delta \mu_{\mathrm{Mn}}$ is defined analogously ($\alpha$-Mn).
Thermodynamic stability of bulk Mn$_x$S$_y$ requires $\Delta \mu_{\mathrm{S}} \leq 0$ and $\Delta \mu_{\mathrm{Mn}} \leq 0$, giving the stability range:
\begin{equation} \label{3rd_eq}
    \frac{\Delta G_{\mathrm{f}}(\mathrm{Mn}_x\mathrm{S}_y)}{y} \leq \Delta \mu_{\mathrm{S}} \leq 0
\end{equation}

For a MnS slab, the combined surface energy of the top and bottom facets is:
\begin{equation} \label{4th_eq}
\begin{split}
    \sigma_{\mathrm{top}} + \sigma_{\mathrm{bottom}} = E_{\mathrm{slab}} &- N_{\mathrm{Mn}} \cdot E_{\mathrm{MnS}} + \Gamma_{\mathrm{s}} \cdot (E_{\mathrm{S}} + \Delta \mu_{\mathrm{S}})
\end{split}
\end{equation}
where $E_{\mathrm{slab}}$ is the slab total energy, $E_{\ce{MnS}}$ and $E_{\ce{S}}$ are the energies per formula unit of bulk MnS and per atom of elemental sulfur, $N_{\ce{Mn}}$ and $N_{\ce{S}}$ are the numbers of Mn and S atoms, and $\Gamma_{\ce{S}} = N_{\ce{Mn}} - N_{\ce{S}}$.
For symmetric slabs with identical terminations, the individual surface energy is:
\begin{equation} \label{5th_eq}
    \gamma = \frac{\sigma_{\mathrm{top}} + \sigma_{\mathrm{bottom}}}{2A}
\end{equation}
where $A$ is the surface area of one side of the slab.

To calculate surface energies of facets where symmetric slabs cannot be generated, wedge calculations introduced by Zhang and Wei\cite{zhang_surface_2004} were employed.
For a wedge, the total energy can be decomposed into bulk, surface, and edge contributions.
Taking the energy difference between two wedges of different sizes eliminates edge contributions, giving the sum of total surface energies ($\sigma_{\mathrm{tot}} = \Delta n_{\mathrm{base}} \cdot \sigma_{\mathrm{base}} + \Delta n_{\mathrm{side}} \cdot \sigma_{\mathrm{side}}$):
\begin{equation} \label{6th_eq}
\begin{split}
    \sigma_{\mathrm{tot}} = \Delta E_{\mathrm{wedge}} &- \Delta N_{\ce{Mn}} \cdot E_{\ce{MnS}} + \Delta \Gamma_{\ce{S}} \cdot (E_{\ce{S}} + \Delta \mu_{\ce{S}})
\end{split}
\end{equation}
where $\Delta E_{\mathrm{wedge}}$ and $\Delta \Gamma_{\ce{S}}$ are the differences in total energy and $\Gamma_{\ce{S}}$ between two wedges, $\Delta n_i$ is the difference in the number of $1 \times 1$ unit facets of type $i$, and $\sigma_i = A_i \cdot \gamma_i$ is the total surface energy of a $1 \times 1$ unit facet with area $A_i$ and surface energy $\gamma_i$.
The individual surface energy of the side facet then follows from combining the wedge and slab results:
\begin{equation} \label{7th_eq}
    \gamma_{\mathrm{side}} = \frac{\sigma_{\mathrm{side}}}{A_{\mathrm{side}}} = \frac{\sigma_{\mathrm{tot}} - \Delta n_{\mathrm{base}} \cdot \sigma_{\mathrm{base}}}{\Delta n_{\mathrm{side}} \cdot A_{\mathrm{side}}}
\end{equation}

To avoid the high computational cost of wedge relaxations, we used fully fixed wedges, fully fixed slabs, and half-fixed slabs (Figure~\ref{fig:fig1}c).
Unlike in previous studies,\cite{zhang_surface_2004, dreyer_absolute_2014, jin_absolute_2021, li_computing_2015} hydrogen passivation was not applied, as convergence tests on clean-surface wedges demonstrated sufficient convergence (Figures~S11d, S12, and S13).

An analogous strategy is applied to evaluate the relative surface energies introduced by Li et al.\cite{li_computing_2015}
As depicted in Figure~\ref{fig:fig1}d, calculations based on fully fixed wedges, half-fixed slabs, and centrally fixed slabs provide four linear combinations of total surface energies using Equations~\ref{4th_eq}~and~\ref{6th_eq}, enabling determination of either $\sigma_{(hkil)} - \sigma_{(0001)}$ or $\sigma_{(hkil)} - \sigma_{(000\overline{1})}$.
The relative surface energy is then:
\begin{equation} \label{8th_eq}
    \Delta \gamma =
    \begin{cases}
        \frac{\gamma_{(hkil)}}{\cos{\theta}} - \gamma_{(0001)} = \frac{\sigma_{(hkil)} - \sigma_{(0001)}}{A_{(0001)}} & l > 0 \\
        \frac{\gamma_{(hkil)}}{\cos{\theta}} - \gamma_{(000\overline{1})} = \frac{\sigma_{(hkil)} - \sigma_{(000\overline{1})}}{A_{(000\overline{1})}} & l < 0
    \end{cases}
\end{equation}
where $\theta$ is the dihedral angle between the two planes.

\subsection{Wulff constructions}

Equilibrium morphologies were constructed using the Gibbs--Wulff theorem as implemented in the \verb|pymatgen.analysis.wulff| module of Pymatgen.\cite{tran_surface_2016}
The \verb|get_all_miller_e| method of the \verb|WulffShape| class was modified to differentiate opposite polar facets in ZB- and WZ-MnS and to assign consistent surface energies to crystallographically equivalent facets.

For RS- and ZB-MnS, individual surface energies were directly calculated and used to construct Wulff shapes.
For WZ-MnS, individual polar surface energies cannot be separated; however, the surface energies of all \{$10\overline{1}0$\} and \{$11\overline{2}0$\} lateral facets are well-defined, fixing the geometric center of the Wulff shape within the horizontal plane and leaving only one degree of freedom along [$0001$] and [$000\overline{1}$].
Following prior work,\cite{li_computing_2015, wang_defining_2022} we assigned a reference value to one of the polar surface energies and derived the remaining individual values by solving the set of linear equations for the relative surface energies under the constraint that all surface energies remain positive.
Adjusting the reference value changes only the absolute position of the center along [0001] in the WZ-MnS Wulff shape, leaving the geometric shape unchanged.
The surface energy of the (0001)--S facet was therefore set to 0.06~eV$\cdot$\AA$^{-2}$ (half of the r$^2$SCAN total surface energy of (0001)--S and (000$\overline{1}$)--Mn) as the reference for constructing Wulff shapes of WZ-MnS.
Complete r$^2$SCAN and r$^2$SCAN+$U$ surface energies and Wulff constructions for WZ-MnS are tabulated in Subsections~S2.5~and~S2.6.

\subsection{Spherical average surface energy approximation}

To evaluate the surface energy relevant to quasi-spherical morphologies, we defined the orientation-averaged surface energy in the spherical limit as:

\begin{equation} \label{9th_eq}
\bar{\gamma} = \frac{1}{4\pi} \int_{\mathcal{S}^2} \gamma(\mathbf{n}) \, d\Omega ,
\end{equation}
where $\mathbf{n}$ is the surface normal direction and $d\Omega$ is the solid angle element on the unit sphere $\mathcal{S}^2$.

Two independent approaches were employed to evaluate $\bar{\gamma}$: a cubic-symmetric polynomial expansion and a spherical Voronoi tessellation.

A cubic-symmetric polynomial expansion was employed to represent the orientation dependence of the surface energy $\gamma(\mathbf{n})$, thereby explicitly enforcing the symmetry of the cubic structure.\cite{nye1985physical}
Specifically, $\gamma(\mathbf{n})$ was expanded as a linear combination of cubic-invariant basis functions,

\begin{equation}  \label{10th_eq}
\gamma(\mathbf{n}) \approx \sum_i c_i B_i(\mathbf{n}),
\end{equation}
where $B_i$ are cubic-invariant basis functions constructed from the direction cosines of $\mathbf{n}$, and $c_i$ are obtained by least-squares fitting to the surface energies of 290 discrete Miller index facets (Table~S4) calculated using r$^2$SCAN+$U$.
The spherical average $\bar{\gamma}$ was then evaluated analytically using the exact angular averages of the basis functions over the unit sphere.

The convergence of the cubic-symmetric polynomial expansion was examined by progressively increasing the truncation order up to tenth order (Table~S18 and Figure~S9).
The spherical average converges by sixth order, beyond which additional terms do not alter the results appreciably.
Consequently, the sixth-order basis was employed.

As an independent discretization-based approach, a spherical Voronoi tessellation\cite{caroli:inria-00405478, 4121581} of the facet normals was constructed on the unit sphere, as implemented in the \verb|scipy.spatial.SphericalVoronoi| module.\cite{virtanen_scipy_2020}
Each Miller index direction $\mathbf{n}_j$ was assigned a weight proportional to the area $A_j$ of its Voronoi cell, yielding

\begin{equation}
\bar{\gamma} \approx \sum_j \left( \frac{A_j}{4\pi} \right) \gamma(\mathbf{n}_j).
\end{equation}

This method provides a geometry-based estimate of the spherical average without assuming a functional form for $\gamma(\mathbf{n})$.

\subsection{Synthesis of RS-MnS NCs}

OLAM (technical grade 70\%), manganese(II) nitrate tetrahydrate (\ce{Mn(NO3)2.4H2O}, $\geq$99.9\% trace metal basis), and sodium diethyldithiocarbamate trihydrate (\ce{Na(S2CNEt2)}, ACS grade) were purchased from Sigma-Aldrich.
Hexanes and ethanol were obtained from Fisher Scientific.
All chemicals were used as received.
The reaction setup consisted of a 3-neck 50~mL round-bottom flask with a condenser on the central neck and rubber septa on the side necks, heated using Glas-Col heating mantles connected to a Gemini temperature controller (J-Kem Scientific) with K-type thermocouples in contact with the reaction mixture.
Safety considerations are detailed in Subsection~S2.9.
Manganese diethyldithiocarbamate (Mn-DDTC) was prepared by precipitation of manganese nitrate tetrahydrate and sodium diethyldithiocarbamate in deionized water under ambient conditions.
The product was washed five times with deionized water by centrifugation at 2000~rpm for 5~min, then dried under vacuum at 60~$^{\circ}$C.
In a typical synthesis,\cite{wang_low-temperature_2008} 20~mmol of Mn-DDTC was dissolved in 30~mL of OLAM and degassed under vacuum at 120~$^{\circ}$C with stirring at 700~rpm.
The atmosphere was then switched to nitrogen, and the reaction was heated to 240~$^{\circ}$C for 0--30~min to yield NCs of varying sizes.
After cooling to room temperature, the NCs were washed three times by precipitation with isopropanol, centrifugation (\num[group-separator = {,}]{10000}~rpm, 2~min), and redispersion in hexane.
Washed samples were stored in OLAM.

\subsection{Characterization}

RS-MnS NC samples were characterized by powder X-ray diffraction (PXRD) and high-resolution transmission electron microscopy (HRTEM).
HRTEM was performed on a Talos F200C microscope at 200~kV; samples were prepared by drop-casting onto carbon type-B, 400 mesh copper (Cu) grids.
PXRD patterns were collected on a Bruker D8 diffractometer with a LynxEye position-sensitive detector and Cu K$\alpha_1$ radiation ($\lambda = 1.54$~\AA).
The OLAM content of each sample was determined by furnace thermogravimetry.
Samples were heated to 1073~K in air, which combusted the OLAM and oxidized MnS to \ce{Mn2O3}.
The initial MnS and OLAM contents were calculated from the mass change using the reaction \ce{4 MnS(s) + 7 O2(g) -> 2 Mn2O3(s) + 4 SO2(g)}.

\subsection{High temperature oxidative solution calorimetry}

All samples were washed several times with a 1:5 hexane:ethanol mixture, centrifuged, and dried at room temperature under vacuum for several days.
Samples with larger surface areas required additional washing cycles and longer drying times to obtain powder suitable for calorimetry.
High-temperature oxidative solution calorimetry was performed at 1073~K in molten sodium molybdate (\ce{3 Na2O}--\ce{4 MoO3}).
Oxygen was flushed over the solvent at 90~mL/min and bubbled through it at 5~mL/min.
The calorimeter was calibrated against the heat of combustion of 5~mg benzoic acid pellets (Parr Instruments).
The final dissolution product was manganese sulfate in the molten oxide solvent.
Further details of this methodology for metal sulfides are described in previous works.\cite{lilova_synergistic_2020, subramani_greigite_2020, subramani_surface_2023, abramchuk_development_2021}
The combustion enthalpy of OLAM was measured under the same conditions.
Since OLAM is a liquid, a small droplet was placed on a sodium molybdate boat and dropped into the calorimeter.
At 1073~K, the sodium molybdate melts and the OLAM combusts, releasing \ce{CO2}, \ce{H2O}, and \ce{N2}.
The heat effect of the sodium molybdate (heat content and heat of melting) was subtracted from the total drop solution enthalpy to obtain the heat of combustion of OLAM at 1073~K.
This technique has been used previously to measure enthalpies of mixing of liquid metals and alloys.\cite{bustamante_enthalpies_2024}

\backmatter

\bmhead{Supplementary information}

Supplementary Information is available online.
The input and output files supporting the computational results of this work are available in the NOMAD repository: \url{https://dx.doi.org/10.17172/NOMAD/2025.12.19-1}. The data and scripts required for analysis, figure generation, and Wulff shape construction are available in the GitHub repository: \url{https://github.com/wexlergroup/ET_CM_MnS-NCs.git}.

\bmhead{Acknowledgements}

R.B.W., K.L., and E.A.H.P.\ acknowledge support from the National Science Foundation under Grant No.\ 2305155. M.H.\ acknowledges support from the National Aeronautics and Space Administration under Grant No.\ 80NSSC22K1640. This work used the Delta GPU system at the National Center for Supercomputing Applications (NCSA) through allocation NNT230005 from the Advanced Cyberinfrastructure Coordination Ecosystem: Services \& Support (ACCESS) program, which is supported by U.S.\ National Science Foundation grants \#2138259, \#2138286, \#2138307, \#2137603, and \#2138296.\cite{boerner_access_2023} A portion of this research was conducted at the Center for Nanophase Materials Sciences, a DOE Office of Science User Facility, with additional computational resources sponsored by the U.S.\ Department of Energy's Office of Critical Minerals and Energy Innovation and located at the National Laboratory of the Rockies.

\section*{Declarations}

\bmhead{Author contributions}

R.B.W.\ conceived the original research idea. R.B.W., K.L., and E.A.H.P.\ provided insight and guidance throughout the project. J.C.\ developed the code, generated the surface models, conducted the simulations, and performed data processing and analysis. J.C., M.K., and R.B.W.\ evaluated the density functional theory benchmarks and surface energy correction. J.C., R.Y., and R.B.W.\ developed the Wulff construction strategies and analyzed the results. J.C.\ and R.B.W.\ assessed the spherical average surface energy approximation. D.M.\ and D.G.\ synthesized and characterized the rock salt MnS nanocrystal samples. T.S., M.H., and A.R.O.\ carried out the high-temperature oxidative solution calorimetry experiments and determined the apparent surface energy. J.C.\ and D.M.\ drafted the sections on computational methods and synthesis/characterization, respectively, and K.L.\ drafted the calorimetry section. R.B.W.\ revised the manuscript. All authors participated in discussions and approved the final manuscript.

\bmhead{Competing interests}

The authors declare no competing interests.

\bibliography{references}

@article{ekimov_quantum_2023,
	title = {Quantum {Size} {Effect} in {Three}-{Dimensional} {Microscopic} {Semiconductor} {Crystals}},
	volume = {118},
	issn = {0021-3640, 1090-6487},
	url = {https://link.springer.com/10.1134/S0021364023130040},
	doi = {10.1134/S0021364023130040},
	number = {S1},
	journal = {Jetp Lett.},
	author = {Ekimov, A. I. and Onushchenko, A. A.},
	month = dec,
	year = {2023},
	pages = {S15--S17},
}

@article{rossetti_quantum_1983,
	title = {Quantum size effects in the redox potentials, resonance {Raman} spectra, and electronic spectra of {CdS} crystallites in aqueous solution},
	volume = {79},
	issn = {0021-9606, 1089-7690},
	url = {https://pubs.aip.org/jcp/article/79/2/1086/776583/Quantum-size-effects-in-the-redox-potentials},
	doi = {10.1063/1.445834},
	number = {2},
	journal = {The Journal of Chemical Physics},
	author = {Rossetti, R. and Nakahara, S. and Brus, L. E.},
	month = jul,
	year = {1983},
	pages = {1086--1088},
}

@article{murray_synthesis_1993,
	title = {Synthesis and characterization of nearly monodisperse {CdE} ({E} = sulfur, selenium, tellurium) semiconductor nanocrystallites},
	volume = {115},
	issn = {0002-7863, 1520-5126},
	url = {https://pubs.acs.org/doi/abs/10.1021/ja00072a025},
	doi = {10.1021/ja00072a025},
	number = {19},
	journal = {J. Am. Chem. Soc.},
	author = {Murray, C. B. and Norris, D. J. and Bawendi, M. G.},
	month = sep,
	year = {1993},
	pages = {8706--8715},
}

@article{alivisatos_semiconductor_1996,
	title = {Semiconductor {Clusters}, {Nanocrystals}, and {Quantum} {Dots}},
	volume = {271},
	url = {http://www.jstor.org/stable/2889983},
	number = {5251},
	journal = {Science, New Series},
	author = {Alivisatos, A. P.},
	year = {1996},
	pages = {933--937},
}

@article{kovalenko_prospects_2015,
	title = {Prospects of {Nanoscience} with {Nanocrystals}},
	volume = {9},
	copyright = {http://pubs.acs.org/page/policy/authorchoice\_termsofuse.html},
	issn = {1936-0851, 1936-086X},
	url = {https://pubs.acs.org/doi/10.1021/nn506223h},
	doi = {10.1021/nn506223h},
	number = {2},
	journal = {ACS Nano},
	author = {Kovalenko, Maksym V. and Manna, Liberato and Cabot, Andreu and Hens, Zeger and Talapin, Dmitri V. and Kagan, Cherie R. and Klimov, Victor I. and Rogach, Andrey L. and Reiss, Peter and Milliron, Delia J. and Guyot-Sionnnest, Philippe and Konstantatos, Gerasimos and Parak, Wolfgang J. and Hyeon, Taeghwan and Korgel, Brian A. and Murray, Christopher B. and Heiss, Wolfgang},
	month = feb,
	year = {2015},
	pages = {1012--1057},
}

@article{ghosh_many_2018,
	title = {The {Many} “{Facets}” of {Halide} {Ions} in the {Chemistry} of {Colloidal} {Inorganic} {Nanocrystals}},
	volume = {118},
	copyright = {http://pubs.acs.org/page/policy/authorchoice\_termsofuse.html},
	issn = {0009-2665, 1520-6890},
	url = {https://pubs.acs.org/doi/10.1021/acs.chemrev.8b00158},
	doi = {10.1021/acs.chemrev.8b00158},
	number = {16},
	journal = {Chem. Rev.},
	author = {Ghosh, Sandeep and Manna, Liberato},
	month = aug,
	year = {2018},
	pages = {7804--7864},
}

@article{nirmal_luminescence_1999,
	title = {Luminescence {Photophysics} in {Semiconductor} {Nanocrystals}},
	volume = {32},
	issn = {0001-4842, 1520-4898},
	url = {https://pubs.acs.org/doi/10.1021/ar9700320},
	doi = {10.1021/ar9700320},
	number = {5},
	journal = {Acc. Chem. Res.},
	author = {Nirmal, Manoj and Brus, Louis},
	month = may,
	year = {1999},
	pages = {407--414},
}

@article{owen_coordination_2015,
	title = {The coordination chemistry of nanocrystal surfaces},
	volume = {347},
	issn = {0036-8075, 1095-9203},
	url = {https://www.science.org/doi/10.1126/science.1259924},
	doi = {10.1126/science.1259924},
	number = {6222},
	journal = {Science},
	author = {Owen, Jonathan},
	month = feb,
	year = {2015},
	pages = {615--616},
}

@article{boles_erratum_2016,
	title = {Erratum: {The} surface science of nanocrystals},
	volume = {15},
	issn = {1476-1122, 1476-4660},
	shorttitle = {Erratum},
	url = {https://www.nature.com/articles/nmat4578},
	doi = {10.1038/nmat4578},
	number = {3},
	journal = {Nature Mater},
	author = {Boles, Michael A. and Ling, Daishun and Hyeon, Taeghwan and Talapin, Dmitri V.},
	month = mar,
	year = {2016},
	pages = {364--364},
}

@article{chen_pure_2013,
	title = {Pure colors from core–shell quantum dots},
	volume = {38},
	copyright = {https://www.cambridge.org/core/terms},
	issn = {0883-7694, 1938-1425},
	url = {http://link.springer.com/10.1557/mrs.2013.179},
	doi = {10.1557/mrs.2013.179},
	number = {9},
	journal = {MRS Bull.},
	author = {Chen, Ou and Wei, He and Maurice, Axel and Bawendi, Moungi and Reiss, Peter},
	month = sep,
	year = {2013},
	pages = {696--702},
}

@article{mirkin_33_2025,
	title = {33 {Unresolved} {Questions} in {Nanoscience} and {Nanotechnology}},
	volume = {19},
	copyright = {https://doi.org/10.15223/policy-001},
	issn = {1936-0851, 1936-086X},
	url = {https://pubs.acs.org/doi/10.1021/acsnano.5c12854},
	doi = {10.1021/acsnano.5c12854},
	number = {36},
	journal = {ACS Nano},
	author = {Mirkin, Chad A. and Petrosko, Sarah Hurst and Artzi, Natalie and Aydin, Koray and Biaggne, Austin and Brinker, C. Jeffrey and Bujold, Katherine E. and Cao, Y. Charles and Chan, Rachel R. and Chen, Chaojian and Chen, Peng-Cheng and Chen, Xiaodong and Chevalier, Olivier J. G. L. and Choi, Chung Hang Jonathan and Crooks, Richard M. and Dravid, Vinayak P. and Du, Jingshan S. and Ebrahimi, Sasha B. and Fan, Hongyou and Farha, Omar K. and Figg, C. Adrian and Fink, Tanner D. and Forsyth, Connor M. and Fuchs, Harald and Geiger, Franz M. and Gianneschi, Nathan C. and Gibson, Kyle J. and Ginger, David S. and Guo, SiShi and Hanes, Justin S. and Hao, Liangliang and Huang, Jin and Hunter, Bryan M. and Huo, Fengwei and Hwang, Jeongmin and Jin, Rongchao and Kelley, Shana O. and Kempa, Thomas J. and Kim, Youngeun and Kudruk, Sergej and Kumari, Sneha and Landy, Kaitlin M. and Lee, Ki-Bum and Leon, Noel J. and Li, Jun and Li, Yuanwei and Li, Zhiwei and Liu, Bin and Liu, Guoliang and Liu, Xiaogang and Liz-Marzán, Luis M. and Lorch, Jochen H. and Luo, Taokun and Macfarlane, Robert J. and Millstone, Jill E. and Mrksich, Milan and Murphy, Catherine J. and Naik, Rajesh R. and Nel, Andre E. and Oetheimer, Christopher and Hedlund Orbeck, Jenny K. and Park, So-Jung and Partridge, Benjamin E. and Peppas, Nicholas A. and Personick, Michelle L. and Raj, Arindam and Ramani, Namrata and Ross, Michael B. and Ross, Stacey Barnaby and Sargent, Edward H. and Sengupta, Tanushri and Schatz, George C. and Seferos, Dwight S. and Seideman, Tamar and Seo, Soyoung Eileen and Shen, Bo and Shim, Wooyoung and Shin, Donghoon and Simon, Ulrich and Sinegra, Andrew J. and Smith, Peter T. and Spokoyny, Alexander M. and Stang, Peter J. and Stegh, Alexander H. and Stoddart, J. Fraser and Swearer, Dayne F. and Tan, Weihong and Teplensky, Michelle H. and Thaxton, C. Shad and Walt, David R. and Wang, Mary X. and Wang, Zhe and Wei, Wei David and Weiss, Paul S. and Winegar, Peter H. and Xia, Younan and Xie, Yi and Xu, Xiaoyang and Yang, Peidong and Yang, Yiming and Ye, Zihao and Yoon, Kuk Ro and Zhang, Cuizheng and Zhang, Hua and Zhang, Ke and Zhang, Liangfang and Zhang, Xiaoyu and Zhang, Ye and Zheng, Zijian and Zhou, Wenjie and Zhu, Shengshuang and Zhu, Wei},
	month = sep,
	year = {2025},
	pages = {31933--31968},
}

@article{ibanez_prospects_2025,
author = {Ibá{\~n}ez, Maria and Boehme, Simon C. and Buonsanti, Raffaella and De Roo, Jonathan and Milliron, Delia J. and Ithurria, Sandrine and Rogach, Andrey L. and Cabot, Andreu and Yarema, Maksym and Cossairt, Brandi M. and Reiss, Peter and Talapin, Dmitri V. and Protesescu, Loredana and Hens, Zeger and Infante, Ivan and Bodnarchuk, Maryna I. and Ye, Xingchen and Wang, Yuanyuan and Zhang, Hao and Lhuillier, Emmanuel and Klimov, Victor I. and Utzat, Hendrik and Rainò, Gabriele and Kagan, Cherie R. and Cargnello, Matteo and Son, Jae Sung and Kovalenko, Maksym V.},
title = {Prospects of Nanoscience with Nanocrystals: 2025 Edition},
journal = {ACS Nano},
volume = {19},
number = {36},
pages = {31969-32051},
year = {2025},
doi = {10.1021/acsnano.5c07838},
URL = {https://doi.org/10.1021/acsnano.5c07838},
}

@article{lokhande_process_1998,
	title = {Process and characterisation of chemical bath deposited manganese sulphide ({MnS}) thin films},
	volume = {330},
	issn = {0040-6090},
	url = {https://www.sciencedirect.com/science/article/pii/S0040609098005008},
	doi = {https://doi.org/10.1016/S0040-6090(98)00500-8},
	number = {2},
	journal = {Thin Solid Films},
	author = {Lokhande, C. D. and Ennaoui, A. and Patil, P. S. and Giersig, M. and Muller, M. and Diesner, K. and Tributsch, H.},
	year = {1998},
	pages = {70--75},
}

@article{alanazi_structural_2021,
	title = {Structural {Investigations} of $\alpha$-{MnS} {Nanocrystals} and {Thin} {Films} {Synthesized} from {Manganese}({II}) {Xanthates} by {Hot} {Injection}, {Solvent}-{Less} {Thermolysis}, and {Doctor} {Blade} {Routes}},
	volume = {6},
	copyright = {https://creativecommons.org/licenses/by/4.0/},
	issn = {2470-1343, 2470-1343},
	url = {https://pubs.acs.org/doi/10.1021/acsomega.1c02907},
	doi = {10.1021/acsomega.1c02907},
	number = {42},
	journal = {ACS Omega},
	author = {Alanazi, Abdulaziz M. and McNaughter, Paul D. and Alam, Firoz and Vitorica-yrezabal, Inigo J. and Whitehead, George F. S. and Tuna, Floriana and O’Brien, Paul and Collison, David and Lewis, David J.},
	month = oct,
	year = {2021},
	pages = {27716--27725},
}

@article{corliss_magnetic_1956,
	title = {Magnetic {Structures} of the {Polymorphic} {Forms} of {Manganous} {Sulfide}},
	volume = {104},
	copyright = {http://link.aps.org/licenses/aps-default-license},
	issn = {0031-899X},
	url = {https://link.aps.org/doi/10.1103/PhysRev.104.924},
	doi = {10.1103/PhysRev.104.924},
	number = {4},
	journal = {Phys. Rev.},
	author = {Corliss, Lester and Elliott, Norman and Hastings, Julius},
	month = nov,
	year = {1956},
	pages = {924--928},
}

@article{chilton_use_1984,
	title = {Use of a {Paramagnetic} {Substance}, {Colloidal} {Manganese} {Sulfide}, as an {NMR} {Contrast} {Material} in {Rats}},
	volume = {25},
	issn = {0161-5505},
	url = {https://jnm.snmjournals.org/content/25/5/604},
	number = {5},
	journal = {Journal of Nuclear Medicine},
	author = {Chilton, Henry M. and Jackets, Susan C. and Hinson, William H. and Ekstrand, Kenneth},
	year = {1984},
	pages = {604--607},
}

@article{meng_phase_2016,
	title = {Phase transfer preparation of ultrasmall {MnS} nanocrystals with a high performance {MRI} contrast agent},
	volume = {6},
	issn = {2046-2069},
	url = {https://xlink.rsc.org/?DOI=C5RA24775F},
	doi = {10.1039/C5RA24775F},
	number = {9},
	journal = {RSC Adv.},
	author = {Meng, Jing and Zhao, Yizhe and Li, Zhongfeng and Wang, Ligang and Tian, Yang},
	year = {2016},
	pages = {6878--6887},
}

@article{zhang_hydrothermal_2008,
	title = {Hydrothermal synthesis and electrochemical properties of alpha-manganese sulfide submicrocrystals as an attractive electrode material for lithium-ion batteries},
	volume = {111},
	copyright = {https://www.elsevier.com/tdm/userlicense/1.0/},
	issn = {02540584},
	url = {https://linkinghub.elsevier.com/retrieve/pii/S0254058408001910},
	doi = {10.1016/j.matchemphys.2008.03.040},
	number = {1},
	journal = {Materials Chemistry and Physics},
	author = {Zhang, Ning and Yi, Ran and Wang, Zhong and Shi, Rongrong and Wang, Haidong and Qiu, Guanzhou and Liu, Xiaohe},
	month = sep,
	year = {2008},
	pages = {13--16},
}

@article{tang_morphology_2015,
	title = {Morphology controlled synthesis of monodispersed manganese sulfide nanocrystals and their primary application in supercapacitors with high performances},
	volume = {51},
	issn = {1359-7345, 1364-548X},
	url = {https://xlink.rsc.org/?DOI=C5CC01700A},
	doi = {10.1039/C5CC01700A},
	number = {43},
	journal = {Chem. Commun.},
	author = {Tang, Yongfu and Chen, Teng and Yu, Shengxue},
	year = {2015},
	pages = {9018--9021},
}

@article{yang_size-controlled_2012,
	title = {Size-{Controlled} {Synthesis} of {Bifunctional} {Magnetic} and {Ultraviolet} {Optical} {Rock}-{Salt} {MnS} {Nanocube} {Superlattices}},
	volume = {28},
	issn = {0743-7463, 1520-5827},
	url = {https://pubs.acs.org/doi/10.1021/la304228w},
	doi = {10.1021/la304228w},
	number = {51},
	journal = {Langmuir},
	author = {Yang, Xinyi and Wang, Yingnan and Sui, Yongming and Huang, Xiaoli and Cui, Tian and Wang, Chunzhong and Liu, Bingbing and Zou, Guangtian and Zou, Bo},
	month = dec,
	year = {2012},
	pages = {17811--17816},
}

@article{chen_low-temperature_2023,
	title = {Low-{Temperature} {Growth} of {Rock} {Salt} {MnS} {Nanocrystals} with {Facet}-{Dependent} {Behaviors}},
	volume = {35},
	copyright = {https://doi.org/10.15223/policy-029},
	issn = {0897-4756, 1520-5002},
	url = {https://pubs.acs.org/doi/10.1021/acs.chemmater.3c01883},
	doi = {10.1021/acs.chemmater.3c01883},
	number = {18},
	journal = {Chem. Mater.},
	author = {Chen, Chien-Kai and Chen, Bo-Hao and Huang, Michael H.},
	month = sep,
	year = {2023},
	pages = {7859--7866},
}

@article{lu_metastable_2001,
	title = {Metastable {MnS} {Crystallites} through {Solvothermal} {Synthesis}},
	volume = {13},
	issn = {0897-4756, 1520-5002},
	url = {https://pubs.acs.org/doi/10.1021/cm010049j},
	doi = {10.1021/cm010049j},
	number = {6},
	journal = {Chem. Mater.},
	author = {Lu, Jun and Qi, Pengfei and Peng, Yiya and Meng, Zhaoyu and Yang, Zhiping and Yu, Weichao and Qian, Yitai},
	month = jun,
	year = {2001},
	pages = {2169--2172},
}

@article{kan_synthesis_2001,
	title = {Synthesis, characterization, and magnetic properties of $\alpha$‐{MnS} nanocrystals},
	volume = {41},
	copyright = {http://onlinelibrary.wiley.com/termsAndConditions\#vor},
	issn = {0021-2148, 1869-5868},
	url = {https://onlinelibrary.wiley.com/doi/10.1560/1FB3-1PF4-72JQ-0AQC},
	doi = {10.1560/1FB3-1PF4-72JQ-0AQC},
	number = {1},
	journal = {Israel Journal of Chemistry},
	author = {Kan, Shihai and Felner, Israel and Banin, Uri},
	month = nov,
	year = {2001},
	pages = {55--62},
}

@article{jun_architectural_2002,
	title = {Architectural {Control} of {Magnetic} {Semiconductor} {Nanocrystals}},
	volume = {124},
	issn = {0002-7863, 1520-5126},
	url = {https://pubs.acs.org/doi/10.1021/ja016887w},
	doi = {10.1021/ja016887w},
	number = {4},
	journal = {J. Am. Chem. Soc.},
	author = {Jun, Young-wook and Jung, Yoon-young and Cheon, Jinwoo},
	month = jan,
	year = {2002},
	pages = {615--619},
}

@article{joo_generalized_2003,
	title = {Generalized and {Facile} {Synthesis} of {Semiconducting} {Metal} {Sulfide} {Nanocrystals}},
	volume = {125},
	issn = {0002-7863, 1520-5126},
	url = {https://pubs.acs.org/doi/10.1021/ja0357902},
	doi = {10.1021/ja0357902},
	number = {36},
	journal = {J. Am. Chem. Soc.},
	author = {Joo, Jin and Na, Hyon Bin and Yu, Taekyung and Yu, Jung Ho and Kim, Young Woon and Wu, Fanxin and Zhang, Jin Z. and Hyeon, Taeghwan},
	month = sep,
	year = {2003},
	pages = {11100--11105},
}

@article{michel_hydrothermal_2006,
	title = {Hydrothermal {Synthesis} of {Pure} $\alpha$-{Phase} {Manganese}({II}) {Sulfide} without the {Use} of {Organic} {Reagents}},
	volume = {18},
	issn = {0897-4756, 1520-5002},
	url = {https://pubs.acs.org/doi/10.1021/cm048320v},
	doi = {10.1021/cm048320v},
	number = {7},
	journal = {Chem. Mater.},
	author = {Michel, F. M. and Schoonen, M. A. A. and Zhang, X. V. and Martin, S. T. and Parise, J. B.},
	month = apr,
	year = {2006},
	pages = {1726--1736},
}

@article{puglisi_monodisperse_2010,
	title = {Monodisperse {Octahedral} $\alpha$-{MnS} and {MnO} {Nanoparticles} by the {Decomposition} of {Manganese} {Oleate} in the {Presence} of {Sulfur}},
	volume = {22},
	issn = {0897-4756, 1520-5002},
	url = {https://pubs.acs.org/doi/10.1021/cm903735e},
	doi = {10.1021/cm903735e},
	number = {9},
	journal = {Chem. Mater.},
	author = {Puglisi, Alessandra and Mondini, Sara and Cenedese, Simone and Ferretti, Anna M. and Santo, Nadia and Ponti, Alessandro},
	month = may,
	year = {2010},
	pages = {2804--2813},
}

@article{gui_hydrothermal_2011,
	title = {Hydrothermal synthesis of uniform rock salt ($\alpha$-) {MnS} transformation from wurtzite ($\gamma$-) {MnS}},
	volume = {125},
	copyright = {https://www.elsevier.com/tdm/userlicense/1.0/},
	issn = {02540584},
	url = {https://linkinghub.elsevier.com/retrieve/pii/S0254058410007984},
	doi = {10.1016/j.matchemphys.2010.09.071},
	number = {3},
	journal = {Materials Chemistry and Physics},
	author = {Gui, Yicai and Qian, Liwu and Qian, Xuefeng},
	month = feb,
	year = {2011},
	pages = {698--703},
}

@article{yang_polymorphism_2012,
	title = {Polymorphism and {Formation} {Mechanism} of {Nanobipods} in {Manganese} {Sulfide} {Nanocrystals} {Induced} by {Temperature} or {Pressure}},
	volume = {116},
	issn = {1932-7447, 1932-7455},
	url = {https://pubs.acs.org/doi/10.1021/jp209591r},
	doi = {10.1021/jp209591r},
	number = {5},
	journal = {J. Phys. Chem. C},
	author = {Yang, Xinyi and Wang, Yingnan and Wang, Kai and Sui, Yongming and Zhang, Meiguang and Li, Bing and Ma, Yanming and Liu, Bingbing and Zou, Guangtian and Zou, Bo},
	month = feb,
	year = {2012},
	pages = {3292--3297},
}

@article{anie.201701087,
    author = {Fenton, Julie L. and Schaak, Raymond E.},
    title = {Structure-Selective Cation Exchange in the Synthesis of Zincblende MnS and CoS Nanocrystals},
    journal = {Angewandte Chemie International Edition},
    volume = {56},
    number = {23},
    pages = {6464-6467},
    doi = {https://doi.org/10.1002/anie.201701087},
    url = {https://onlinelibrary.wiley.com/doi/abs/10.1002/anie.201701087},
    year = {2017}
}

@article{HAO20179,
    title = {Studies on intrinsic phase-dependent electrochemical properties of MnS nanocrystals as anodes for lithium-ion batteries},
    journal = {Journal of Power Sources},
    volume = {338},
    pages = {9-16},
    year = {2017},
    issn = {0378-7753},
    doi = {https://doi.org/10.1016/j.jpowsour.2016.11.032},
    url = {https://www.sciencedirect.com/science/article/pii/S0378775316315555},
    author = {Yong Hao and Chunhui Chen and Xinyi Yang and Guanjun Xiao and Bo Zou and Jianwen Yang and Chunlei Wang},
}

@article{gendler_halide-driven_2023,
	title = {Halide-driven polymorph selectivity in the synthesis of {MnX} ({X} = {S}, {Se}) nanoparticles},
	volume = {15},
	issn = {2040-3364, 2040-3372},
	url = {http://xlink.rsc.org/?DOI=D2NR05854E},
	doi = {10.1039/D2NR05854E},
	number = {6},
	journal = {Nanoscale},
	author = {Gendler, Danielle and Bi, Jiaying and Mekan, Deep and Warokomski, Ashley and Armstrong, Cameron and Hernandez-Pagan, Emil A.},
	year = {2023},
	pages = {2650--2658},
}

@article{vitos_surface_1998,
	title = {The surface energy of metals},
	volume = {411},
	copyright = {https://www.elsevier.com/tdm/userlicense/1.0/},
	issn = {00396028},
	url = {https://linkinghub.elsevier.com/retrieve/pii/S003960289800363X},
	doi = {10.1016/S0039-6028(98)00363-X},
	number = {1-2},
	journal = {Surface Science},
	author = {Vitos, L. and Ruban, A.V. and Skriver, H.L. and Kollár, J.},
	month = aug,
	year = {1998},
	pages = {186--202},
}

@article{xia_shape-controlled_2015,
	title = {Shape-{Controlled} {Synthesis} of {Colloidal} {Metal} {Nanocrystals}: {Thermodynamic} versus {Kinetic} {Products}},
	volume = {137},
	issn = {0002-7863, 1520-5126},
	shorttitle = {Shape-{Controlled} {Synthesis} of {Colloidal} {Metal} {Nanocrystals}},
	url = {https://pubs.acs.org/doi/10.1021/jacs.5b04641},
	doi = {10.1021/jacs.5b04641},
	number = {25},
	journal = {J. Am. Chem. Soc.},
	author = {Xia, Younan and Xia, Xiaohu and Peng, Hsin-Chieh},
	month = jul,
	year = {2015},
	pages = {7947--7966},
}

@article{tran_surface_2016,
	title = {Surface energies of elemental crystals},
	volume = {3},
	issn = {2052-4463},
	url = {https://doi.org/10.1038/sdata.2016.80},
	doi = {10.1038/sdata.2016.80},
	number = {1},
	journal = {Scientific Data},
	author = {Tran, Richard and Xu, Zihan and Radhakrishnan, Balachandran and Winston, Donald and Sun, Wenhao and Persson, Kristin A. and Ong, Shyue Ping},
	month = sep,
	year = {2016},
	pages = {160080},
}

@article{fichthorn_theory_2023,
	title = {Theory of {Anisotropic} {Metal} {Nanostructures}},
	volume = {123},
	copyright = {https://doi.org/10.15223/policy-029},
	issn = {0009-2665, 1520-6890},
	url = {https://pubs.acs.org/doi/10.1021/acs.chemrev.2c00831},
	doi = {10.1021/acs.chemrev.2c00831},
	number = {7},
	journal = {Chem. Rev.},
	author = {Fichthorn, Kristen A.},
	month = apr,
	year = {2023},
	pages = {4146--4183},
}

@article{xiao_high-index-facet-_2020,
	title = {High-{Index}-{Facet}- and {High}-{Surface}-{Energy} {Nanocrystals} of {Metals} and {Metal} {Oxides} as {Highly} {Efficient} {Catalysts}},
	volume = {4},
	issn = {25424351},
	url = {https://linkinghub.elsevier.com/retrieve/pii/S2542435120304591},
	doi = {10.1016/j.joule.2020.10.002},
	number = {12},
	journal = {Joule},
	author = {Xiao, Chi and Lu, Bang-An and Xue, Peng and Tian, Na and Zhou, Zhi-You and Lin, Xiao and Lin, Wen-Feng and Sun, Shi-Gang},
	month = dec,
	year = {2020},
	pages = {2562--2598},
}

@article{boukouvala_approaches_2021,
	title = {Approaches to modelling the shape of nanocrystals},
	volume = {8},
	issn = {2196-5404},
	url = {https://nanoconvergencejournal.springeropen.com/articles/10.1186/s40580-021-00275-6},
	doi = {10.1186/s40580-021-00275-6},
	number = {1},
	journal = {Nano Convergence},
	author = {Boukouvala, Christina and Daniel, Joshua and Ringe, Emilie},
	month = sep,
	year = {2021},
	pages = {26},
}

@article{sanspeur_wherewulff_2023,
	title = {\textit{{WhereWulff}} : {A} {Semiautonomous} {Workflow} for {Systematic} {Catalyst} {Surface} {Reactivity} under {Reaction} {Conditions}},
	volume = {63},
	copyright = {https://creativecommons.org/licenses/by/4.0/},
	issn = {1549-9596, 1549-960X},
	shorttitle = {\textit{{WhereWulff}}},
	url = {https://pubs.acs.org/doi/10.1021/acs.jcim.3c00142},
	doi = {10.1021/acs.jcim.3c00142},
	number = {8},
	journal = {J. Chem. Inf. Model.},
	author = {Sanspeur, Rohan Yuri and Heras-Domingo, Javier and Kitchin, John R. and Ulissi, Zachary},
	month = apr,
	year = {2023},
	pages = {2427--2437},
}

@article{sun_efficient_2013,
	title = {Efficient creation and convergence of surface slabs},
	volume = {617},
	issn = {00396028},
	url = {https://linkinghub.elsevier.com/retrieve/pii/S003960281300160X},
	doi = {10.1016/j.susc.2013.05.016},
	journal = {Surface Science},
	author = {Sun, Wenhao and Ceder, Gerbrand},
	month = nov,
	year = {2013},
	pages = {53--59},
}

@article{jp0445573,
    author = {Manna, Liberato and Wang and Cingolani, Roberto and Alivisatos, A. Paul},
    title = {First-Principles Modeling of Unpassivated and Surfactant-Passivated Bulk Facets of Wurtzite CdSe: A Model System for Studying the Anisotropic Growth of CdSe Nanocrystals},
    journal = {The Journal of Physical Chemistry B},
    volume = {109},
    number = {13},
    pages = {6183-6192},
    year = {2005},
    doi = {10.1021/jp0445573},
    URL = {https://doi.org/10.1021/jp0445573},
}

@article{tian_dft_2018,
	title = {A {DFT} based method for calculating the surface energies of asymmetric {MoP} facets},
	volume = {427},
	issn = {01694332},
	url = {https://linkinghub.elsevier.com/retrieve/pii/S0169433217325485},
	doi = {10.1016/j.apsusc.2017.08.172},
	journal = {Applied Surface Science},
	author = {Tian, Xinxin and Wang, Tao and Fan, Lifang and Wang, Yuekui and Lu, Haigang and Mu, Yuewen},
	month = jan,
	year = {2018},
	pages = {357--362},
}

@article{yoo_efficient_2021,
	title = {Efficient electronic passivation scheme for computing low-symmetry compound semiconductor surfaces in density-functional theory slab calculations},
	volume = {5},
	issn = {2475-9953},
	url = {https://link.aps.org/doi/10.1103/PhysRevMaterials.5.044605},
	doi = {10.1103/PhysRevMaterials.5.044605},
	number = {4},
	journal = {Phys. Rev. Materials},
	author = {Yoo, Su-Hyun and Lymperakis, Liverios and Neugebauer, Jörg},
	month = apr,
	year = {2021},
	pages = {044605},
}

@article{stuart_method_2023,
	title = {A method of calculating surface energies for asymmetric slab models},
	volume = {25},
	issn = {1463-9076, 1463-9084},
	url = {http://xlink.rsc.org/?DOI=D2CP04460A},
	doi = {10.1039/D2CP04460A},
	number = {19},
	journal = {Phys. Chem. Chem. Phys.},
	author = {Stuart, Natalie M. and Sohlberg, Karl},
	year = {2023},
	pages = {13351--13358},
}

@article{zhang_surface_2004,
	title = {Surface {Energy} and the {Common} {Dangling} {Bond} {Rule} for {Semiconductors}},
	volume = {92},
	copyright = {http://link.aps.org/licenses/aps-default-license},
	issn = {0031-9007, 1079-7114},
	url = {https://link.aps.org/doi/10.1103/PhysRevLett.92.086102},
	doi = {10.1103/PhysRevLett.92.086102},
	number = {8},
	journal = {Phys. Rev. Lett.},
	author = {Zhang, S. B. and Wei, Su-Huai},
	month = feb,
	year = {2004},
	pages = {086102},
}

@article{dreyer_absolute_2014,
	title = {Absolute surface energies of polar and nonpolar planes of {GaN}},
	volume = {89},
	copyright = {http://link.aps.org/licenses/aps-default-license},
	issn = {1098-0121, 1550-235X},
	url = {https://link.aps.org/doi/10.1103/PhysRevB.89.081305},
	doi = {10.1103/PhysRevB.89.081305},
	number = {8},
	journal = {Phys. Rev. B},
	author = {Dreyer, C. E. and Janotti, A. and Van De Walle, C. G.},
	month = feb,
	year = {2014},
	pages = {081305},
}

@article{akiyama_modified_2019,
	title = {Modified approach for calculating individual energies of polar and semipolar surfaces of group-{III} nitrides},
	volume = {3},
	issn = {2475-9953},
	url = {https://link.aps.org/doi/10.1103/PhysRevMaterials.3.023401},
	doi = {10.1103/PhysRevMaterials.3.023401},
	number = {2},
	journal = {Phys. Rev. Materials},
	author = {Akiyama, Toru and Seta, Yuki and Nakamura, Kohji and Ito, Tomonori},
	month = feb,
	year = {2019},
	pages = {023401},
}

@article{jin_absolute_2021,
	title = {Absolute surface energies of wurtzite ($101\overline{1}$) surfaces and the instability of the cation-adsorbed surfaces of {II}–{VI} semiconductors},
	volume = {119},
	issn = {0003-6951, 1077-3118},
	url = {https://pubs.aip.org/apl/article/119/20/201603/40503/Absolute-surface-energies-of-wurtzite-101-1},
	doi = {10.1063/5.0068226},
	number = {20},
	journal = {Applied Physics Letters},
	author = {Jin, Wentao and Chen, Guangde and Duan, Xiangyang and Araujo, C. Moyses and Jia, Xubo and Yin, Yuan and Wu, Yelong},
	month = nov,
	year = {2021},
	pages = {201603},
}

@article{li_computing_2015,
	title = {Computing {Equilibrium} {Shapes} of {Wurtzite} {Crystals}: {The} {Example} of {GaN}},
	volume = {115},
	copyright = {http://link.aps.org/licenses/aps-default-license},
	issn = {0031-9007, 1079-7114},
	shorttitle = {Computing {Equilibrium} {Shapes} of {Wurtzite} {Crystals}},
	url = {https://link.aps.org/doi/10.1103/PhysRevLett.115.085503},
	doi = {10.1103/PhysRevLett.115.085503},
	number = {8},
	journal = {Phys. Rev. Lett.},
	author = {Li, Hong and Geelhaar, Lutz and Riechert, Henning and Draxl, Claudia},
	month = aug,
	year = {2015},
	pages = {085503},
}

@article{wang_defining_2022,
	title = {Defining shapes of two-dimensional crystals with undefinable edge energies},
	volume = {2},
	issn = {2662-8457},
	url = {https://www.nature.com/articles/s43588-022-00347-5},
	doi = {10.1038/s43588-022-00347-5},
	number = {11},
	journal = {Nat Comput Sci},
	author = {Wang, Luqing and Shirodkar, Sharmila N. and Zhang, Zhuhua and Yakobson, Boris I.},
	month = nov,
	year = {2022},
	pages = {729--735},
}

@article{Perdew:1996pki,
	title = {Generalized {Gradient} {Approximation} {Made} {Simple}},
	volume = {77},
	copyright = {http://link.aps.org/licenses/aps-default-license},
	issn = {0031-9007, 1079-7114},
	url = {https://link.aps.org/doi/10.1103/PhysRevLett.77.3865},
	doi = {10.1103/PhysRevLett.77.3865},
	number = {18},
	journal = {Phys. Rev. Lett.},
	author = {Perdew, John P. and Burke, Kieron and Ernzerhof, Matthias},
	month = oct,
	year = {1996},
	pages = {3865},
}

@article{sun_strongly_2015,
	title = {Strongly {Constrained} and {Appropriately} {Normed} {Semilocal} {Density} {Functional}},
	volume = {115},
	issn = {0031-9007, 1079-7114},
	url = {https://link.aps.org/doi/10.1103/PhysRevLett.115.036402},
	doi = {10.1103/PhysRevLett.115.036402},
	number = {3},
	journal = {Phys. Rev. Lett.},
	author = {Sun, Jianwei and Ruzsinszky, Adrienn and Perdew, John P.},
	month = jul,
	year = {2015},
	pages = {036402},
}

@article{furness_accurate_2020,
	title = {Accurate and {Numerically} {Efficient} r$^2${SCAN} {Meta}-{Generalized} {Gradient} {Approximation}},
	volume = {11},
	issn = {1948-7185, 1948-7185},
	url = {https://pubs.acs.org/doi/10.1021/acs.jpclett.0c02405},
	doi = {10.1021/acs.jpclett.0c02405},
	number = {19},
	journal = {J. Phys. Chem. Lett.},
	author = {Furness, James W. and Kaplan, Aaron D. and Ning, Jinliang and Perdew, John P. and Sun, Jianwei},
	month = oct,
	year = {2020},
	pages = {8208--8215},
}

@article{peng_versatile_2016,
	title = {Versatile van der {Waals} {Density} {Functional} {Based} on a {Meta}-{Generalized} {Gradient} {Approximation}},
	volume = {6},
	copyright = {http://creativecommons.org/licenses/by/3.0/},
	issn = {2160-3308},
	url = {https://link.aps.org/doi/10.1103/PhysRevX.6.041005},
	doi = {10.1103/PhysRevX.6.041005},
	number = {4},
	journal = {Phys. Rev. X},
	author = {Peng, Haowei and Yang, Zeng-Hui and Perdew, John P. and Sun, Jianwei},
	month = oct,
	year = {2016},
	pages = {041005},
}

@article{ning_workhorse_2022,
	title = {Workhorse minimally empirical dispersion-corrected density functional with tests for weakly bound systems: r$^2${SCAN} + {rVV} 10},
	volume = {106},
	issn = {2469-9950, 2469-9969},
	shorttitle = {Workhorse minimally empirical dispersion-corrected density functional with tests for weakly bound systems},
	url = {https://link.aps.org/doi/10.1103/PhysRevB.106.075422},
	doi = {10.1103/PhysRevB.106.075422},
	number = {7},
	journal = {Phys. Rev. B},
	author = {Ning, Jinliang and Kothakonda, Manish and Furness, James W. and Kaplan, Aaron D. and Ehlert, Sebastian and Brandenburg, Jan Gerit and Perdew, John P. and Sun, Jianwei},
	month = aug,
	year = {2022},
	pages = {075422},
}

@article{kothakonda_testing_2023,
	title = {Testing the r$^{\textrm{2}}$ {SCAN} {Density} {Functional} for the {Thermodynamic} {Stability} of {Solids} with and without a van der {Waals} {Correction}},
	volume = {3},
	copyright = {https://creativecommons.org/licenses/by/4.0/},
	issn = {2694-2461, 2694-2461},
	url = {https://pubs.acs.org/doi/10.1021/acsmaterialsau.2c00059},
	doi = {10.1021/acsmaterialsau.2c00059},
	number = {2},
	journal = {ACS Mater. Au},
	author = {Kothakonda, Manish and Kaplan, Aaron D. and Isaacs, Eric B. and Bartel, Christopher J. and Furness, James W. and Ning, Jinliang and Wolverton, Chris and Perdew, John P. and Sun, Jianwei},
	month = mar,
	year = {2023},
	pages = {102--111},
}

@article{kingsbury_performance_2022,
	title = {Performance comparison of r 2 {SCAN} and {SCAN} {metaGGA} density functionals for solid materials via an automated, high-throughput computational workflow},
	volume = {6},
	issn = {2475-9953},
	url = {https://link.aps.org/doi/10.1103/PhysRevMaterials.6.013801},
	doi = {10.1103/PhysRevMaterials.6.013801},
	number = {1},
	journal = {Phys. Rev. Materials},
	author = {Kingsbury, Ryan and Gupta, Ayush S. and Bartel, Christopher J. and Munro, Jason M. and Dwaraknath, Shyam and Horton, Matthew and Persson, Kristin A.},
	month = jan,
	year = {2022},
	pages = {013801},
}

@article{horton_accelerated_2025,
	title = {Accelerated data-driven materials science with the {Materials} {Project}},
	issn = {1476-1122, 1476-4660},
	url = {https://www.nature.com/articles/s41563-025-02272-0},
	doi = {10.1038/s41563-025-02272-0},
	journal = {Nat. Mater.},
	author = {Horton, Matthew K. and Huck, Patrick and Yang, Ruo Xi and Munro, Jason M. and Dwaraknath, Shyam and Ganose, Alex M. and Kingsbury, Ryan S. and Wen, Mingjian and Shen, Jimmy X. and Mathis, Tyler S. and Kaplan, Aaron D. and Berket, Karlo and Riebesell, Janosh and George, Janine and Rosen, Andrew S. and Spotte-Smith, Evan W. C. and McDermott, Matthew J. and Cohen, Orion A. and Dunn, Alex and Kuner, Matthew C. and Rignanese, Gian-Marco and Petretto, Guido and Waroquiers, David and Griffin, Sinead M. and Neaton, Jeffrey B. and Chrzan, Daryl C. and Asta, Mark and Hautier, Geoffroy and Cholia, Shreyas and Ceder, Gerbrand and Ong, Shyue Ping and Jain, Anubhav and Persson, Kristin A.},
	month = jul,
	year = {2025},
}

@misc{barroso-luque_open_2024,
	title = {Open {Materials} 2024 ({OMat24}) {Inorganic} {Materials} {Dataset} and {Models}},
	url = {http://arxiv.org/abs/2410.12771},
	doi = {10.48550/arXiv.2410.12771},
	publisher = {arXiv},
	author = {Barroso-Luque, Luis and Shuaibi, Muhammed and Fu, Xiang and Wood, Brandon M. and Dzamba, Misko and Gao, Meng and Rizvi, Ammar and Zitnick, C. Lawrence and Ulissi, Zachary W.},
	month = oct,
	year = {2024},
}

@misc{kuner_mp-aloe_2025,
	title = {{MP}-{ALOE}: {An} {r$^2$SCAN} dataset for universal machine learning interatomic potentials},
	shorttitle = {{MP}-{ALOE}},
	url = {http://arxiv.org/abs/2507.05559},
	doi = {10.48550/arXiv.2507.05559},
	publisher = {arXiv},
	author = {Kuner, Matthew C. and Kaplan, Aaron D. and Persson, Kristin A. and Asta, Mark and Chrzan, Daryl C.},
	month = jul,
	year = {2025},
}

@misc{kaplan_foundational_2025,
	title = {A {Foundational} {Potential} {Energy} {Surface} {Dataset} for {Materials}},
	url = {http://arxiv.org/abs/2503.04070},
	doi = {10.48550/arXiv.2503.04070},
	publisher = {arXiv},
	author = {Kaplan, Aaron D. and Liu, Runze and Qi, Ji and Ko, Tsz Wai and Deng, Bowen and Riebesell, Janosh and Ceder, Gerbrand and Persson, Kristin A. and Ong, Shyue Ping},
	month = mar,
	year = {2025},
}

@article{10.1063/5.0297006,
    author = {Batatia, Ilyes and Benner, Philipp and Chiang, Yuan and Elena, Alin M. and Kovács, Dávid P. and Riebesell, Janosh and Advincula, Xavier R. and Asta, Mark and Avaylon, Matthew and Baldwin, William J. and Berger, Fabian and Bernstein, Noam and Bhowmik, Arghya and Bigi, Filippo and Blau, Samuel M. and Cărare, Vlad and Ceriotti, Michele and Chong, Sanggyu and Darby, James P. and De, Sandip and Della Pia, Flaviano and Deringer, Volker L. and Elijošius, Rokas and El-Machachi, Zakariya and Fako, Edvin and Falcioni, Fabio and Ferrari, Andrea C. and Gardner, John L. A. and Gawkowski, Mikołaj J. and Genreith-Schriever, Annalena and George, Janine and Goodall, Rhys E. A. and Grandel, Jonas and Grey, Clare P. and Grigorev, Petr and Han, Shuang and Handley, Will and Heenen, Hendrik H. and Hermansson, Kersti and Ho, Cheuk Hin and Hofmann, Stephan and Holm, Christian and Jaafar, Jad and Jakob, Konstantin S. and Jung, Hyunwook and Kapil, Venkat and Kaplan, Aaron D. and Karimitari, Nima and Kermode, James R. and Kourtis, Panagiotis and Kroupa, Namu and Kullgren, Jolla and Kuner, Matthew C. and Kuryla, Domantas and Liepuoniute, Guoda and Lin, Chen and Margraf, Johannes T. and Magdău, Ioan-Bogdan and Michaelides, Angelos and Moore, J. Harry and Naik, Aakash A. and Niblett, Samuel P. and Norwood, Sam Walton and O’Neill, Niamh and Ortner, Christoph and Persson, Kristin A. and Reuter, Karsten and Rosen, Andrew S. and Rosset, Louise A. M. and Schaaf, Lars L. and Schran, Christoph and Shi, Benjamin X. and Sivonxay, Eric and Stenczel, Tamás K. and Sutton, Christopher and Svahn, Viktor and Swinburne, Thomas D. and Tilly, Jules and van der Oord, Cas and Vargas, Santiago and Varga-Umbrich, Eszter and Vegge, Tejs and Vondrák, Martin and Wang, Yangshuai and Witt, William C. and Wolf, Thomas and Zills, Fabian and Csányi, Gábor},
    title = {A foundation model for atomistic materials chemistry},
    journal = {The Journal of Chemical Physics},
    volume = {163},
    number = {18},
    pages = {184110},
    year = {2025},
    month = {11},
    issn = {0021-9606},
    doi = {10.1063/5.0297006},
    url = {https://doi.org/10.1063/5.0297006},
}

@misc{wood_uma_2025,
	title = {{UMA}: {A} {Family} of {Universal} {Models} for {Atoms}},
	shorttitle = {{UMA}},
	url = {http://arxiv.org/abs/2506.23971},
	doi = {10.48550/arXiv.2506.23971},
	publisher = {arXiv},
	author = {Wood, Brandon M. and Dzamba, Misko and Fu, Xiang and Gao, Meng and Shuaibi, Muhammed and Barroso-Luque, Luis and Abdelmaqsoud, Kareem and Gharakhanyan, Vahe and Kitchin, John R. and Levine, Daniel S. and Michel, Kyle and Sriram, Anuroop and Cohen, Taco and Das, Abhishek and Rizvi, Ammar and Sahoo, Sushree Jagriti and Ulissi, Zachary W. and Zitnick, C. Lawrence},
	month = jun,
	year = {2025},
}

@article{grimme_consistent_2010,
	title = {A consistent and accurate \textit{ab initio} parametrization of density functional dispersion correction ({DFT}-{D}) for the 94 elements {H}-{Pu}},
	volume = {132},
	issn = {0021-9606, 1089-7690},
	url = {https://pubs.aip.org/jcp/article/132/15/154104/926936/A-consistent-and-accurate-ab-initio},
	doi = {10.1063/1.3382344},
	number = {15},
	journal = {The Journal of Chemical Physics},
	author = {Grimme, Stefan and Antony, Jens and Ehrlich, Stephan and Krieg, Helge},
	month = apr,
	year = {2010},
	pages = {154104},
}

@article{grimme_effect_2011,
	title = {Effect of the damping function in dispersion corrected density functional theory},
	volume = {32},
	copyright = {http://onlinelibrary.wiley.com/termsAndConditions\#vor},
	issn = {0192-8651, 1096-987X},
	url = {https://onlinelibrary.wiley.com/doi/10.1002/jcc.21759},
	doi = {10.1002/jcc.21759},
	number = {7},
	journal = {J Comput Chem},
	author = {Grimme, Stefan and Ehrlich, Stephan and Goerigk, Lars},
	month = may,
	year = {2011},
	pages = {1456--1465},
}

@article{Zagorac:in5024,
    author = "Zagorac, D. and M{\"{u}}ller, H. and Ruehl, S. and Zagorac, J. and Rehme, S.",
    title = "{Recent developments in the Inorganic Crystal Structure Database: theoretical crystal structure data and related features}",
    journal = "Journal of Applied Crystallography",
    year = "2019",
    volume = "52",
    number = "5",
    pages = "918--925",
    month = "Oct",
    doi = {10.1107/S160057671900997X},
    url = {https://doi.org/10.1107/S160057671900997X},
}

@misc{thomas_c_allison_nist-janaf_2013,
	title = {{NIST}-{JANAF} {Thermochemical} {Tables} - {SRD} 13},
	copyright = {License Information for NIST data},
	url = {https://janaf.nist.gov/},
	doi = {10.18434/T42S31},
	publisher = {National Institute of Standards and Technology},
	author = {Thomas C. Allison},
	collaborator = {Allison, Thomas C.},
	month = jan,
	year = {2013},
}

@article{krukau_influence_2006,
	title = {Influence of the exchange screening parameter on the performance of screened hybrid functionals},
	volume = {125},
	issn = {0021-9606, 1089-7690},
	url = {https://pubs.aip.org/jcp/article/125/22/224106/953719/Influence-of-the-exchange-screening-parameter-on},
	doi = {10.1063/1.2404663},
	number = {22},
	journal = {The Journal of Chemical Physics},
	author = {Krukau, Aliaksandr V. and Vydrov, Oleg A. and Izmaylov, Artur F. and Scuseria, Gustavo E.},
	month = dec,
	year = {2006},
	pages = {224106},
}

@article{moellmann_dft-d3_2014,
	title = {{DFT}-{D3} {Study} of {Some} {Molecular} {Crystals}},
	volume = {118},
	issn = {1932-7447, 1932-7455},
	url = {https://pubs.acs.org/doi/10.1021/jp501237c},
	doi = {10.1021/jp501237c},
	number = {14},
	journal = {J. Phys. Chem. C},
	author = {Moellmann, Jonas and Grimme, Stefan},
	month = apr,
	year = {2014},
	pages = {7615--7621},
}

@inbook{thermochemical_data_barin,
    publisher = {John Wiley \& Sons, Ltd},
    isbn = {9783527619825},
    title = {Mg-Mo5Si3},
    booktitle = {Thermochemical Data of Pure Substances},
    chapter = {12},
    pages = {993-1079},
    doi = {https://doi.org/10.1002/9783527619825.ch12l},
    url = {https://onlinelibrary.wiley.com/doi/abs/10.1002/9783527619825.ch12l},
    eprint = {https://onlinelibrary.wiley.com/doi/pdf/10.1002/9783527619825.ch12l},
    year = {1995},
    author = {Barin, Ihsan}
}

@article{methfessel_high-precision_1989,
	title = {High-precision sampling for {Brillouin}-zone integration in metals},
	volume = {40},
	copyright = {http://link.aps.org/licenses/aps-default-license},
	issn = {0163-1829},
	url = {https://link.aps.org/doi/10.1103/PhysRevB.40.3616},
	doi = {10.1103/PhysRevB.40.3616},
	number = {6},
	journal = {Phys. Rev. B},
	author = {Methfessel, M. and Paxton, A. T.},
	month = aug,
	year = {1989},
	pages = {3616--3621},
}

@article{boettger_nonconvergence_1994,
	title = {Nonconvergence of surface energies obtained from thin-film calculations},
	volume = {49},
	copyright = {http://link.aps.org/licenses/aps-default-license},
	issn = {0163-1829, 1095-3795},
	url = {https://link.aps.org/doi/10.1103/PhysRevB.49.16798},
	doi = {10.1103/PhysRevB.49.16798},
	number = {23},
	journal = {Phys. Rev. B},
	author = {Boettger, J. C.},
	month = jun,
	year = {1994},
	pages = {16798--16800},
}

@article{fiorentini_extracting_1996,
	title = {Extracting convergent surface energies from slab calculations},
	volume = {8},
	issn = {0953-8984, 1361-648X},
	url = {https://iopscience.iop.org/article/10.1088/0953-8984/8/36/005},
	doi = {10.1088/0953-8984/8/36/005},
	number = {36},
	journal = {J. Phys.: Condens. Matter},
	author = {Fiorentini, Vincenzo and Methfessel, M},
	month = sep,
	year = {1996},
	pages = {6525--6529},
}

@article{sai_gautam_evaluating_2018,
	title = {Evaluating transition metal oxides within {DFT}-{SCAN} and {SCAN} + {U} frameworks for solar thermochemical applications},
	volume = {2},
	issn = {2475-9953},
	url = {https://link.aps.org/doi/10.1103/PhysRevMaterials.2.095401},
	doi = {10.1103/PhysRevMaterials.2.095401},
	number = {9},
	journal = {Phys. Rev. Materials},
	author = {Sai Gautam, Gopalakrishnan and Carter, Emily A.},
	month = sep,
	year = {2018},
	pages = {095401},
}

@article{hayun_enthalpies_2020,
	title = {Enthalpies of formation of high entropy and multicomponent alloys using oxide melt solution calorimetry},
	volume = {125},
	issn = {09669795},
	url = {https://linkinghub.elsevier.com/retrieve/pii/S0966979519311768},
	doi = {10.1016/j.intermet.2020.106897},
	journal = {Intermetallics},
	author = {Hayun, S. and Lilova, K. and Salhov, S. and Navrotsky, A.},
	month = oct,
	year = {2020},
	pages = {106897},
}

@article{lilova_synergistic_2020,
	title = {A {Synergistic} {Approach} to {Unraveling} the {Thermodynamic} {Stability} of {Binary} and {Ternary} {Chevrel} {Phase} {Sulfides}},
	volume = {32},
	copyright = {https://doi.org/10.15223/policy-029},
	issn = {0897-4756, 1520-5002},
	url = {https://pubs.acs.org/doi/10.1021/acs.chemmater.0c02648},
	doi = {10.1021/acs.chemmater.0c02648},
	number = {16},
	journal = {Chem. Mater.},
	author = {Lilova, Kristina and Perryman, Joseph T. and Singstock, Nicholas R. and Abramchuk, Mykola and Subramani, Tamilarasan and Lam, Andy and Yoo, Ray and Ortiz-Rodríguez, Jessica C. and Musgrave, Charles B. and Navrotsky, Alexandra and Velázquez, Jesús M.},
	month = aug,
	year = {2020},
	pages = {7044--7051},
}

@article{subramani_greigite_2020,
	title = {Greigite ({Fe}$_{\textrm{3}}$ {S}$_{\textrm{4}}$ ) is thermodynamically stable: {Implications} for its terrestrial and planetary occurrence},
	volume = {117},
	issn = {0027-8424, 1091-6490},
	shorttitle = {Greigite ({Fe}$_{\textrm{3}}$ {S}$_{\textrm{4}}$ ) is thermodynamically stable},
	url = {https://pnas.org/doi/full/10.1073/pnas.2017312117},
	doi = {10.1073/pnas.2017312117},
	number = {46},
	journal = {Proc. Natl. Acad. Sci. U.S.A.},
	author = {Subramani, Tamilarasan and Lilova, Kristina and Abramchuk, Mykola and Leinenweber, Kurt D. and Navrotsky, Alexandra},
	month = nov,
	year = {2020},
	pages = {28645--28648},
}

@article{subramani_surface_2023,
	title = {Surface energetics of wurtzite and sphalerite polymorphs of zinc sulfide and implications for their formation in nature},
	volume = {340},
	issn = {00167037},
	url = {https://linkinghub.elsevier.com/retrieve/pii/S0016703722006020},
	doi = {10.1016/j.gca.2022.11.003},
	journal = {Geochimica et Cosmochimica Acta},
	author = {Subramani, Tamilarasan and Lilova, Kristina and Householder, Megan and Yang, Shuhao and Lyons, James and Navrotsky, Alexandra},
	month = jan,
	year = {2023},
	pages = {99--107},
}

@article{barnard_model_2004,
	title = {A model for the phase stability of arbitrary nanoparticles as a function of size and shape},
	volume = {121},
	issn = {0021-9606, 1089-7690},
	url = {https://pubs.aip.org/jcp/article/121/9/4276/186794/A-model-for-the-phase-stability-of-arbitrary},
	doi = {10.1063/1.1775770},
	number = {9},
	journal = {The Journal of Chemical Physics},
	author = {Barnard, A. S. and Zapol, P.},
	month = sep,
	year = {2004},
	pages = {4276--4283},
}

@article{zhang_surface_2007,
	title = {Surface {Enthalpies} of {Nanophase} {ZnO} with {Different} {Morphologies}},
	volume = {19},
	issn = {0897-4756, 1520-5002},
	url = {https://pubs.acs.org/doi/10.1021/cm0711919},
	doi = {10.1021/cm0711919},
	number = {23},
	journal = {Chem. Mater.},
	author = {Zhang, Peng and Xu, Fen and Navrotsky, Alexandra and Lee, Jong Soo and Kim, Sangtae and Liu, Jun},
	month = nov,
	year = {2007},
	pages = {5687--5693},
}

@article{lilova_energetics_2014,
	title = {Energetics of {Spinels} in the {Fe}—{Ti}—{O} {System} at the {Nanoscale}},
	volume = {15},
	issn = {1439-4235, 1439-7641},
	url = {https://chemistry-europe.onlinelibrary.wiley.com/doi/10.1002/cphc.201402441},
	doi = {10.1002/cphc.201402441},
	number = {16},
	journal = {ChemPhysChem},
	author = {Lilova, Kristina I. and Pearce, Carolyn I. and Rosso, Kevin M. and Navrotsky, Alexandra},
	month = nov,
	year = {2014},
	pages = {3655--3662},
}

@article{birkner_thermodynamics_2012,
	title = {Thermodynamics of manganese oxides: {Effects} of particle size and hydration on oxidation-reduction equilibria among hausmannite, bixbyite, and pyrolusite},
	volume = {97},
	issn = {0003-004X},
	shorttitle = {Thermodynamics of manganese oxides},
	url = {https://www.degruyter.com/document/doi/10.2138/am.2012.3982/html},
	doi = {10.2138/am.2012.3982},
	number = {8-9},
	journal = {American Mineralogist},
	author = {Birkner, N. and Navrotsky, A.},
	month = aug,
	year = {2012},
	pages = {1291--1298},
}

@article{kresse_ultrasoft_1999,
	title = {From ultrasoft pseudopotentials to the projector augmented-wave method},
	volume = {59},
	copyright = {http://link.aps.org/licenses/aps-default-license},
	issn = {0163-1829, 1095-3795},
	url = {https://link.aps.org/doi/10.1103/PhysRevB.59.1758},
	doi = {10.1103/PhysRevB.59.1758},
	number = {3},
	journal = {Phys. Rev. B},
	author = {Kresse, G. and Joubert, D.},
	month = jan,
	year = {1999},
	pages = {1758--1775},
}

@article{kresse_ab_1993,
	title = {\textit{{Ab} initio} molecular dynamics for liquid metals},
	volume = {47},
	copyright = {http://link.aps.org/licenses/aps-default-license},
	issn = {0163-1829, 1095-3795},
	url = {https://link.aps.org/doi/10.1103/PhysRevB.47.558},
	doi = {10.1103/PhysRevB.47.558},
	number = {1},
	journal = {Phys. Rev. B},
	author = {Kresse, G. and Hafner, J.},
	month = jan,
	year = {1993},
	pages = {558--561},
}

@article{kresse_efficient_1996,
	title = {Efficient iterative schemes for \textit{ab initio} total-energy calculations using a plane-wave basis set},
	volume = {54},
	copyright = {http://link.aps.org/licenses/aps-default-license},
	issn = {0163-1829, 1095-3795},
	url = {https://link.aps.org/doi/10.1103/PhysRevB.54.11169},
	doi = {10.1103/PhysRevB.54.11169},
	number = {16},
	journal = {Phys. Rev. B},
	author = {Kresse, G. and Furthmüller, J.},
	month = oct,
	year = {1996},
	pages = {11169--11186},
}

@article{kresse_efficiency_1996,
	title = {Efficiency of ab-initio total energy calculations for metals and semiconductors using a plane-wave basis set},
	volume = {6},
	copyright = {https://www.elsevier.com/tdm/userlicense/1.0/},
	issn = {09270256},
	url = {https://linkinghub.elsevier.com/retrieve/pii/0927025696000080},
	doi = {10.1016/0927-0256(96)00008-0},
	number = {1},
	journal = {Computational Materials Science},
	author = {Kresse, G. and Furthmüller, J.},
	month = jul,
	year = {1996},
	pages = {15--50},
}

@article{neugebauer_adsorbate-substrate_1992,
	title = {Adsorbate-substrate and adsorbate-adsorbate interactions of {Na} and {K} adlayers on {Al}(111)},
	volume = {46},
	copyright = {http://link.aps.org/licenses/aps-default-license},
	issn = {0163-1829, 1095-3795},
	url = {https://link.aps.org/doi/10.1103/PhysRevB.46.16067},
	doi = {10.1103/PhysRevB.46.16067},
	number = {24},
	journal = {Phys. Rev. B},
	author = {Neugebauer, Jörg and Scheffler, Matthias},
	month = dec,
	year = {1992},
	pages = {16067--16080},
}

@article{makov_periodic_1995,
	title = {Periodic boundary conditions in \textit{ab initio} calculations},
	volume = {51},
	copyright = {http://link.aps.org/licenses/aps-default-license},
	issn = {0163-1829, 1095-3795},
	url = {https://link.aps.org/doi/10.1103/PhysRevB.51.4014},
	doi = {10.1103/PhysRevB.51.4014},
	number = {7},
	journal = {Phys. Rev. B},
	author = {Makov, G. and Payne, M. C.},
	month = feb,
	year = {1995},
	pages = {4014--4022},
}

@article{dudarev_electron-energy-loss_1998,
	title = {Electron-energy-loss spectra and the structural stability of nickel oxide: {An} {LSDA}+{U} study},
	volume = {57},
	copyright = {http://link.aps.org/licenses/aps-default-license},
	issn = {0163-1829, 1095-3795},
	shorttitle = {Electron-energy-loss spectra and the structural stability of nickel oxide},
	url = {https://link.aps.org/doi/10.1103/PhysRevB.57.1505},
	doi = {10.1103/PhysRevB.57.1505},
	number = {3},
	journal = {Phys. Rev. B},
	author = {Dudarev, S. L. and Botton, G. A. and Savrasov, S. Y. and Humphreys, C. J. and Sutton, A. P.},
	month = jan,
	year = {1998},
	pages = {1505--1509},
}

@article{wang_low-temperature_2008,
	title = {Low-temperature synthesis of pure rock-salt structure manganese sulfide using a single-source molecular precursor},
	volume = {144},
	issn = {13858947},
	url = {https://linkinghub.elsevier.com/retrieve/pii/S1385894708001794},
	doi = {10.1016/j.cej.2008.03.017},
	number = {1},
	journal = {Chemical Engineering Journal},
	author = {Wang, Tian Xi and Chen, Wei Wei},
	month = oct,
	year = {2008},
	pages = {146--148},
}

@article{abramchuk_development_2021,
	title = {Development of high-temperature oxide melt solution calorimetry for p-block element containing materials – {CORRIGENDUM}},
	volume = {36},
	issn = {0884-2914, 2044-5326},
	url = {https://link.springer.com/10.1557/s43578-020-00057-6},
	doi = {10.1557/s43578-020-00057-6},
	number = {3},
	journal = {Journal of Materials Research},
	author = {Abramchuk, Mykola and Lilova, Kristina and Subramani, Tamilarasan and Yoo, Ray and Navrostky, Alexandra},
	month = feb,
	year = {2021},
	pages = {785--785},
}

@article{bustamante_enthalpies_2024,
	title = {Enthalpies of mixing for alloys liquid below room temperature determined by oxidative solution calorimetry},
	volume = {149},
	issn = {1588-2926},
	url = {https://doi.org/10.1007/s10973-024-13035-5},
	doi = {10.1007/s10973-024-13035-5},
	number = {10},
	journal = {Journal of Thermal Analysis and Calorimetry},
	author = {Bustamante, Michael and Lilova, Kristina and Navrotsky, Alexandra and Harvey, Jean-Philippe and Oishi, Kentaro},
	month = may,
	year = {2024},
	pages = {4817--4826},
}

@book{nye1985physical,
  title={Physical properties of crystals: their representation by tensors and matrices},
  author={Nye, John Frederick},
  year={1985},
  publisher={Oxford University Press},
  address = {Oxford [Oxfordshire] : New York},
}

@techreport{caroli:inria-00405478,
  TITLE = {{Robust and Efficient Delaunay triangulations of points on or close to a sphere}},
  AUTHOR = {Caroli, Manuel and Machado Manh{\~a}es de Castro, Pedro and Loriot, Sebastien and Rouiller, Olivier and Teillaud, Monique and Wormser, Camille},
  URL = {https://inria.hal.science/inria-00405478},
  TYPE = {Research Report},
  NUMBER = {RR-7004},
  INSTITUTION = {{INRIA}},
  YEAR = {2009},
  HAL_ID = {inria-00405478},
  HAL_VERSION = {v4},
}

@ARTICLE{4121581,
  author={Van Oosterom, A. and Strackee, J.},
  journal={IEEE Transactions on Biomedical Engineering}, 
  title={The Solid Angle of a Plane Triangle}, 
  year={1983},
  volume={BME-30},
  number={2},
  pages={125-126},
  doi={10.1109/TBME.1983.325207}
}

@article{virtanen_scipy_2020,
	title = {{SciPy} 1.0: fundamental algorithms for scientific computing in {Python}},
	volume = {17},
	issn = {1548-7105},
	url = {https://doi.org/10.1038/s41592-019-0686-2},
	doi = {10.1038/s41592-019-0686-2},
	number = {3},
	journal = {Nature Methods},
	author = {Virtanen, Pauli and Gommers, Ralf and Oliphant, Travis E. and Haberland, Matt and Reddy, Tyler and Cournapeau, David and Burovski, Evgeni and Peterson, Pearu and Weckesser, Warren and Bright, Jonathan and van der Walt, Stéfan J. and Brett, Matthew and Wilson, Joshua and Millman, K. Jarrod and Mayorov, Nikolay and Nelson, Andrew R. J. and Jones, Eric and Kern, Robert and Larson, Eric and Carey, C. J. and Polat, {\.I}lhan and Feng, Yu and Moore, Eric W. and VanderPlas, Jake and Laxalde, Denis and Perktold, Josef and Cimrman, Robert and Henriksen, Ian and Quintero, E. A. and Harris, Charles R. and Archibald, Anne M. and Ribeiro, Antônio H. and Pedregosa, Fabian and van Mulbregt, Paul and Vijaykumar, Aditya and Bardelli, Alessandro Pietro and Rothberg, Alex and Hilboll, Andreas and Kloeckner, Andreas and Scopatz, Anthony and Lee, Antony and Rokem, Ariel and Woods, C. Nathan and Fulton, Chad and Masson, Charles and Häggström, Christian and Fitzgerald, Clark and Nicholson, David A. and Hagen, David R. and Pasechnik, Dmitrii V. and Olivetti, Emanuele and Martin, Eric and Wieser, Eric and Silva, Fabrice and Lenders, Felix and Wilhelm, Florian and Young, G. and Price, Gavin A. and Ingold, Gert-Ludwig and Allen, Gregory E. and Lee, Gregory R. and Audren, Hervé and Probst, Irvin and Dietrich, Jörg P. and Silterra, Jacob and Webber, James T. and Slavič, Janko and Nothman, Joel and Buchner, Johannes and Kulick, Johannes and Schönberger, Johannes L. and de Miranda Cardoso, José Vinícius and Reimer, Joscha and Harrington, Joseph and Rodríguez, Juan Luis Cano and Nunez-Iglesias, Juan and Kuczynski, Justin and Tritz, Kevin and Thoma, Martin and Newville, Matthew and Kümmerer, Matthias and Bolingbroke, Maximilian and Tartre, Michael and Pak, Mikhail and Smith, Nathaniel J. and Nowaczyk, Nikolai and Shebanov, Nikolay and Pavlyk, Oleksandr and Brodtkorb, Per A. and Lee, Perry and McGibbon, Robert T. and Feldbauer, Roman and Lewis, Sam and Tygier, Sam and Sievert, Scott and Vigna, Sebastiano and Peterson, Stefan and More, Surhud and Pudlik, Tadeusz and Oshima, Takuya and Pingel, Thomas J. and Robitaille, Thomas P. and Spura, Thomas and Jones, Thouis R. and Cera, Tim and Leslie, Tim and Zito, Tiziano and Krauss, Tom and Upadhyay, Utkarsh and Halchenko, Yaroslav O. and Vázquez-Baeza, Yoshiki and {SciPy 1.0 Contributors}},
	month = mar,
	year = {2020},
	pages = {261--272},
}

@inproceedings{boerner_access_2023,
author = {Boerner, Timothy J. and Deems, Stephen and Furlani, Thomas R. and Knuth, Shelley L. and Towns, John},
title = {ACCESS: Advancing Innovation: NSF’s Advanced Cyberinfrastructure Coordination Ecosystem: Services \& Support},
year = {2023},
isbn = {9781450399852},
publisher = {Association for Computing Machinery},
address = {New York, NY, USA},
doi = {10.1145/3569951.3597559},
booktitle = {Practice and Experience in Advanced Research Computing 2023: Computing for the Common Good},
pages = {173–176},
numpages = {4},
location = {Portland, OR, USA},
series = {PEARC '23}
}

@article{Qing_JPCL_2019,
    author = {Zhao, Qing and Kulik, Heather J.},
    title = {Stable Surfaces That Bind Too Tightly: Can Range-Separated Hybrids or DFT+U Improve Paradoxical Descriptions of Surface Chemistry?},
    journal = {The Journal of Physical Chemistry Letters},
    volume = {10},
    number = {17},
    pages = {5090-5098},
    year = {2019},
    month = {09},
    issn = {1948-7185},
    doi = {10.1021/acs.jpclett.9b01650},
    url = {https://doi.org/10.1021/acs.jpclett.9b01650},
}

@article{Heather_PRL_2006,
      title = {Density Functional Theory in Transition-Metal Chemistry: A Self-Consistent Hubbard $U$ Approach},
      author = {Kulik, Heather J. and Cococcioni, Matteo and Scherlis, Damian A. and Marzari, Nicola},
      journal = {Phys. Rev. Lett.},
      volume = {97},
      issue = {10},
      pages = {103001},
      numpages = {4},
      year = {2006},
      month = {Sep},
      publisher = {American Physical Society},
      doi = {10.1103/PhysRevLett.97.103001},
      url = {https://link.aps.org/doi/10.1103/PhysRevLett.97.103001}
}

@article{Iurii_PRB_2018,
      title = {Hubbard parameters from density-functional perturbation theory},
      author = {Timrov, Iurii and Marzari, Nicola and Cococcioni, Matteo},
      journal = {Phys. Rev. B},
      volume = {98},
      issue = {8},
      pages = {085127},
      numpages = {15},
      year = {2018},
      month = {Aug},
      publisher = {American Physical Society},
      doi = {10.1103/PhysRevB.98.085127},
      url = {https://link.aps.org/doi/10.1103/PhysRevB.98.085127}
}

@misc{nomad2025dataset,
  title        = {NOMAD dataset: ET\_CM\_MnS-NCs},
  author       = {Chen, Junchi and Wexler, Robert Bruce},
  year         = {2025},
  howpublished = {\url{https://dx.doi.org/10.17172/NOMAD/2025.12.19-1}},
  doi          = {10.17172/NOMAD/2025.12.19-1},
}

@article{10.1063/1.4812323,
    author = {Jain, Anubhav and Ong, Shyue Ping and Hautier, Geoffroy and Chen, Wei and Richards, William Davidson and Dacek, Stephen and Cholia, Shreyas and Gunter, Dan and Skinner, David and Ceder, Gerbrand and Persson, Kristin A.},
    title = {Commentary: The Materials Project: A materials genome approach to accelerating materials innovation},
    journal = {APL Materials},
    volume = {1},
    number = {1},
    pages = {011002},
    year = {2013},
    month = {07},
    issn = {2166-532X},
    doi = {10.1063/1.4812323},
    url = {https://doi.org/10.1063/1.4812323},
}

@misc{noauthor_materials_2020_MnI2,
	address = {United States},
    key = {Materials Project MnI2},
	title = {Materials {Data} on {MnI2} by {Materials} {Project}},
	url = {https://www.osti.gov/dataexplorer/servlets/purl/1202148},
	doi = {10.17188/1202148},
	month = jul,
	year = {2020},
}

@misc{noauthor_materials_2020_MnCl2,
	address = {United States},
    key = {Materials Project MnCl2},
	title = {Materials {Data} on {MnCl2} by {Materials} {Project}},
	url = {https://www.osti.gov/dataexplorer/servlets/purl/1202359},
	doi = {10.17188/1202359},
	month = jul,
	year = {2020},
}

@misc{noauthor_materials_2020_MnS2,
	address = {United States},
    key = {Materials Project MnS2},
	title = {Materials {Data} on {MnS2} by {Materials} {Project}},
	url = {https://www.osti.gov/dataexplorer/servlets/purl/1190721},
	doi = {10.17188/1190721},
	month = jul,
	year = {2020},
}

@misc{noauthor_materials_2020_ZB-MnS,
	address = {United States},
    key = {Materials Project ZB-MnS},
	title = {Materials {Data} on {MnS} by {Materials} {Project}},
	url = {https://www.osti.gov/dataexplorer/servlets/purl/1192718},
	doi = {10.17188/1192718},
	month = jul,
	year = {2020},
}

@misc{noauthor_materials_2020_WZ-MnS,
	address = {United States},
    key = {Materials Project WZ-MnS},
	title = {Materials {Data} on {MnS} by {Materials} {Project}},
	url = {https://www.osti.gov/dataexplorer/servlets/purl/1200816},
	doi = {10.17188/1200816},
	month = jul,
	year = {2020},
}

@misc{noauthor_materials_2020_RS-MnS,
	address = {United States},
    key = {Materials Project RS-MnS},
	title = {Materials {Data} on {MnS} by {Materials} {Project}},
	url = {https://www.osti.gov/dataexplorer/servlets/purl/1195781},
	doi = {10.17188/1195781},
	month = jul,
	year = {2020},
}

@misc{noauthor_materials_2020_I,
	address = {United States},
    key = {Materials Project I},
	title = {Materials {Data} on {I} by {Materials} {Project}},
	url = {https://www.osti.gov/dataexplorer/servlets/purl/1199273},
	doi = {10.17188/1199273},
	month = jul,
	year = {2020},
}

@misc{noauthor_materials_2020_Mn,
	address = {United States},
    key = {Materials Project Mn},
	title = {Materials {Data} on {Mn} by {Materials} {Project}},
	url = {https://www.osti.gov/dataexplorer/servlets/purl/1206920},
	doi = {10.17188/1206920},
	month = jul,
	year = {2020},
}

@misc{noauthor_materials_2020_S,
	address = {United States},
    key = {Materials Project S},
	title = {Materials {Data} on {S} by {Materials} {Project}},
	url = {https://www.osti.gov/dataexplorer/servlets/purl/1299376},
	doi = {10.17188/1299376},
	month = apr,
	year = {2020},
}

@article{fuad_magnetochemische_1938,
author = {Mehmed, Fuad and Haraldsen, Haakon},
title = {Magnetochemische Untersuchungen. XXVIII. Das magnetische Verhalten der allotropen Modifikationen des Mangan(II)-Sulfids},
journal = {Zeitschrift für anorganische und allgemeine Chemie},
volume = {235},
number = {3},
pages = {193-200},
doi = {https://doi.org/10.1002/zaac.19382350305},
url = {https://onlinelibrary.wiley.com/doi/abs/10.1002/zaac.19382350305},
eprint = {https://onlinelibrary.wiley.com/doi/pdf/10.1002/zaac.19382350305},
year = {1938}
}

@article{makoto_crystal_2019,
url = {https://doi.org/10.1515/zkri-2018-2134},
title = {Crystal structure refinement of MnTe2, MnSe2, and MnS2: cation-anion and anion–anion bonding distances in pyrite-type structures},
author = {Makoto Tokuda and Akira Yoshiasa and Tsutomu Mashimo and Hiroshi Arima and Hidetomo Hongu and Tsubasa Tobase and Akihiko Nakatsuka and Kazumasa Sugiyama},
pages = {371--377},
volume = {234},
number = {6},
journal = {Zeitschrift für Kristallographie - Crystalline Materials},
doi = {doi:10.1515/zkri-2018-2134},
year = {2019},
lastchecked = {2025-12-08}
}

@article{ONG2013314,
    title = {Python Materials Genomics (pymatgen): A robust, open-source python library for materials analysis},
    journal = {Computational Materials Science},
    volume = {68},
    pages = {314-319},
    year = {2013},
    issn = {0927-0256},
    doi = {https://doi.org/10.1016/j.commatsci.2012.10.028},
    url = {https://www.sciencedirect.com/science/article/pii/S0927025612006295},
    author = {Shyue Ping Ong and William Davidson Richards and Anubhav Jain and Geoffroy Hautier and Michael Kocher and Shreyas Cholia and Dan Gunter and Vincent L. Chevrier and Kristin A. Persson and Gerbrand Ceder},
}

@article{perdew_rationale_1996,
	title = {Rationale for mixing exact exchange with density functional approximations},
	volume = {105},
	doi = {10.1063/1.472933},
	number = {22},
	journal = {The Journal of Chemical Physics},
	author = {Perdew, John P. and Ernzerhof, Matthias and Burke, Kieron},
	year = {1996},
	pages = {9982--9985},
}

@article{adamo_toward_1999,
	title = {Toward reliable density functional methods without adjustable parameters: {The} {PBE0} model},
	volume = {110},
	doi = {10.1063/1.478522},
	number = {13},
	journal = {The Journal of Chemical Physics},
	author = {Adamo, Carlo and Barone, Vincenzo},
	year = {1999},
	pages = {6158--6170},
}

\end{document}


\title[Article Title]{Supplementary Information for
Equilibrium Thermochemistry and Crystallographic Morphology of Manganese Sulfide Nanocrystals
}

\author[1]{\fnm{Junchi} \sur{Chen}}

\author[2,3]{\fnm{Tamilarasan} \sur{Subramani}}

\author[4]{\fnm{Deep} \sur{Mekan}}

\author[4]{\fnm{Danielle} \sur{Gendler}}

\author[1,5]{\fnm{Ray} \sur{Yang}}

\author[1]{\fnm{Manish} \sur{Kumar}}

\author[2,6]{\fnm{Megan} \sur{Householder}}

\author[2,3,7]{\fnm{Alexis} \sur{Rosado Ortiz}}

\author[4]{\fnm{Emil A.}\ \sur{Hernandez-Pagan}}

\author[2,3]{\fnm{Kristina} \sur{Lilova}}

\author*[1]{\fnm{Robert B.}\ \sur{Wexler}}
\email{wexler@wustl.edu}

\affil*[1]{\orgdiv{Department of Chemistry and Institute of Materials Science and Engineering}, \orgname{Washington University in St. Louis}, \orgaddress{\city{St.\ Louis}, \postcode{63130}, \state{MO}, \country{USA}}}

\affil[2]{\orgdiv{Center for Materials of the Universe}, \orgname{Arizona State University}, \orgaddress{\city{Tempe}, \postcode{85287}, \state{AZ}, \country{USA}}}

\affil[3]{\orgdiv{School of Molecular Sciences}, \orgname{Arizona State University}, \orgaddress{\city{Tempe}, \postcode{85287}, \state{AZ}, \country{USA}}}

\affil[4]{\orgdiv{Department of Chemistry and Biochemistry}, \orgname{University of Delaware}, \orgaddress{\city{Newark}, \postcode{19711}, \state{DE}, \country{USA}}}

\affil[5]{\orgdiv{Department of Computer Science and Engineering, McKelvey School of Engineering}, \orgname{Washington University in St. Louis}, \orgaddress{\city{St.\ Louis}, \postcode{63130}, \state{MO}, \country{USA}}}

\affil[6]{\orgdiv{School of Earth and Space Exploration}, \orgname{Arizona State University}, \orgaddress{\city{Tempe}, \postcode{85287}, \state{AZ}, \country{USA}}}

\affil[7]{\orgdiv{School of Pharmacy}, \orgname{Massachusetts College of Pharmacy and Health Sciences}, \orgaddress{\city{Boston}, \postcode{02115}, \state{MA}, \country{USA}}}

\maketitle

\newpage
\tableofcontents


\clearpage

\section{Computational methods benchmarking}

This section provides supplementary data supporting the benchmarking of exchange-correlation (XC) functionals for density functional theory (DFT) calculations (\ref{subsection:s1.1}), the bulk energies used in surface energy calculations (\ref{subsection:s1.2}), and the choice of the Heyd--Scuseria--Ernzerhof (HSE06) hybrid functional as the reference and the transferability of the r$^2$SCAN+$U$ correction (\ref{subsection:s1.3}).
These topics correspond to the Computational framework validation section of the main text.

\subsection{Lattice constants comparison}
\label{subsection:s1.1}

The set of XC functionals tested includes the Perdew--Burke--Ernzerhof (PBE) generalized gradient approximation (GGA)\cite{Perdew:1996pki} and its dispersion-corrected counterpart with Becke--Johnson damping, PBE-D3+BJ;\cite{grimme_consistent_2010, grimme_effect_2011} the strongly constrained and appropriately normed (SCAN)\cite{sun_strongly_2015} meta-GGA functional and its regularized form r$^2$SCAN;\cite{furness_accurate_2020} as well as their van der Waals dispersion-corrected variants, SCAN-rVV10\cite{peng_versatile_2016} and r$^2$SCAN-rVV10.\cite{ning_workhorse_2022}
Table~\ref{tab:tabs1} summarizes the lattice constants of rock salt (RS), wurtzite (WZ), and zinc blende (ZB) MnS polymorphs and pyrite \ce{MnS2} predicted by DFT calculations with the above functionals, along with the corresponding experimental references.

\begin{table}[htbp]
\centering
\caption{
Lattice constants from density functional theory (DFT) predictions and experimental references for rock salt (RS) MnS, wurtzite (WZ) MnS, zinc blende (ZB) MnS, and pyrite (PY) \ce{MnS2}.
}
\label{tab:tabs1}
\sisetup{round-mode=places, round-precision=4}
\begin{tabular}{
    l
    l
    S[table-format=1.4]
    S[table-format=1.4]
    S[table-format=1.4]
}
\toprule
    {Crystal} & {Data source} & \multicolumn{3}{c}{Lattice constants} \\
    \cmidrule(lr){3-5}
              &               & {$a$ / \unit{\angstrom}} & {$b$ / \unit{\angstrom}} & {$c$ / \unit{\angstrom}} \\
\midrule
    \multirow{7}{*}{RS-MnS}
        & ICSD-148200\cite{wang_low-temperature_2008}  & 5.2232 & 5.2232 & 5.2232 \\
        & PBE               & 5.1371 & 5.0981 & 5.1371 \\
        & PBE-D3+BJ         & 5.0028 & 5.0588 & 5.0028 \\
        & SCAN              & 5.1860 & 5.1878 & 5.1860 \\
        & SCAN-rVV10        & 5.1714 & 5.1730 & 5.1714 \\
        & r$^2$SCAN         & 5.2000 & 5.2027 & 5.2000 \\
        & r$^2$SCAN-rVV10   & 5.1765 & 5.1782 & 5.1765 \\
\midrule
    \multirow{7}{*}{WZ-MnS}
        & ICSD-44765\cite{corliss_magnetic_1956}   & 3.9870 & 3.9870 & 6.4380 \\
        & PBE               & 3.9559 & 3.9559 & 6.2894 \\
        & PBE-D3+BJ         & 3.8819 & 3.8819 & 6.2010 \\
        & SCAN              & 3.9519 & 3.9519 & 6.3401 \\
        & SCAN-rVV10        & 3.9458 & 3.9458 & 6.3322 \\
        & r$^2$SCAN         & 3.9721 & 3.9721 & 6.3747 \\
        & r$^2$SCAN-rVV10   & 3.9527 & 3.9527 & 6.3415 \\
\midrule
    \multirow{7}{*}{ZB-MnS}
        & ICSD-76205\cite{fuad_magnetochemische_1938}   & 5.5900 & 5.5900 & 5.5900 \\
        & PBE               & 5.6515 & 5.6515 & 5.6392 \\
        & PBE-D3+BJ         & 5.5338 & 5.5338 & 5.5219 \\
        & SCAN              & 5.6330 & 5.6330 & 5.6181 \\
        & SCAN-rVV10        & 5.6197 & 5.6197 & 5.6052 \\
        & r$^2$SCAN         & 5.6521 & 5.6521 & 5.6392 \\
        & r$^2$SCAN-rVV10   & 5.6229 & 5.6300 & 5.6171 \\
\midrule
    \multirow{7}{*}{PY-\ce{MnS2}}
        & ICSD-12958\cite{makoto_crystal_2019}   & 6.1013 & 6.1013 & 6.1013 \\
        & PBE               & 5.5148 & 5.5148 & 5.5148 \\
        & PBE-D3+BJ         & 5.4595 & 5.4595 & 5.4595 \\
        & SCAN              & 5.8738 & 5.8738 & 5.8738 \\
        & SCAN-rVV10        & 5.5223 & 5.5223 & 5.5222 \\
        & r$^2$SCAN         & 6.1054 & 6.1054 & 6.1054 \\
        & r$^2$SCAN-rVV10   & 5.5067 & 5.5067 & 5.5067 \\
\bottomrule
\end{tabular}
\end{table}

\clearpage

\subsection{Bulk energies for surface energy calculations}
\label{subsection:s1.2}
A well-known issue in DFT surface modeling is that inconsistent Brillouin zone sampling between bulk and slab calculations can yield unconverged surface energies.\cite{sun_efficient_2013, methfessel_high-precision_1989, boettger_nonconvergence_1994}
We addressed this by fitting slab energies as a linear function of the number of atoms, with the slope corresponding to the bulk energy per atom.\cite{fiorentini_extracting_1996}

\begin{figure*}[htbp]
    \centering
    \includegraphics[width=1.0\linewidth]{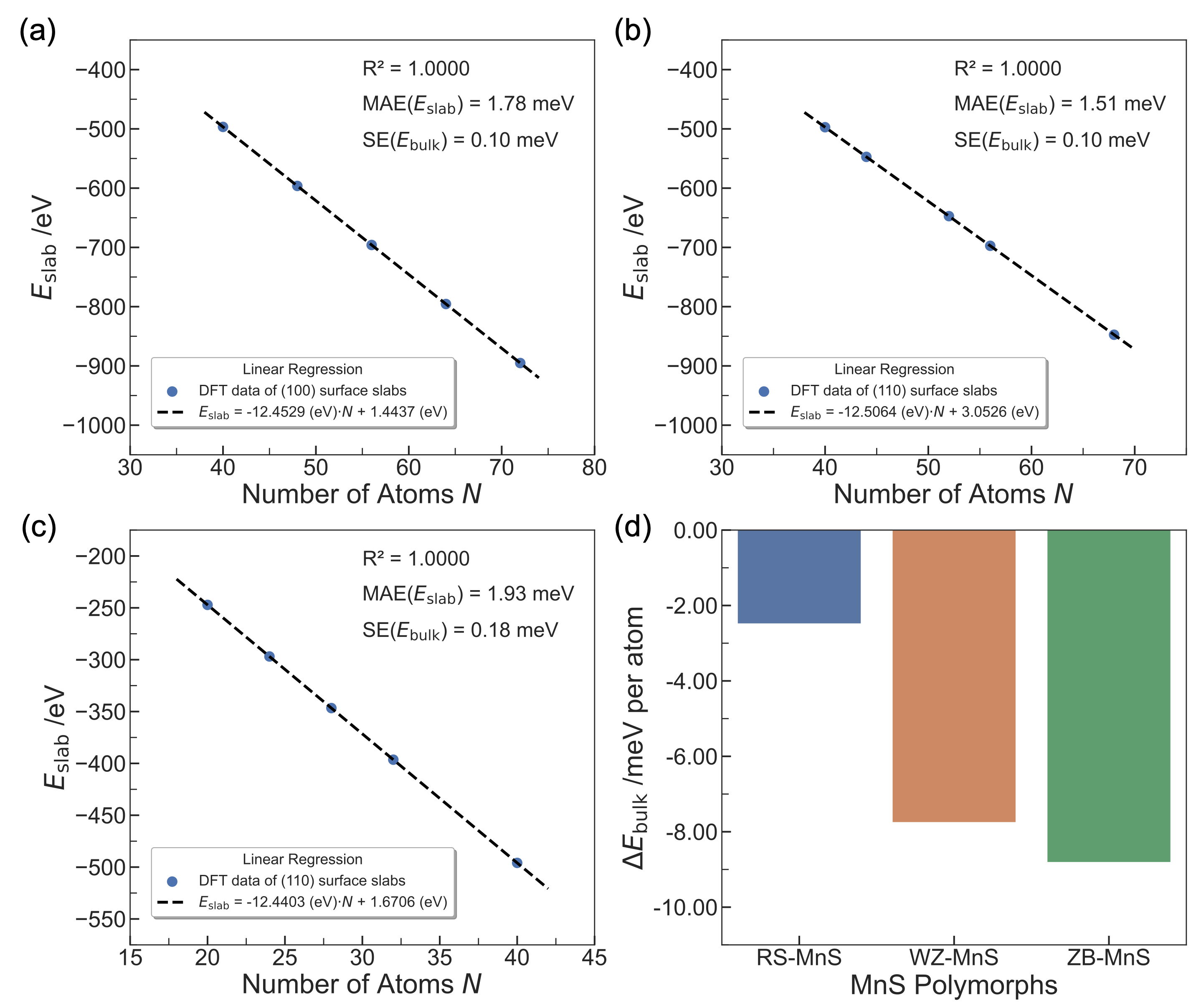}
    \caption{
Bulk energies derived from slab model calculations for (a) rock salt (RS), (b) wurtzite (WZ), and (c) zinc blende (ZB) MnS.
(d) Comparison of bulk energies obtained from slab-fitted and direct bulk calculations for RS-, WZ-, and ZB-MnS.
All results are from r$^2$SCAN calculations.
}
    \label{fig:figs1}
\end{figure*}

Figure~\ref{fig:figs1} shows the resulting fits for RS-, WZ-, and ZB-MnS based on r$^2$SCAN-calculated slab energies; the extracted bulk energies agree with those from direct bulk optimizations to within 10~meV$\cdot$atom$^{-1}$.
The values for each polymorph, model, and DFT method are listed in Table~\ref{tab:tabs2}.

\begin{table}[htbp]
\centering
\caption{
Calculated bulk energies of MnS polymorphs obtained from different structural models and DFT methods.
}
\label{tab:tabs2}
\sisetup{round-mode=places, round-precision=5}
\begin{tabular}{
    l
    l
    l
    S[table-format=-2.5]
}
\toprule
    {Crystal structure} & {Model} & {DFT method} & {$E_{\ce{MnS}}$ / \unit{eV\per atom}} \\
\midrule
    \multirow{7}{*}{Rock salt}
        & Bulk              & r$^2$SCAN              & -12.45043 \\
        & (100) surface slab & r$^2$SCAN              & -12.45292 \\
        & (131) surface slab & r$^2$SCAN              & -12.45188 \\
        & Bulk              & r$^2$SCAN+$U$ (2.7 eV) & -12.11335 \\
        & (100) surface slab & r$^2$SCAN+$U$ (2.7 eV) & -12.11343 \\
        & Bulk              & HSE06                  & -10.26874 \\
        & (100) surface slab & HSE06                  & -10.26897 \\
\midrule
    \multirow{6}{*}{Wurtzite}
        & Bulk                             & r$^2$SCAN              & -12.49862 \\
        & (11$\overline{2}$0) surface slab & r$^2$SCAN              & -12.50638 \\
        & Bulk                             & r$^2$SCAN+$U$ (2.7 eV) & -12.10663 \\
        & (11$\overline{2}$0) surface slab & r$^2$SCAN+$U$ (2.7 eV) & -12.10864 \\
        & Bulk                             & HSE06                  & -10.31285 \\
        & (11$\overline{2}$0) surface slab & HSE06                  & -10.31532 \\
\midrule
    \multirow{6}{*}{Zinc blende}
        & Bulk              & r$^2$SCAN              & -12.43146 \\
        & (110) surface slab & r$^2$SCAN              & -12.44027 \\
        & Bulk              & r$^2$SCAN+$U$ (2.7 eV) & -12.08443 \\
        & (110) surface slab & r$^2$SCAN+$U$ (2.7 eV) & -12.08451 \\
        & Bulk              & HSE06                  & -10.27230 \\
        & (110) surface slab & HSE06                  & -10.27340 \\
\bottomrule
\end{tabular}
\end{table}

\clearpage

\subsection{Choice of reference functional and transferability of r$^2$SCAN+$U$}
\label{subsection:s1.3}

\begin{figure*}[htbp]
    \centering
    \includegraphics[width=1.0\linewidth]{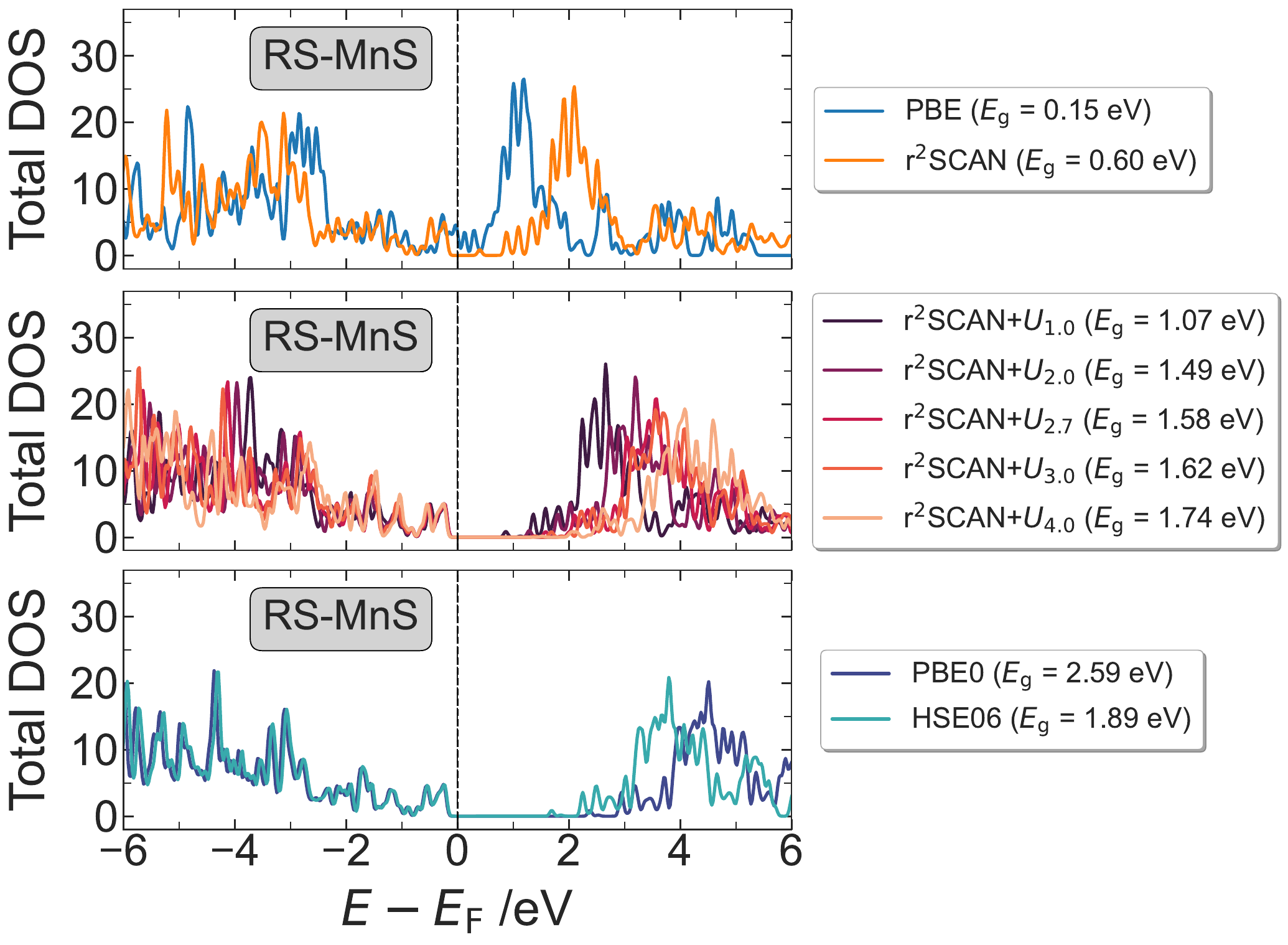}
    \caption{
Total density of states (DOS) and band gap ($E_\mathrm{g}$) of r$^2$SCAN-optimized rock salt (RS) MnS from single-point density functional theory (DFT) calculations using PBE, r$^2$SCAN, r$^2$SCAN+$U$, PBE0, and HSE06.
In r$^2$SCAN+$U_N$, $N$ is the Hubbard $U$ value (in eV) applied to the Mn 3d states.
All self-consistent calculations used the settings described in the Computational methods subsection of the main text, and the DOS was computed on a grid of 3000 energy points.
The experimentally reported band gap is 2.7--3.7 eV.\cite{lokhande_process_1998,alanazi_structural_2021}
}
    \label{fig:dos_Eg}
\end{figure*}

Figure~\ref{fig:dos_Eg} shows that, for the bulk electronic structure, PBE0\cite{perdew_rationale_1996,adamo_toward_1999} predicts an RS-MnS band gap ($E_\mathrm{g}$) closer to the experimentally reported range of 2.7--3.7 eV\cite{lokhande_process_1998,alanazi_structural_2021} than HSE06.\cite{krukau_influence_2006}
However, accurate electronic structure properties do not necessarily translate to accuracy in thermodynamic quantities such as reaction or surface energies.
Because HSE06 reaction energies agree with the 0 K experimental values to within 13.5 kJ$\cdot$mol$^{-1}$ for all three reactions in Figure 3b of the main text, we retain HSE06 as the reference for calibrating the r$^2$SCAN+$U$ surface energy correction.

The two S-terminated facets used to benchmark the $U$ value span different undercoordination degrees of surface sulfur atoms, from three dangling bonds on RS-MnS (111)--S to one or two on (131)--S (Figure 4 of the main text), comparable to the polar ZB- and WZ-MnS facets considered in this study.
However, directly benchmarking individual polar surface energies of ZB-MnS, or relative polar surface energies of WZ-MnS, would require computationally prohibitive HSE06 calculations on wedge models.
Therefore, as a practical test, we selected representative ZB-MnS (111) and WZ-MnS (0001) slabs and compared the combined surface energies of their two opposing surfaces to assess the applicability of r$^2$SCAN+$U$ with $U = 2.7$ eV.
As summarized in Table~\ref{tab:surface_energy_corrections}, the r$^2$SCAN+$U$ correction brings the surface energies of r$^2$SCAN-optimized facets closer to HSE06 for two representative polar ZB- and WZ-MnS slabs.
Although these combined surface energies do not allow us to quantify the error of each termination separately, the RS-MnS benchmarks (Figure 4 of the main text) suggest that r$^2$SCAN+$U$ primarily corrects the strong underestimation of S-terminated surface energies while slightly overestimating Mn-terminated surface energies.
If the residual deviations in Table~\ref{tab:surface_energy_corrections} likewise arise mainly from the Mn-terminated surfaces, they should not affect the ZB-MnS Wulff shapes, because (111)--Mn lies at least 0.45 J$\cdot$m$^{-2}$ above the exposed (110) and ($\overline{111}$)--S facets (Subsection~\ref{subsection:s2.4}).
In WZ-MnS, however, Mn-terminated facets are part of the predicted morphology: $\{10\overline{1}1\}$--Mn$^3$ bounds the top of the rod and ($000\overline{1}$)--Mn forms the base under Mn-rich conditions (Figure 9 of the main text).
The $\Delta \mu_{\ce{S}}$ at which S-terminated facets begin to truncate the WZ-MnS base should therefore be regarded as approximate.

\begin{table}[tb]
\renewcommand{\arraystretch}{1.5}
\caption{
Combined surface energies $\gamma_\mathrm{tot}$ of representative stoichiometric polar slabs of zinc blende (ZB) and wurtzite (WZ) MnS from r$^2$SCAN, r$^2$SCAN+$U$, and HSE06, all evaluated on r$^2$SCAN-optimized structures.
$\gamma_\mathrm{tot}$ is the sum of the surface energies of the two opposite terminations of the slab, and $\gamma_\mathrm{tot} - \gamma_\mathrm{tot}(\mathrm{HSE06})$ is its deviation from HSE06.
For r$^2$SCAN+$U$ calculations, $U = 2.7$ eV was applied to the Mn 3d states.
All three methods use the same slab models (Subsection~\ref{subsection:s3.2}) and the settings described in the Computational methods subsection of the main text.
}
\label{tab:surface_energy_corrections}
\centering
\small
\begin{tabular}{@{} l cccc @{}}
\toprule
    Crystal structure & Facets & DFT method & $\gamma_\mathrm{tot}$ /J$\cdot$m$^{-2}$ & $\gamma_\mathrm{tot} - \gamma_\mathrm{tot}(\mathrm{HSE06})$ /J$\cdot$m$^{-2}$\\
\midrule
\multirow{3}{*}{ZB-MnS}
    & \multirow{3}{*}{(111)--Mn and ($\overline{111}$)--S}
        & r$^2$SCAN      & 1.87 & $-0.49$ \\
    &   & r$^2$SCAN+$U$  & 2.48 & 0.12 \\
    &   & HSE06          & 2.36 & 0.00 \\
\midrule
    \multirow{3}{*}{WZ-MnS} & \multirow{3}{*}{(0001)--S and (000$\overline{1}$)--Mn}
        & r$^2$SCAN      & 1.92 & $-0.50$ \\
    &   & r$^2$SCAN+$U$  & 2.64 & 0.22 \\
    &   & HSE06          & 2.42 & 0.00 \\
\bottomrule
\end{tabular}
\end{table}


\clearpage

\section{Surface energies and morphologies of MnS NCs}

In this work, ($hkl$) denotes an individual crystallographic facet, [$hkl$] represents the direction normal to the ($hkl$) facet (perpendicular in cubic systems), and \{$hkl$\} refers to the family of symmetry-equivalent facets.
As established in the main text, r$^2$SCAN provides the most accurate lattice constants and thermochemical reaction energies among the functionals tested.
However, it substantially underestimates S-terminated polar surface energies; applying a Hubbard $U$ correction (r$^2$SCAN+$U$, $U = 2.7$~eV) to the Mn 3d states brings the results into close agreement with HSE06 (see Computational framework validation in the main text).
This section compares r$^2$SCAN and r$^2$SCAN+$U$ results, providing detailed surface energy data, Wulff constructions, and surface area fractions for RS-MnS (\ref{subsection:s2.1} and \ref{subsection:s2.2}), ZB-MnS (\ref{subsection:s2.3} and \ref{subsection:s2.4}), and WZ-MnS (\ref{subsection:s2.5} and \ref{subsection:s2.6}) nanocrystals (NCs).
It also presents supplementary discussion of the r$^2$SCAN+$U$ surface energy correction (\ref{subsection:s2.7}), thermodynamic morphology comparisons across MnS polymorphs (\ref{subsection:s2.8}), experimental methods and results for RS-MnS NCs (\ref{subsection:s2.9}), and the spherical average surface energy analysis used to bridge the theory--experiment comparison (\ref{subsection:s2.10}).

\subsection{\texorpdfstring{r$^2$SCAN}{r²SCAN} results for RS-MnS NCs}
\label{subsection:s2.1}

\begin{figure*}[htbp]
    \centering
    \includegraphics[width=1.0\linewidth]{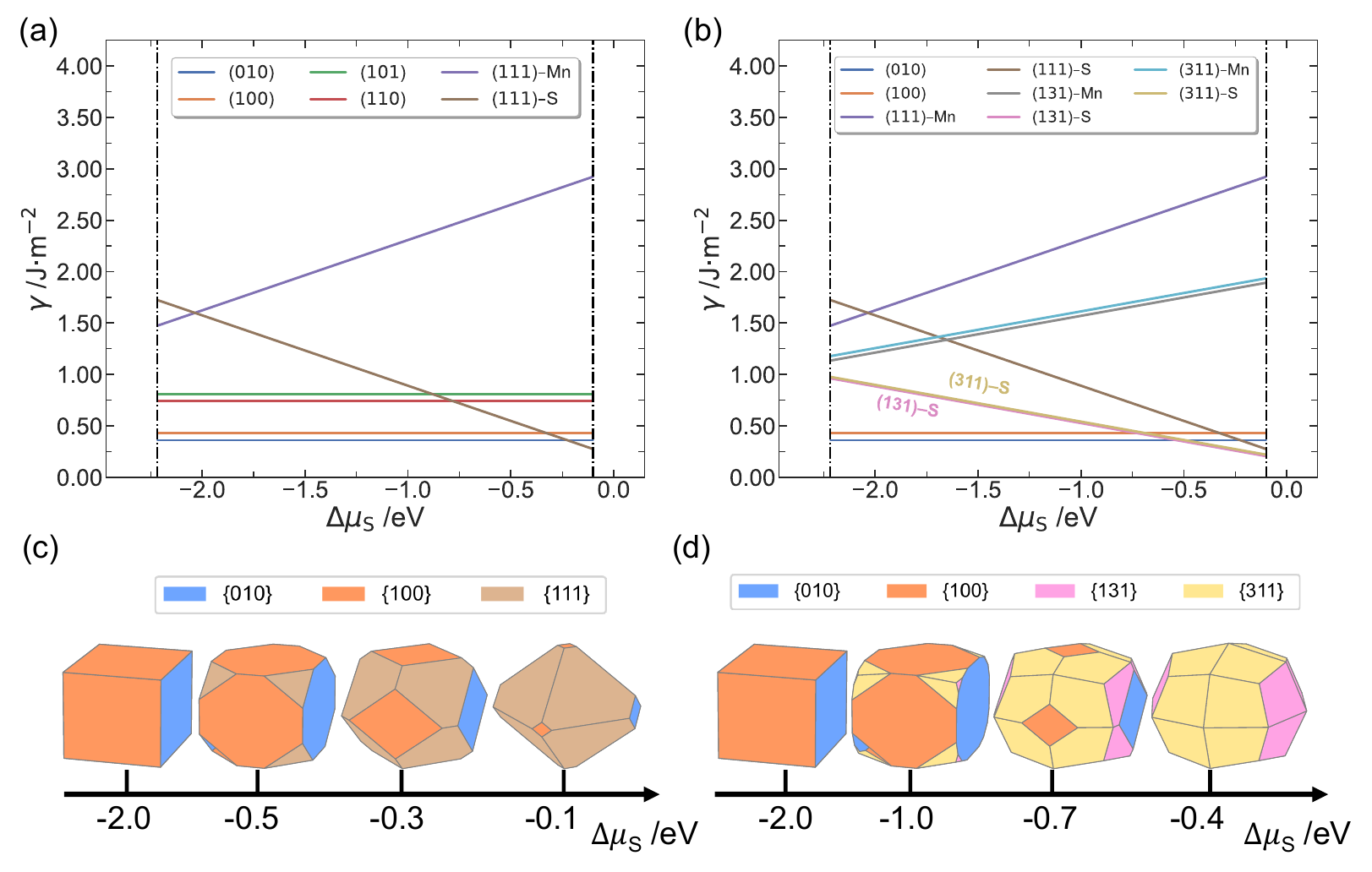}
    \caption{
Dependence of rock salt (RS) MnS surface energies on the relative chemical potential of sulfur, $\Delta \mu_{\ce{S}}$, for (a) low-index facets only and (b) the six lowest-energy facets among all $\{h,k,l\} \leq 3$ facets.
The $\Delta \mu_{\ce{S}}$ range between the two vertical dash-dotted lines ($-2.22$~eV $\leq \Delta \mu_{\ce{S}} \leq -0.10$~eV) indicates the thermodynamically stable region of bulk RS-MnS.
Wulff constructions for RS-MnS nanocrystals considering (c) low-index facets only and (d) all facets with Miller indices up to 3.
All results are from r$^2$SCAN calculations.
}
    \label{fig:figs2}
\end{figure*}

Figure~\ref{fig:figs2}a shows the dependence of low-index RS-MnS surface energies on the relative chemical potential of sulfur, $\Delta \mu_{\ce{S}}$, calculated using r$^2$SCAN.
Throughout most of the thermodynamically stable region of bulk RS-MnS, the (100) and (010) facets exhibit the lowest surface energies.
Only when $\Delta \mu_{\ce{S}}$ exceeds $-0.25$~eV do the (111)--S facets become the most stable surfaces.
The corresponding Wulff constructions (Figure~\ref{fig:figs2}c) show that RS-MnS NCs adopt cubic morphologies under Mn-rich conditions; with increasing $\Delta \mu_{\ce{S}}$, the nanocubes are progressively truncated by (111)--S facets, eventually evolving into nearly octahedral morphologies at the S-rich limit ($\Delta \mu_{\ce{S}} = -0.10$~eV).
When all facets with Miller indices up to 3 are included, the r$^2$SCAN-predicted morphology changes significantly.
Figure~\ref{fig:figs2}b shows the six lowest-energy facets among all $\{h,k,l\} \leq 3$ surfaces.
Over more than half of the stable region, the (100) and (010) facets remain lowest in energy; however, the (131)--S and (311)--S facets exhibit consistently lower surface energies than (111)--S and gradually replace (100) and (010) as $\Delta \mu_{\ce{S}}$ increases.
The corresponding Wulff constructions (Figure~\ref{fig:figs2}d) show that nanocubes are favored under Mn-rich conditions, whereas nano-trapezohedra are preferred under S-rich conditions.
The symmetry-equivalent facets in each $\{hkl\}$ family for RS-MnS are summarized in Table~\ref{tab:tabs3}.
Table~\ref{tab:tabs4} lists the surface energies and surface area fractions of the exposed facets, along with the weighted surface energies of the Wulff constructions shown in Figure~\ref{fig:figs2}d.

\clearpage

\renewcommand{\arraystretch}{1.15}
\begin{table}[htbp]
    \centering
    \caption{
Symmetry-equivalent $\{hkl\}$ facets of rock salt (RS) MnS.
    }
    \label{tab:tabs3}
    \begin{tabular}{lll}
        \toprule
        $\{hkl\}$ & Count & Symmetry-equivalent $(hkl)$ \\
        \midrule

        \{010\} & 2 & (0$\overline{1}$0); (010) \\
        \{100\} & 4 & ($\overline{1}$00); (00$\overline{1}$); (001); (100) \\
        \{101\} & 4 & ($\overline{1}$0$\overline{1}$); ($\overline{1}$01); (10$\overline{1}$); (101) \\

        \{110\} & 8 & ($\overline{11}$0); ($\overline{1}$10); (0$\overline{11}$); (0$\overline{1}$1); (01$\overline{1}$); (011); (1$\overline{1}$0); (110) \\
        \{111\} & 8 & ($\overline{111}$); ($\overline{11}$1); ($\overline{1}$1$\overline{1}$); ($\overline{1}$11); (1$\overline{11}$); (1$\overline{1}$1); (11$\overline{1}$); (111) \\

        \{120\} & 8 & ($\overline{12}$0); ($\overline{1}$20); (0$\overline{21}$); (0$\overline{2}$1); (02$\overline{1}$); (021); (1$\overline{2}$0); (120) \\
        \{121\} & 8 & ($\overline{121}$); ($\overline{12}$1); ($\overline{1}$2$\overline{1}$); ($\overline{1}$21); (1$\overline{21}$); (1$\overline{2}$1); (12$\overline{1}$); (121) \\

        \{130\} & 8 & ($\overline{13}$0); ($\overline{1}$30); (0$\overline{31}$); (0$\overline{3}$1); (03$\overline{1}$); (031); (1$\overline{3}$0); (130) \\
        \{131\} & 8 & ($\overline{131}$); ($\overline{13}$1); ($\overline{1}$3$\overline{1}$); ($\overline{1}$31); (1$\overline{31}$); (1$\overline{3}$1); (13$\overline{1}$); (131) \\

        \{201\} & 8 & ($\overline{2}$0$\overline{1}$); ($\overline{2}$01); ($\overline{1}$0$\overline{2}$); ($\overline{1}$02); (10$\overline{2}$); (102); (20$\overline{1}$); (201) \\
        \{210\} & 8 & ($\overline{21}$0); ($\overline{2}$10); (0$\overline{12}$); (0$\overline{1}$2); (01$\overline{2}$); (012); (2$\overline{1}$0); (210) \\

        \multirow{2}{*}{\{211\}}
        & \multirow{2}{*}{16}
        & \rule{0pt}{3.2ex}%
          ($\overline{211}$); ($\overline{21}$1); ($\overline{2}$1$\overline{1}$); ($\overline{2}$11); 
          ($\overline{112}$); ($\overline{11}$2); ($\overline{1}$1$\overline{2}$); ($\overline{1}$12); 
          (1$\overline{12}$); (1$\overline{1}$2); (11$\overline{2}$); (112) \\
        &
        & (2$\overline{11}$); (2$\overline{1}$1); (21$\overline{1}$); (211) \\

        \rule{0pt}{3.2ex}%
        \{212\} & 8 & ($\overline{212}$); ($\overline{21}$2); ($\overline{2}$1$\overline{2}$); ($\overline{2}$12); (2$\overline{12}$); (2$\overline{1}$2); (21$\overline{2}$); (212) \\

        \multirow{2}{*}{\{221\}}
        & \multirow{2}{*}{16}
        & \rule{0pt}{3.2ex}%
          ($\overline{221}$); ($\overline{22}$1); ($\overline{2}$2$\overline{1}$); ($\overline{2}$21); 
          ($\overline{122}$); ($\overline{12}$2); ($\overline{1}$2$\overline{2}$); ($\overline{1}$22); 
          (1$\overline{22}$); (1$\overline{2}$2); (12$\overline{2}$); (122) \\
        &
        & (2$\overline{21}$); (2$\overline{2}$1); (22$\overline{1}$); (221) \\

        \rule{0pt}{3.2ex}%
        \{230\} & 8 & ($\overline{23}$0); ($\overline{2}$30); (0$\overline{32}$); (0$\overline{3}$2); (03$\overline{2}$); (032); (2$\overline{3}$0); (230) \\

        \multirow{2}{*}{\{231\}}
        & \multirow{2}{*}{16}
        & \rule{0pt}{3.2ex}%
          ($\overline{231}$); ($\overline{23}$1); ($\overline{2}$3$\overline{1}$); ($\overline{2}$31); 
          ($\overline{132}$); ($\overline{13}$2); ($\overline{1}$3$\overline{2}$); ($\overline{1}$32); 
          (1$\overline{32}$); (1$\overline{3}$2); (13$\overline{2}$); (132) \\
        &
        & (2$\overline{31}$); (2$\overline{3}$1); (23$\overline{1}$); (231) \\

        \rule{0pt}{3.2ex}%
        \{232\} & 8 & ($\overline{232}$); ($\overline{23}$2); ($\overline{2}$3$\overline{2}$); ($\overline{2}$32); (2$\overline{32}$); (2$\overline{3}$2); (23$\overline{2}$); (232) \\

        \{301\} & 8 & ($\overline{3}$0$\overline{1}$); ($\overline{3}$01); ($\overline{1}$0$\overline{3}$); ($\overline{1}$03); (10$\overline{3}$); (103); (30$\overline{1}$); (301) \\
        \{302\} & 8 & ($\overline{3}$0$\overline{2}$); ($\overline{3}$02); ($\overline{2}$0$\overline{3}$); ($\overline{2}$03); (20$\overline{3}$); (203); (30$\overline{2}$); (302) \\
        \{310\} & 8 & ($\overline{31}$0); ($\overline{3}$10); (0$\overline{13}$); (0$\overline{1}$3); (01$\overline{3}$); (013); (3$\overline{1}$0); (310) \\

        \multirow{2}{*}{\{311\}}
        & \multirow{2}{*}{16}
        & \rule{0pt}{3.2ex}%
          ($\overline{311}$); ($\overline{31}$1); ($\overline{3}$1$\overline{1}$); ($\overline{3}$11); 
          ($\overline{113}$); ($\overline{11}$3); ($\overline{1}$1$\overline{3}$); ($\overline{1}$13); 
          (1$\overline{13}$); (1$\overline{1}$3); (11$\overline{3}$); (113) \\
        &
        & (3$\overline{11}$); (3$\overline{1}$1); (31$\overline{1}$); (311) \\

        \multirow{2}{*}{\{312\}}
        & \multirow{2}{*}{16}
        & \rule{0pt}{3.2ex}%
          ($\overline{312}$); ($\overline{31}$2); ($\overline{3}$1$\overline{2}$); ($\overline{3}$12); ($\overline{213}$); ($\overline{21}$3); ($\overline{2}$1$\overline{3}$); ($\overline{2}$13); (2$\overline{13}$); (2$\overline{1}$3); (21$\overline{3}$); (213) \\
        &
        & (3$\overline{12}$); (3$\overline{1}$2); (31$\overline{2}$); (312) \\

        \rule{0pt}{3.2ex}%
        \{313\} & 8 & ($\overline{313}$); ($\overline{31}$3); ($\overline{3}$1$\overline{3}$); ($\overline{3}$13); (3$\overline{13}$); (3$\overline{1}$3); (31$\overline{3}$); (313) \\
        \{320\} & 8 & ($\overline{32}$0); ($\overline{3}$20); (0$\overline{23}$); (0$\overline{2}$3); (02$\overline{3}$); (023); (3$\overline{2}$0); (320) \\

        \multirow{2}{*}{\{321\}}
        & \multirow{2}{*}{16}
        & \rule{0pt}{3.2ex}%
          ($\overline{321}$); ($\overline{32}$1); ($\overline{3}$2$\overline{1}$); ($\overline{3}$21); ($\overline{123}$); ($\overline{12}$3); ($\overline{1}$2$\overline{3}$); ($\overline{1}$23); (1$\overline{23}$); (1$\overline{2}$3); (12$\overline{3}$); (123) \\
        &
        & (3$\overline{21}$); (3$\overline{2}$1); (32$\overline{1}$); (321) \\

        \multirow{2}{*}{\{322\}}
        & \multirow{2}{*}{16}
        & \rule{0pt}{3.2ex}%
          ($\overline{322}$); ($\overline{32}$2); ($\overline{3}$2$\overline{2}$); ($\overline{3}$22); ($\overline{223}$); ($\overline{22}$3); ($\overline{2}$2$\overline{3}$); ($\overline{2}$23); (2$\overline{23}$); (2$\overline{2}$3); (22$\overline{3}$); (223) \\
        &
        & (3$\overline{22}$); (3$\overline{2}$2); (32$\overline{2}$); (322) \\

        \rule{0pt}{3.2ex}%
        \{323\} & 8 & ($\overline{323}$); ($\overline{32}$3); ($\overline{3}$2$\overline{3}$); ($\overline{3}$23); (3$\overline{23}$); (3$\overline{2}$3); (32$\overline{3}$); (323) \\

        \multirow{2}{*}{\{331\}}
        & \multirow{2}{*}{16}
        & \rule{0pt}{3.2ex}%
          ($\overline{331}$); ($\overline{33}$1); ($\overline{3}$3$\overline{1}$); ($\overline{3}$31); ($\overline{133}$); ($\overline{13}$3); ($\overline{1}$3$\overline{3}$); ($\overline{1}$33); (1$\overline{33}$); (1$\overline{3}$3); (13$\overline{3}$); (133) \\
        &
        & (3$\overline{31}$); (3$\overline{3}$1); (33$\overline{1}$); (331) \\

        \multirow{2}{*}{\{332\}}
        & \multirow{2}{*}{16}
        & \rule{0pt}{3.2ex}%
          ($\overline{332}$); ($\overline{33}$2); ($\overline{3}$3$\overline{2}$); ($\overline{3}$32); ($\overline{233}$); ($\overline{23}$3); ($\overline{2}$3$\overline{3}$); ($\overline{2}$33); (2$\overline{33}$); (2$\overline{3}$3); (23$\overline{3}$); (233) \\
        &
        & (3$\overline{32}$); (3$\overline{3}$2); (33$\overline{2}$); (332) \\

        \bottomrule
    \end{tabular}
\end{table}

\begin{table}[htbp]
\centering
\caption{
Surface energies and surface area fractions of exposed facets in rock salt (RS) MnS Wulff constructions, along with the weighted surface energies under different thermodynamic equilibrium conditions.
All results are from r$^2$SCAN calculations.
Data correspond to the Wulff constructions shown in Figure~\ref{fig:figs2}d.
}
\label{tab:tabs4}
\begin{tabular}{
    l
    S[table-format=1.2] S[table-format=2.2]
    S[table-format=1.2] S[table-format=2.2]
    S[table-format=1.2] S[table-format=2.2]
    S[table-format=1.2] S[table-format=2.2]
}
\toprule
    & \multicolumn{2}{c}{$\Delta \mu_{\ce{S}} = -2.0$ eV}
    & \multicolumn{2}{c}{$\Delta \mu_{\ce{S}} = -1.0$ eV}
    & \multicolumn{2}{c}{$\Delta \mu_{\ce{S}} = -0.7$ eV}
    & \multicolumn{2}{c}{$\Delta \mu_{\ce{S}} = -0.4$ eV} \\
    \cmidrule(lr){2-3} \cmidrule(lr){4-5} \cmidrule(lr){6-7} \cmidrule(lr){8-9}
    {$\{hkl\}$}
    & {$\gamma$} & {$f_A$}
    & {$\gamma$} & {$f_A$}
    & {$\gamma$} & {$f_A$}
    & {$\gamma$} & {$f_A$} \\
\midrule
    $\{010\}$  & 0.36 & 37.18 & 0.36 & 35.15 & 0.36 & 14.70 & 0.36 &  0.00 \\
    $\{100\}$  & 0.43 & 62.82 & 0.43 & 51.00 & 0.43 &  7.78 & 0.43 &  0.00 \\
    $\{131\}$  & 0.88 &  0.00 & 0.53 &  0.78 & 0.42 & 19.83 & 0.31 & 35.90 \\
    $\{311\}$  & 0.90 &  0.00 & 0.54 & 13.07 & 0.43 & 57.69 & 0.33 & 64.10 \\
\midrule
    Wulff      & 0.40 & 100.00 & 0.42 & 100.00 & 0.42 & 100.00 & 0.32 & 100.00 \\
\bottomrule
\addlinespace[2pt]
\multicolumn{9}{l}{\footnotesize $\gamma$: surface energy / \unit{J.m^{-2}}; \quad $f_A$: area fraction / \unit{\percent}}
\end{tabular}
\end{table}

\clearpage

\subsection{\texorpdfstring{r$^2$SCAN+$U$}{r²SCAN+U} and HSE06 results for RS-MnS NCs}
\label{subsection:s2.2}

\begin{figure*}[htbp]
    \centering
    \includegraphics[width=0.7\linewidth]{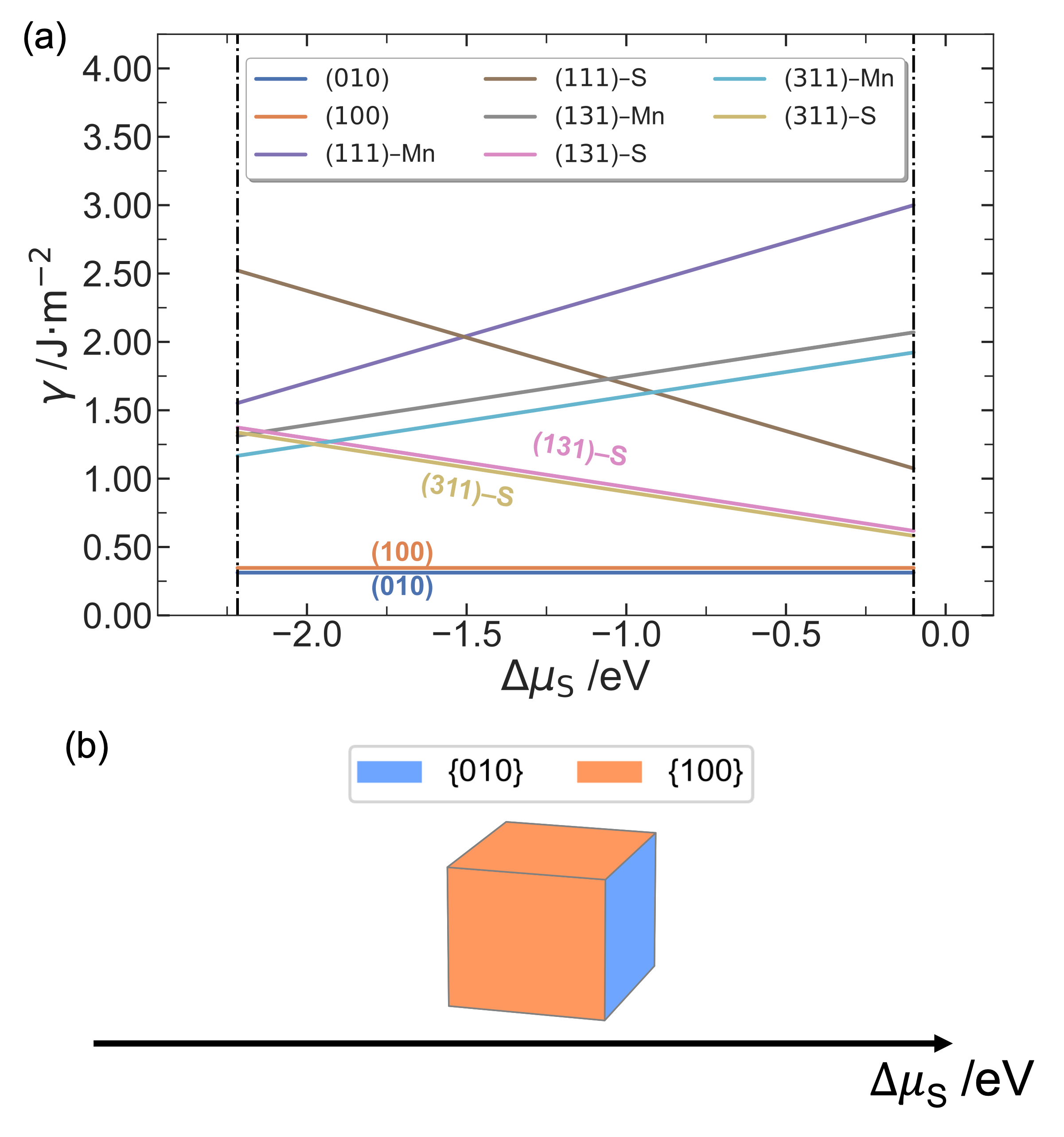}
    \caption{
(a) Dependence of rock salt (RS) MnS surface energies on the relative chemical potential of sulfur, $\Delta \mu_{\ce{S}}$, for eight facets: (010), (100), (111)--Mn, (111)--S, (131)--Mn, (131)--S, (311)--Mn, and (311)--S.
The $\Delta \mu_{\ce{S}}$ range between the two vertical dash-dotted lines indicates the thermodynamically stable region of bulk RS-MnS.
(b) Wulff constructions derived from the surface energies in (a).
All results are from HSE06 single-point calculations on r$^2$SCAN-optimized slab structures.
}
    \label{fig:figs3}
\end{figure*}

To benchmark the ability of r$^2$SCAN to describe surface energies, eight representative RS-MnS facets were selected: the nonpolar (010) and (100) facets, the low-index polar (111) facets, and the high-index polar (131) and (311) facets.
Slab models were structurally optimized using r$^2$SCAN, followed by single-point self-consistent field (SCF) calculations using HSE06 to obtain reference energies.
As shown in Figure~\ref{fig:figs3}a, throughout the entire thermodynamically stable region of RS-MnS, the (100) and (010) facets consistently exhibit the lowest surface energies.
The remaining six polar facets have sufficiently high surface energies that RS-MnS NCs consistently favor cubic morphologies (Figure~\ref{fig:figs3}b).
Table~\ref{tab:tabs5} lists the surface energies and surface area fractions of the exposed facets, along with the weighted surface energies of the Wulff constructions under Mn-rich and S-rich conditions.

\begin{table}[htbp]
\centering
\caption{
Surface energies and surface area fractions of exposed facets in rock salt (RS) MnS Wulff constructions, along with the weighted surface energies under Mn-rich ($\Delta \mu_{\ce{S}} = -2.0$~eV) and S-rich ($\Delta \mu_{\ce{S}} = -0.1$~eV) thermodynamic equilibrium conditions.
Results are from r$^2$SCAN+$U$ and HSE06 calculations.
Data correspond to the Wulff constructions presented in Figure~5d of the main text and Figure~\ref{fig:figs3}b, respectively.
}
\label{tab:tabs5}
\begin{tabular}{
    l
    S[table-format=1.2] S[table-format=2.2]
    S[table-format=1.2] S[table-format=2.2]
    S[table-format=1.2] S[table-format=2.2]
    S[table-format=1.2] S[table-format=2.2]
}
\toprule
    & \multicolumn{4}{c}{r$^2$SCAN+$U$}
    & \multicolumn{4}{c}{HSE06} \\
    \cmidrule(lr){2-5} \cmidrule(lr){6-9}
    & \multicolumn{2}{c}{$\Delta \mu_{\text{S}} = -2.0$~eV}
    & \multicolumn{2}{c}{$\Delta \mu_{\text{S}} = -0.1$~eV}
    & \multicolumn{2}{c}{$\Delta \mu_{\text{S}} = -2.0$~eV}
    & \multicolumn{2}{c}{$\Delta \mu_{\text{S}} = -0.1$~eV} \\
    \cmidrule(lr){2-3} \cmidrule(lr){4-5} \cmidrule(lr){6-7} \cmidrule(lr){8-9}
    {$\{hkl\}$}
    & {$\gamma$} & {$f_A$}
    & {$\gamma$} & {$f_A$}
    & {$\gamma$} & {$f_A$}
    & {$\gamma$} & {$f_A$} \\
\midrule
    $\{010\}$  & 0.41 & 34.20 & 0.41 & 34.05 & 0.31 & 35.70 & 0.31 & 35.70 \\
    $\{100\}$  & 0.43 & 65.80 & 0.43 & 64.74 & 0.35 & 64.30 & 0.35 & 64.30 \\
    $\{131\}$  & 1.33 &  0.00 & 0.65 &  0.00 & 1.30 &  0.00 & 0.62 &  0.00 \\
    $\{311\}$  & 1.29 &  0.00 & 0.61 &  1.21 & 1.26 &  0.00 & 0.58 &  0.00 \\
\midrule
    Wulff      & 0.42 & 100.00 & 0.43 & 100.00 & 0.33 & 100.00 & 0.33 & 100.00 \\
\bottomrule
\addlinespace[2pt]
\multicolumn{9}{l}{\footnotesize $\gamma$: surface energy / \unit{J.m^{-2}}; \quad $f_A$: area fraction / \unit{\percent}}
\end{tabular}
\end{table}

\clearpage

\subsection{\texorpdfstring{r$^2$SCAN}{r²SCAN} results for ZB-MnS NCs}
\label{subsection:s2.3}

\begin{figure*}[htbp]
    \centering
    \includegraphics[width=1.0\linewidth]{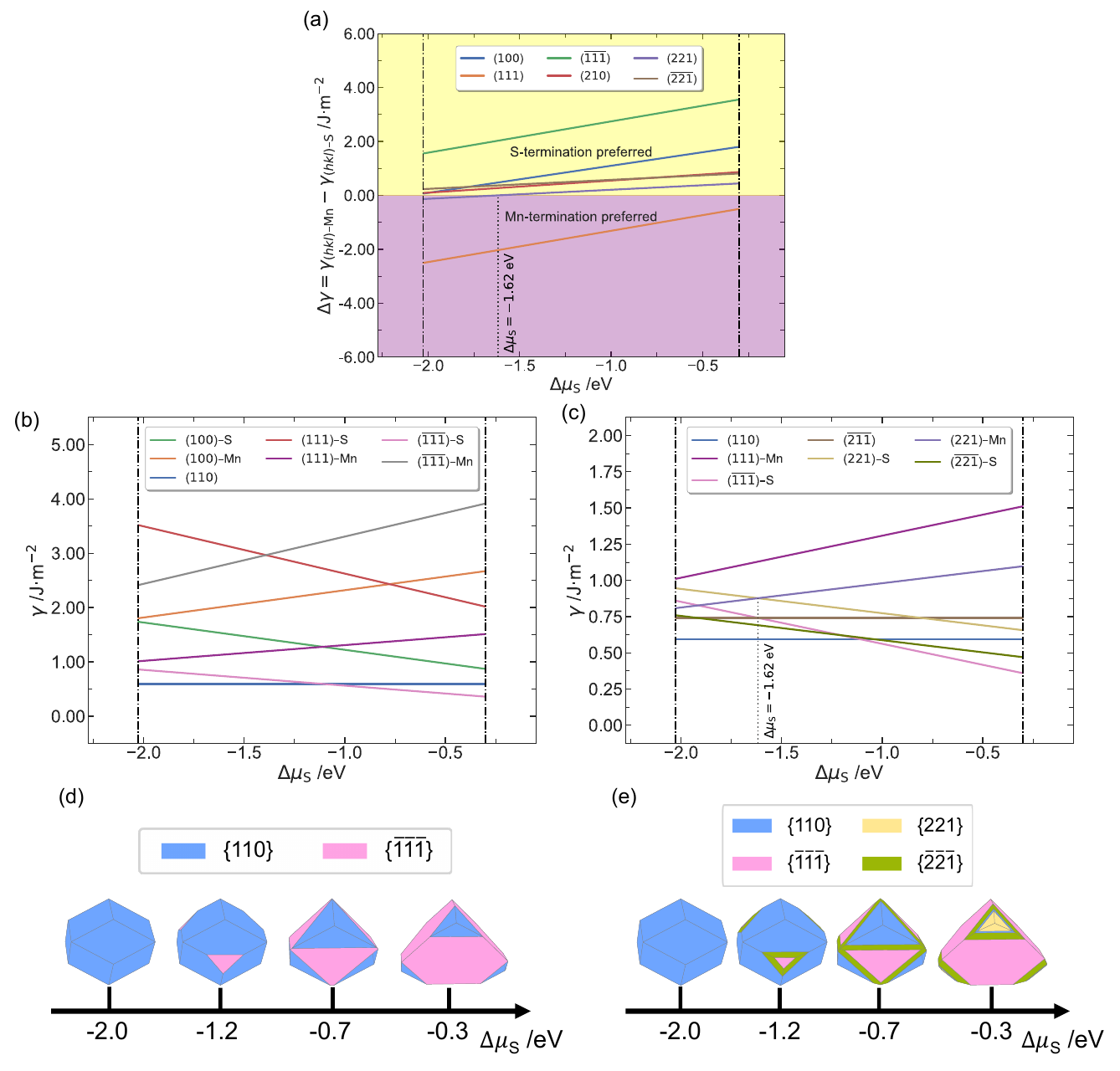}
    \caption{
(a) Termination stability comparison for six zinc blende (ZB) MnS facets with two possible terminations.
Dependence of ZB-MnS surface energies on the relative chemical potential of sulfur, $\Delta \mu_{\ce{S}}$, for (b) seven low-index facets and (c) the seven lowest-energy facets among all $\{h,k,l\} \leq 2$ facets.
The $\Delta \mu_{\ce{S}}$ range between the two vertical dash-dotted lines ($-2.03$~eV $\leq \Delta \mu_{\ce{S}} \leq -0.30$~eV) indicates the thermodynamically stable region of bulk ZB-MnS.
Wulff constructions for ZB-MnS nanocrystals considering (d) low-index facets only and (e) all facets with Miller indices up to 2.
All results are from r$^2$SCAN calculations.
}
    \label{fig:figs4}
\end{figure*}

Figure~\ref{fig:figs4}a summarizes termination stability for ZB-MnS facets with two possible terminations as a function of $\Delta \mu_{\ce{S}}$.
The (221) facet prefers Mn-termination at low $\Delta \mu_{\ce{S}}$ but transitions to S-termination once $\Delta \mu_{\ce{S}}$ exceeds $-1.62$~eV.
Within the thermodynamically stable region of ZB-MnS, (111)--Mn is always more stable than (111)--S, whereas the opposite trend holds for the (100), ($\overline{111}$), (210), and ($\overline{221}$) facets.
Figure~\ref{fig:figs4}b shows the surface energies of low-index ZB-MnS facets as a function of $\Delta \mu_{\ce{S}}$, with the corresponding Wulff constructions in Figure~\ref{fig:figs4}d.
Under Mn-rich conditions, the (110) facet exhibits the lowest surface energy, followed by ($\overline{111}$)--S and (111)--Mn, and the predicted NCs adopt rhombic dodecahedral morphologies.
As $\Delta \mu_{\ce{S}}$ increases, the ($\overline{111}$)--S surface energy decreases continuously, enlarging its exposed area; at the S-rich limit ($\Delta \mu_{\ce{S}} = -0.30$~eV), the NCs evolve into a 16-faced polyhedron bounded by 4 \{$\overline{111}$\}--S and 12 \{110\} facets.
When all facets with Miller indices up to 2 are included (Figures~\ref{fig:figs4}c~and~\ref{fig:figs4}e), the Mn-rich morphology remains a rhombic dodecahedron.
With increasing $\Delta \mu_{\ce{S}}$, the NCs evolve into a morphology in which each of the 4 \{$\overline{111}$\} tetrahedral vertices is truncated by 3 \{110\}, 3 \{221\}--S, and 3 \{$\overline{221}$\}--S facets.
The symmetry-equivalent facets in each $\{hkl\}$ family for ZB-MnS are summarized in Table~\ref{tab:tabs6}.
Table~\ref{tab:tabs7} lists the surface energies and surface area fractions of exposed facets, along with the weighted surface energies of the Wulff constructions shown in Figure~\ref{fig:figs4}e.

\clearpage

\begin{table}[htbp]
    \centering
    \caption{
Symmetry-equivalent $\{hkl\}$ facets of zinc blende (ZB) MnS.
    }
    \label{tab:tabs6}
    \begin{tabular}{lll}
        \toprule
        $\{hkl\}$ & Count & Symmetry-equivalent $(hkl)$ \\
        \midrule
        \{$\overline{221}$\}  & 12 & ($\overline{221}$); ($\overline{212}$); ($\overline{2}$12); ($\overline{2}$21); 
                                   ($\overline{122}$); ($\overline{1}$22); (1$\overline{2}$2); (12$\overline{2}$);
                                   (2$\overline{2}$1); (2$\overline{1}$2); (21$\overline{2}$); (22$\overline{1}$) \\
                                   
        \{$\overline{211}$\}  & 12 & ($\overline{211}$); ($\overline{2}$11); ($\overline{121}$); ($\overline{112}$); 
                                   ($\overline{1}$12); ($\overline{1}$21); (1$\overline{2}$1); (1$\overline{1}$2);
                                   (11$\overline{2}$); (12$\overline{1}$); (2$\overline{1}$1); (21$\overline{1}$) \\
                                   
        \{$\overline{111}$\}  & 4  & ($\overline{111}$); ($\overline{1}$11); (1$\overline{1}$1); (11$\overline{1}$) \\
        
        \{100\}               & 6  & ($\overline{1}$00); (0$\overline{1}$0); (00$\overline{1}$); (001); (010); (100) \\
        
        \{110\}               & 12 & ($\overline{11}$0); ($\overline{1}$0$\overline{1}$); ($\overline{1}$01); ($\overline{1}$10); 
                                   (0$\overline{11}$); (0$\overline{1}$1); (01$\overline{1}$); (011);
                                   (1$\overline{1}$0); (10$\overline{1}$); (101); (110) \\
                                   
        \{111\}               & 4  & ($\overline{11}$1); ($\overline{1}$1$\overline{1}$); (1$\overline{11}$); (111) \\
        
        \multirow{2}{*}{\{210\}} 
        & \multirow{2}{*}{24} 
        & \rule{0pt}{3.2ex}%
            ($\overline{2}\overline{1}$0); ($\overline{2}$0$\overline{1}$); ($\overline{2}$01); ($\overline{2}$10);
            ($\overline{12}$0); ($\overline{1}$0$\overline{2}$); ($\overline{1}$02); ($\overline{1}$20);
            (0$\overline{2}\overline{1}$); (0$\overline{2}$1); (0$\overline{1}\overline{2}$); (0$\overline{1}$2) \\
        &
        & 
            (01$\overline{2}$); (012); (02$\overline{1}$); (021);
            (1$\overline{2}$0); (10$\overline{2}$); (102); (120);
            (2$\overline{1}$0); (20$\overline{1}$); (201); (210) \\

        \rule{0pt}{3.2ex}%
        \{211\}               & 12 & ($\overline{21}$1); ($\overline{2}$1$\overline{1}$); ($\overline{12}$1); ($\overline{11}$2); 
                                   ($\overline{1}$1$\overline{2}$); ($\overline{1}$2$\overline{1}$); (1$\overline{21}$); (1$\overline{12}$); 
                                   (112); (121); (2$\overline{1}\overline{1}$); (211) \\
        
        \{221\}               & 12 & ($\overline{22}$1); ($\overline{21}$2); ($\overline{2}$1$\overline{2}$); ($\overline{2}$2$\overline{1}$); 
                                     ($\overline{12}$2); ($\overline{1}$2$\overline{2}$); (1$\overline{22}$); (122); 
                                     (2$\overline{21}$); (2$\overline{12}$); (212); (221) \\
        
        \bottomrule
    \end{tabular}
\end{table}

\begin{table}[htbp]
\centering
\caption{
Surface energies and surface area fractions of exposed facets in zinc blende (ZB) MnS Wulff constructions, along with the weighted surface energies under different thermodynamic equilibrium conditions.
All results are from r$^2$SCAN calculations.
Data correspond to the Wulff constructions shown in Figure~\ref{fig:figs4}e.
}
\label{tab:tabs7}
\begin{tabular}{
    l
    S[table-format=1.2] S[table-format=2.2]
    S[table-format=1.2] S[table-format=2.2]
    S[table-format=1.2] S[table-format=2.2]
    S[table-format=1.2] S[table-format=2.2]
}
\toprule
    & \multicolumn{2}{c}{$\Delta \mu_{\ce{S}} = -2.0$~eV}
    & \multicolumn{2}{c}{$\Delta \mu_{\ce{S}} = -1.2$~eV}
    & \multicolumn{2}{c}{$\Delta \mu_{\ce{S}} = -0.7$~eV}
    & \multicolumn{2}{c}{$\Delta \mu_{\ce{S}} = -0.3$~eV} \\
    \cmidrule(lr){2-3} \cmidrule(lr){4-5} \cmidrule(lr){6-7} \cmidrule(lr){8-9}
    {$\{hkl\}$}
    & {$\gamma$} & {$f_A$}
    & {$\gamma$} & {$f_A$}
    & {$\gamma$} & {$f_A$}
    & {$\gamma$} & {$f_A$} \\
\midrule
    $\{110\}$                    & 0.59 & 100.00 & 0.59 & 86.39 & 0.59 & 39.82 & 0.59 &  2.47 \\
    $\{\overline{111}\}$         & 0.85 & 0.00 & 0.62 & 2.76 & 0.48 & 33.75 & 0.36 & 75.36 \\
    $\{221\}$                    & 0.81 &  0.00 & 0.81 &  0.00 & 0.72 &  0.00 & 0.66 &  5.65 \\
    $\{\overline{221}\}$         & 0.75 &  0.00 & 0.62 &  10.85 & 0.54 &  26.43 & 0.47 &  16.51 \\
\midrule
    Wulff                        & 0.59 & 100.00 & 0.60 & 100.00 & 0.54 & 100.00 & 0.40 & 100.00 \\
\bottomrule
\addlinespace[2pt]
\multicolumn{9}{l}{\footnotesize $\gamma$: surface energy / \unit{J.m^{-2}}; \quad $f_A$: area fraction / \unit{\percent}}
\end{tabular}
\end{table}

\clearpage

\subsection{\texorpdfstring{r$^2$SCAN+$U$}{r²SCAN+U} results for ZB-MnS NCs}
\label{subsection:s2.4}

As shown in Figure~7a of the main text, the stoichiometric (110) facet has a $\Delta \mu_{\ce{S}}$-independent surface energy of 0.61~J$\cdot$m$^{-2}$, while the ($\overline{111}$)--S facet decreases continuously with increasing $\Delta \mu_{\ce{S}}$, dropping below (110) only near the S-rich stability limit.
The next-lowest-energy low-index facets, (100)--S and (111)--Mn in their most stable terminations, are still at least 0.45~J$\cdot$m$^{-2}$ above (110) and ($\overline{111}$)--S, and consequently do not appear in the Wulff shapes.
When including high-index facets  (Figure~7b of the main text), the (110) and ($\overline{111}$)--S facets remain the most stable, while the (211) facet has the next lowest surface energy at 0.77~J$\cdot$m$^{-2}$, independent of $\Delta \mu_{\ce{S}}$ (0.16~J$\cdot$m$^{-2}$ above (110)).
In contrast, the surface energy of the (221)--S facet decreases with increasing $\Delta \mu_{\ce{S}}$, gradually approaching that of (110). 
Although the stable configuration switches from Mn-termination to S-termination at $\Delta \mu_{\ce{S}} = -1.62$~eV, the (221) facet is not expected to appear in the equilibrium morphology due to the relatively high surface energies of both terminations.
Table~\ref{tab:tabs8} lists the surface energies and surface area fractions of exposed facets, along with the weighted surface energies, for ZB-MnS Wulff constructions considering all facets with Miller indices up to 2.

\begin{table}[htbp]
\centering
\caption{
Surface energies and surface area fractions of exposed facets in zinc blende (ZB) MnS Wulff constructions, along with the weighted surface energies under different thermodynamic equilibrium conditions.
All results are from r$^2$SCAN+$U$ calculations.
Data correspond to the Wulff constructions presented in Figure~7c of the main text.
}
\label{tab:tabs8}
\begin{tabular}{
    l
    S[table-format=1.2] S[table-format=2.2]
    S[table-format=1.2] S[table-format=2.2]
    S[table-format=1.2] S[table-format=2.2]
    S[table-format=1.2] S[table-format=2.2]
    S[table-format=1.2] S[table-format=2.2]
}
\toprule
    & \multicolumn{2}{c}{$\Delta \mu_{\ce{S}} = -2.0$~eV}
    & \multicolumn{2}{c}{$\Delta \mu_{\ce{S}} = -1.4$~eV}
    & \multicolumn{2}{c}{$\Delta \mu_{\ce{S}} = -1.0$~eV}
    & \multicolumn{2}{c}{$\Delta \mu_{\ce{S}} = -0.5$~eV}
    & \multicolumn{2}{c}{$\Delta \mu_{\ce{S}} = -0.3$~eV} \\
    \cmidrule(lr){2-3} \cmidrule(lr){4-5} \cmidrule(lr){6-7} \cmidrule(lr){8-9} \cmidrule(lr){10-11}
    {$\{hkl\}$}
    & {$\gamma$} & {$f_A$}
    & {$\gamma$} & {$f_A$}
    & {$\gamma$} & {$f_A$}
    & {$\gamma$} & {$f_A$}
    & {$\gamma$} & {$f_A$} \\
\midrule
    $\{110\}$                    & 0.61 & 100.00 & 0.61 & 100.00 & 0.61 & 98.50 & 0.61 & 74.25 & 0.61 & 54.61 \\
    $\{\overline{111}\}$         & 0.99 &  0.00 & 0.82 & 0.00 & 0.70 & 1.50 & 0.56 & 25.75 & 0.50 & 45.39 \\
\midrule
    Wulff                        & 0.61 & 100.00 & 0.61 & 100.00 & 0.61 & 100.00 & 0.60 & 100.00 & 0.56 & 100.00 \\
\bottomrule
\addlinespace[2pt]
\multicolumn{9}{l}{\footnotesize $\gamma$: surface energy / \unit{J.m^{-2}}; \quad $f_A$: area fraction / \unit{\percent}}
\end{tabular}
\end{table}

\clearpage

\subsection{\texorpdfstring{r$^2$SCAN}{r²SCAN} results for WZ-MnS NCs}
\label{subsection:s2.5}

\begin{figure*}[htbp]
    \centering
    \includegraphics[width=1.0\linewidth]{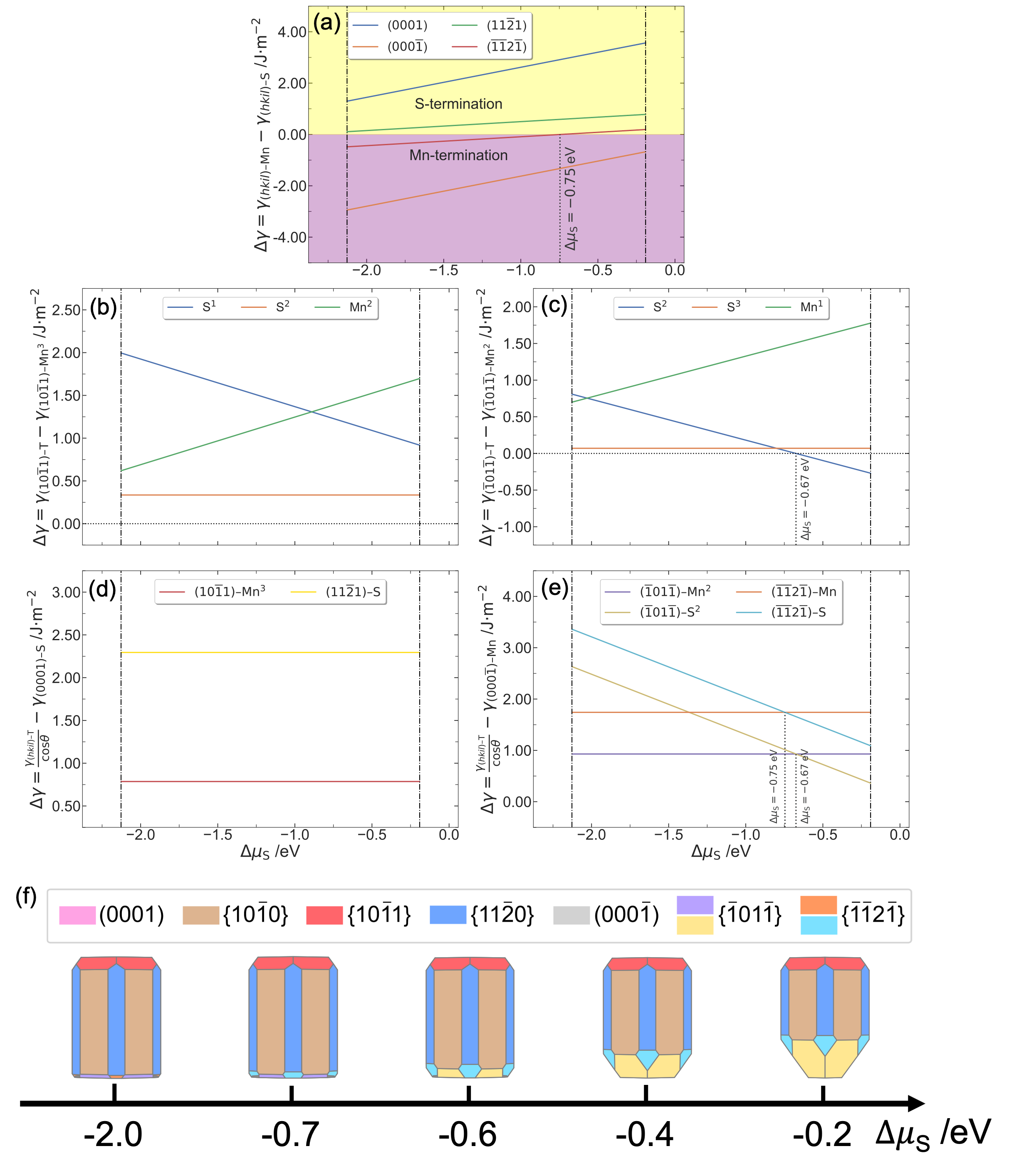}
    \caption{
(a) Termination stability comparison for the (0001), ($000\overline{1}$), ($11\overline{2}1$), and ($\overline{11}2\overline{1}$) facets in wurtzite (WZ) MnS.
(b) Termination stability comparison for four terminations of the ($10\overline{1}1$) facet.
(c) Termination stability comparison for four terminations of the ($\overline{1}01\overline{1}$) facet.
(d, e) Relative surface energies $\Delta \gamma$ of facets along the (d) [0001] and (e) [$000\overline{1}$] directions as a function of the relative chemical potential of sulfur, $\Delta \mu_{\ce{S}}$.
(f) Wulff constructions of WZ-MnS nanocrystals at five representative $\Delta \mu_{\ce{S}}$ values.
All results are from r$^2$SCAN calculations.
}
    \label{fig:figs5}
\end{figure*}

Figure~\ref{fig:figs5}a summarizes termination stability between Mn- and S-terminations for the (0001), ($11\overline{2}1$), and their opposite facets in WZ-MnS as a function of $\Delta \mu_{\ce{S}}$.
The ($\overline{11}2\overline{1}$) facet favors Mn-termination when $\Delta \mu_{\ce{S}} \leq -0.75$~eV, with S-termination becoming more stable at higher $\Delta \mu_{\ce{S}}$.
Within the thermodynamically stable region of WZ-MnS, ($000\overline{1}$)--Mn is always more stable than ($000\overline{1}$)--S, while the opposite trend holds for the (0001) and ($11\overline{2}1$) facets.
The ($10\overline{1}1$) and ($\overline{1}01\overline{1}$) facets each admit four distinct terminations.
For ($10\overline{1}1$), the three-coordinated Mn termination, ($10\overline{1}1$)--Mn$^3$, is most stable throughout (Figure~\ref{fig:figs5}b).
For ($\overline{1}01\overline{1}$), the most stable termination changes from ($\overline{1}01\overline{1}$)--Mn$^2$ to ($\overline{1}01\overline{1}$)--S$^2$ at $\Delta \mu_{\ce{S}} = -0.67$~eV (Figure~\ref{fig:figs5}c).

Figures~\ref{fig:figs5}d~and~\ref{fig:figs5}e show the relative surface energies along the [0001] and [$000\overline{1}$] directions, referenced to the (0001) and ($000\overline{1}$) facets, respectively; only the most stable termination for each facet is used.
Along [0001], the relative surface energies of ($10\overline{1}1$)--Mn$^3$ and ($11\overline{2}1$)--S are independent of $\Delta \mu_{\ce{S}}$, so the NC morphology is insensitive to $\Delta \mu_{\ce{S}}$ in this direction.
Along [$000\overline{1}$], the morphology is likewise nearly unchanged at low $\Delta \mu_{\ce{S}}$ because the relative surface energies of ($\overline{1}01\overline{1}$)--Mn$^2$ and ($\overline{11}2\overline{1}$)--Mn are invariant.
As $\Delta \mu_{\ce{S}}$ increases, the ($\overline{11}2\overline{1}$) facet transitions to S-termination at $\Delta \mu_{\ce{S}} = -0.75$~eV and the ($\overline{1}01\overline{1}$) facet at $-0.67$~eV, and the base of the WZ-MnS NCs becomes increasingly dominated by exposed ($\overline{11}2\overline{1}$)--S and ($\overline{1}01\overline{1}$)--S$^2$ facets.

The resulting Wulff constructions (Figure~\ref{fig:figs5}f) predict that WZ-MnS NCs exhibit hexagonal nanorod morphologies under Mn-rich conditions, transitioning to bullet-like morphologies under S-rich conditions.
The symmetry-equivalent facets in each $\{hkil\}$ family for WZ-MnS are summarized in Table~\ref{tab:tabs9}.
Table~\ref{tab:tabs10} lists the surface energies, surface area fractions, and weighted surface energies.
Note that the individual polar surface energies, except those of $\{10\overline{1}0\}$ and $\{11\overline{2}0\}$, are not physically meaningful because the (0001)--S surface energy was set to a reference value of 0.06~eV$\cdot$\AA$^{-2}$, corresponding to half of the total surface energy of the (0001)--S and (000$\overline{1}$)--Mn surfaces.
The surface area fractions and the weighted surface energies are well defined, as demonstrated by the two sets of data at $\Delta \mu_{\ce{S}}=-0.20$~eV shown in Table~\ref{tab:tabs10}.
These two sets correspond to reference (0001)--S surface energies of 0.06~eV$\cdot$\AA$^{-2}$ and 0.10~eV$\cdot$\AA$^{-2}$.

\clearpage

\begin{table}[htbp]
    \centering
    \caption{
Symmetry-equivalent $\{hkil\}$ facets of wurtzite (WZ) MnS.
    }
    \label{tab:tabs9}
    \begin{tabular}{lll}
        \toprule
        $\{hkil\}$ & Count & Symmetry-equivalent $(hkil)$ \\
        \midrule
        \{$\overline{11}$2$\overline{1}$\}  & 6 & ($\overline{2}$11$\overline{1}$); ($\overline{11}$2$\overline{1}$); ($\overline{1}$2$\overline{11}$);
                                                  (1$\overline{2}$1$\overline{1}$); (11$\overline{21}$); (2$\overline{111}$) \\
                                                  
        \{$\overline{1}$01$\overline{1}$\}  & 6 & ($\overline{1}$01$\overline{1}$); ($\overline{1}$10$\overline{1}$); (0$\overline{1}$1$\overline{1}$);
                                                  (01$\overline{11}$); (1$\overline{1}$0$\overline{1}$); (10$\overline{11}$) \\
                                   
        \{000$\overline{1}$\}               & 1 & (000$\overline{1}$) \\

        \{0001\}                            & 1 & (0001) \\

        \{10$\overline{1}$0\}               & 6 & ($\overline{1}$010); ($\overline{1}$100); (0$\overline{1}$10); 
                                                  (01$\overline{1}$0); (1$\overline{1}$00); (10$\overline{1}$0)    \\

        \{10$\overline{1}$1\}               & 6 & ($\overline{1}$011); ($\overline{1}$101); (0$\overline{1}$11); 
                                                  (01$\overline{1}$1); (1$\overline{1}$01); (10$\overline{1}$1)    \\

        \{11$\overline{2}$0\}               & 6 & ($\overline{2}$110); ($\overline{11}$20); ($\overline{1}$2$\overline{1}$0); 
                                                  (1$\overline{2}$10); (11$\overline{2}$0); (2$\overline{11}$0)    \\

        \{11$\overline{2}$1\}               & 6 & ($\overline{2}$111); ($\overline{11}$21); ($\overline{1}$2$\overline{1}$1); 
                                                  (1$\overline{2}$11); (11$\overline{2}$1); (2$\overline{11}$1)    \\

        \bottomrule
    \end{tabular}
\end{table}

\begin{table}[htbp]
\centering
\caption{
Surface energies and surface area fractions of exposed facets in wurtzite (WZ) MnS Wulff constructions, along with weighted surface energies under different thermodynamic equilibrium conditions.
Individual polar surface energies are referenced to an arbitrary (0001)--S value (see Subsection~\ref{subsection:s2.5}) and are not physically meaningful; the surface area fractions and the weighted surface energies are well defined.
All results are from r$^2$SCAN calculations.
Data correspond to the Wulff constructions presented in Figure~\ref{fig:figs5}f.
}
\label{tab:tabs10}

\vspace{4pt}
\begin{tabular}{
    l
    S[table-format=1.2] S[table-format=2.2]
    S[table-format=1.2] S[table-format=2.2]
    S[table-format=1.2] S[table-format=2.2]
}
\toprule
    & \multicolumn{2}{c}{$\Delta \mu_{\ce{S}} = -2.00$~eV}
    & \multicolumn{2}{c}{$\Delta \mu_{\ce{S}} = -0.70$~eV}
    & \multicolumn{2}{c}{$\Delta \mu_{\ce{S}} = -0.60$~eV} \\
    \cmidrule(lr){2-3} \cmidrule(lr){4-5} \cmidrule(lr){6-7}
    {$\{hkl\}$}
    & {$\gamma$} & {$f_A$}
    & {$\gamma$} & {$f_A$}
    & {$\gamma$} & {$f_A$} \\
\midrule
    $(0001)$                               & 0.96 &  7.54 & 0.96 &  7.55 & 0.96 &  7.69 \\
    $(000\overline{1})$                    & 0.96 & 10.37 & 0.96 & 10.21 & 0.96 &  8.66 \\
    $\{10\overline{1}0\}$                  & 0.53 & 46.47 & 0.53 & 46.49 & 0.53 & 44.66 \\
    $\{10\overline{1}1\}$                  & 0.83 &  8.25 & 0.83 &  8.27 & 0.83 &  8.42 \\
    $\{\overline{1}01\overline{1}\}$       & 0.90 &  1.94 & 0.90 &  1.64 & 0.86 &  4.10 \\
    $\{11\overline{2}0\}$                  & 0.56 & 24.88 & 0.56 & 24.17 & 0.56 & 22.86 \\
    $\{\overline{11}2\overline{1}\}$       & 0.80 &  0.54 & 0.79 &  1.67 & 0.75 &  3.60 \\
\midrule
    Wulff                                  & 0.65 & 100.00 & 0.65 & 100.00 & 0.65 & 100.00 \\
\bottomrule
\addlinespace[2pt]
\multicolumn{7}{l}{\footnotesize $\gamma$: surface energy / \unit{J.m^{-2}}; \quad $f_A$: area fraction / \unit{\percent}}
\end{tabular}

\vspace{12pt}

\begin{tabular}{
    l
    S[table-format=1.2] S[table-format=2.2]
    S[table-format=1.2] S[table-format=2.2]
    S[table-format=1.2] S[table-format=2.2]
}
\toprule
    & \multicolumn{2}{c}{$\Delta \mu_{\ce{S}} = -0.40$~eV}
    & \multicolumn{2}{c}{$\Delta \mu_{\ce{S}} = -0.20$~eV}
    & \multicolumn{2}{c}{$\Delta \mu_{\ce{S}} = -0.20$~eV} \\
    \cmidrule(lr){2-3} \cmidrule(lr){4-5} \cmidrule(lr){6-7}
    {$\{hkl\}$}
    & {$\gamma$} & {$f_A$}
    & {$\gamma$} & {$f_A$}
    & {$\gamma$} & {$f_A$} \\
\midrule
    $(0001)$                               & 0.96 &  8.13 & 0.96 &  8.78 & 1.60 &  8.78 \\
    $(000\overline{1})$                    & 0.96 &  4.86 & 0.96 &  1.98 & 0.32 &  1.98 \\
    $\{10\overline{1}0\}$                  & 0.53 & 40.17 & 0.53 & 35.74 & 0.53 &  35.74  \\
    $\{10\overline{1}1\}$                  & 0.83 &  8.90 & 0.83 &  9.61 & 1.13 &  9.61 \\
    $\{\overline{1}01\overline{1}\}$       & 0.74 & 13.04 & 0.63 & 20.96 & 0.33 &  20.96 \\
    $\{11\overline{2}0\}$                  & 0.56 & 20.52 & 0.56 & 18.20 & 0.56 &  18.20 \\
    $\{\overline{11}2\overline{1}\}$       & 0.68 &  4.38 & 0.61 &  4.73 & 0.42 &  4.73  \\
\midrule
    Wulff                                  & 0.65 & 100.00 & 0.64 & 100.00 & 0.64 & 100.00 \\
\bottomrule
\addlinespace[2pt]
\multicolumn{5}{l}{\footnotesize $\gamma$: surface energy / \unit{J.m^{-2}}; \quad $f_A$: area fraction / \unit{\percent}}
\end{tabular}
\end{table}

\clearpage

\subsection{\texorpdfstring{r$^2$SCAN+$U$}{r²SCAN+U} results for WZ-MnS NCs}
\label{subsection:s2.6}

Table~\ref{tab:tabs11} lists the surface energies and surface area fractions of exposed facets, along with the weighted surface energies, for WZ-MnS Wulff constructions based on r$^2$SCAN+$U$ calculations.
As in Subsection~\ref{subsection:s2.5}, the individual polar surface energies are not physically meaningful because the (0001)--S surface energy was set to an arbitrary reference value of 0.06~eV$\cdot$\AA$^{-2}$; the nonpolar $\{10\overline{1}0\}$ and $\{11\overline{2}0\}$ surface energies are unaffected by this choice.
The surface area fractions and the weighted surface energies are well defined, as demonstrated by the two sets of data at $\Delta \mu_{\ce{S}}=-0.20$~eV shown in Table~\ref{tab:tabs11}, corresponding to reference (0001)--S surface energies of 0.06~eV$\cdot$\AA$^{-2}$ and 0.10~eV$\cdot$\AA$^{-2}$.

\begin{table}[htbp]
\centering
\caption{
Surface energies and surface area fractions of exposed facets in wurtzite (WZ) MnS Wulff constructions, along with weighted surface energies under different thermodynamic equilibrium conditions.
Individual polar surface energies are referenced to an arbitrary (0001)--S value (see Subsection~\ref{subsection:s2.6}) and are not physically meaningful; the surface area fractions and the weighted surface energies are well defined.
All results are from r$^2$SCAN+$U$ calculations.
Data correspond to the Wulff constructions presented in Figure~9c of the main text.
}
\label{tab:tabs11}

\vspace{4pt}
\begin{tabular}{
    l
    S[table-format=1.2] S[table-format=2.2]
    S[table-format=1.2] S[table-format=2.2]
    S[table-format=1.2] S[table-format=2.2]
}
\toprule
    & \multicolumn{2}{c}{$\Delta \mu_{\ce{S}} = -2.00$~eV}
    & \multicolumn{2}{c}{$\Delta \mu_{\ce{S}} = -0.70$~eV}
    & \multicolumn{2}{c}{$\Delta \mu_{\ce{S}} = -0.50$~eV} \\
    \cmidrule(lr){2-3} \cmidrule(lr){4-5} \cmidrule(lr){6-7}
    {$\{hkl\}$}
    & {$\gamma$} & {$f_A$}
    & {$\gamma$} & {$f_A$}
    & {$\gamma$} & {$f_A$} \\
\midrule
    $(0001)$                               & 0.96 &  3.60 & 0.96 &  3.60 & 0.96 &  3.61 \\
    $(000\overline{1})$                    & 1.68 &  9.97 & 1.68 &  9.97 & 1.68 &  9.73 \\
    $\{10\overline{1}0\}$                  & 0.60 & 39.34 & 0.60 & 39.34 & 0.60 & 39.35 \\
    $\{10\overline{1}1\}$                  & 0.76 & 13.42 & 0.76 & 13.42 & 0.76 & 13.45 \\
    $\{\overline{1}01\overline{1}\}$       & 1.37 &  0.00 & 1.37 &  0.00 & 1.37 &  0.00 \\
    $\{11\overline{2}0\}$                  & 0.61 & 33.67 & 0.61 & 33.67 & 0.61 & 32.98 \\
    $\{\overline{11}2\overline{1}\}$       & 1.14 &  0.00 & 1.14 &  0.00 & 1.07 &  0.89 \\
\midrule
    Wulff                                  & 0.75 & 100.00 & 0.75 & 100.00 & 0.75 & 100.00 \\
\bottomrule
\addlinespace[2pt]
\multicolumn{7}{l}{\footnotesize $\gamma$: surface energy / \unit{J.m^{-2}}; \quad $f_A$: area fraction / \unit{\percent}}
\end{tabular}

\vspace{12pt}

\begin{tabular}{
    l
    S[table-format=1.2] S[table-format=2.2]
    S[table-format=1.2] S[table-format=2.2]
    S[table-format=1.2] S[table-format=2.2]
}
\toprule
    & \multicolumn{2}{c}{$\Delta \mu_{\ce{S}} = -0.35$~eV}
    & \multicolumn{2}{c}{$\Delta \mu_{\ce{S}} = -0.20$~eV}
    & \multicolumn{2}{c}{$\Delta \mu_{\ce{S}} = -0.20$~eV} \\
    \cmidrule(lr){2-3} \cmidrule(lr){4-5} \cmidrule(lr){6-7}
    {$\{hkl\}$}
    & {$\gamma$} & {$f_A$}
    & {$\gamma$} & {$f_A$}
    & {$\gamma$} & {$f_A$} \\
\midrule
    $(0001)$                               & 0.96 &  3.66 & 0.96 &  3.75 & 1.60 &  3.75 \\
    $(000\overline{1})$                    & 1.68 &  8.50 & 1.68 &  6.68 & 1.04 &  6.68 \\
    $\{10\overline{1}0\}$                  & 0.60 & 38.13 & 0.60 & 35.76 & 0.60 & 35.76 \\
    $\{10\overline{1}1\}$                  & 0.76 & 13.63 & 0.76 & 13.97 & 1.07 & 13.97 \\
    $\{\overline{1}01\overline{1}\}$       & 1.31 &  0.25 & 1.23 &  2.27 & 0.92 &  2.27 \\
    $\{11\overline{2}0\}$                  & 0.61 & 30.75 & 0.61 & 28.76 & 0.61 & 28.76 \\
    $\{\overline{11}2\overline{1}\}$       & 1.01 &  5.07 & 0.96 &  8.81 & 0.77 &  8.81 \\
\midrule
    Wulff                                  & 0.75 & 100.00 & 0.76 & 100.00 & 0.76 & 100.00 \\
\bottomrule
\addlinespace[2pt]
\multicolumn{5}{l}{\footnotesize $\gamma$: surface energy / \unit{J.m^{-2}}; \quad $f_A$: area fraction / \unit{\percent}}
\end{tabular}
\end{table}

\clearpage

\subsection{Refining MnS surface modeling with \texorpdfstring{r$^{\text{2}}$SCAN+$U$}{r²SCAN+U}}
\label{subsection:s2.7}

The agreement between our surface-energy-derived $U = 2.7$~eV and the value that Gautam and Carter fitted to experimental oxidation enthalpies of Mn oxides\cite{sai_gautam_evaluating_2018} suggests that this correction may be transferable to other Mn chalcogenide systems, though benchmarking against higher-level theory remains advisable for each new material.
Here, we discuss why S-terminated polar surfaces may be disproportionately affected.

For RS- and ZB-MnS, r$^2$SCAN describes nonpolar and Mn-terminated facets reasonably well but consistently underestimates S-terminated polar surface energies (Figures~5~and~7 of the main text).
This behavior may be associated with the known challenges of semilocal DFT in describing localized transition-metal \(d\) states in coordinatively unsaturated surface environments.\cite{Qing_JPCL_2019}
The substantially improved agreement with HSE06 upon applying a Hubbard \(U\) correction to the Mn 3d states is consistent with this interpretation, although the detailed electronic origin of the functional dependence is beyond the scope of the present work.\cite{Heather_PRL_2006, Iurii_PRB_2018}
Without the $U$ correction, the resulting underestimation of S-terminated surface energies leads to qualitatively incorrect Wulff constructions, notably trapezohedral RS-MnS NCs that have not been observed experimentally (Figure~\ref{fig:figs2}d).
With $U = 2.7$~eV, the predicted morphologies are consistent across low- and high-index facet sets and nearly coincide with HSE06 benchmarks.

We therefore recommend that computational studies of transition-metal chalcogenide surfaces validate DFT methods against higher-level theory for both bulk thermochemistry and surface energies.

\clearpage

\subsection{Comparison of thermodynamic morphologies across MnS polymorphs}
\label{subsection:s2.8}

Coordination geometry determines morphological sensitivity to $\Delta\mu_{\ce{S}}$.
In octahedrally coordinated RS-MnS (space group $Fm\overline{3}m$), centrosymmetry permits symmetric slab construction along every cleavage direction, and the nonpolar \{100\} and \{010\} facets, which are slightly inequivalent due to antiferromagnetic ordering, maintain the lowest surface energies across the entire $\Delta\mu_{\ce{S}}$ stability window, rendering the nanocube morphology essentially invariant to $\Delta\mu_{\ce{S}}$ (Figure~5 of the main text).
Only at the extreme S-rich boundary ($\Delta\mu_{\ce{S}} \rightarrow -0.10$~eV, approaching the RS-MnS $\rightarrow$ PY-\ce{MnS2} transition) do \{311\}--S facets marginally truncate the cubes.
Moreover, polar solvents may further reduce the surface energy of polar facets, thereby modifying the equilibrium morphology of RS-MnS NCs beyond the nanocube predicted in vacuum.

In contrast, the tetrahedrally coordinated polymorphs exhibit pronounced morphological sensitivity: ZB-MnS transitions from rhombic dodecahedra to 16-faced polyhedra with increasing $\Delta \mu_{\ce{S}}$, while WZ-MnS maintains a $\Delta\mu_{\ce{S}}$-invariant rod top but develops an increasingly faceted base under S-rich conditions (Figures~7~and~9 of the main text).

Facet exposure areas differ substantially between polymorphs.
For RS-MnS nanocubes, the \{100\} and \{010\} families contribute $\sim$100\% of the surface area under virtually all conditions (Table~\ref{tab:tabs5}).
For ZB-MnS, the dominant facets shift from 100\% \{110\} under Mn-rich conditions to $\sim$55\% \{110\} and $\sim$45\% \{$\overline{111}$\}--S under S-rich conditions (Table~\ref{tab:tabs8}).
For WZ-MnS nanorods, the nonpolar \{10$\overline{1}$0\} and \{11$\overline{2}$0\} prism facets comprise 65--73\% of total surface area across the $\Delta\mu_{\ce{S}}$ range, with the morphological evolution confined primarily to the rod termini (Table~\ref{tab:tabs11}).

\clearpage

\subsection{Experimental details and results for RS-MnS NCs}
\label{subsection:s2.9}

When conducting the synthesis of RS-MnS NCs on a Schlenk line, care must be taken to ensure that the system is never sealed during elevated-temperature steps, as sealing the system could result in overpressurization.
Heating mantles are extremely hot when removed from the setup; caution should be exercised when handling and placing them.
All chemical manipulations should be performed in a fume hood.

\begin{figure*}[htbp]
    \centering
    \includegraphics[width=0.8\linewidth]{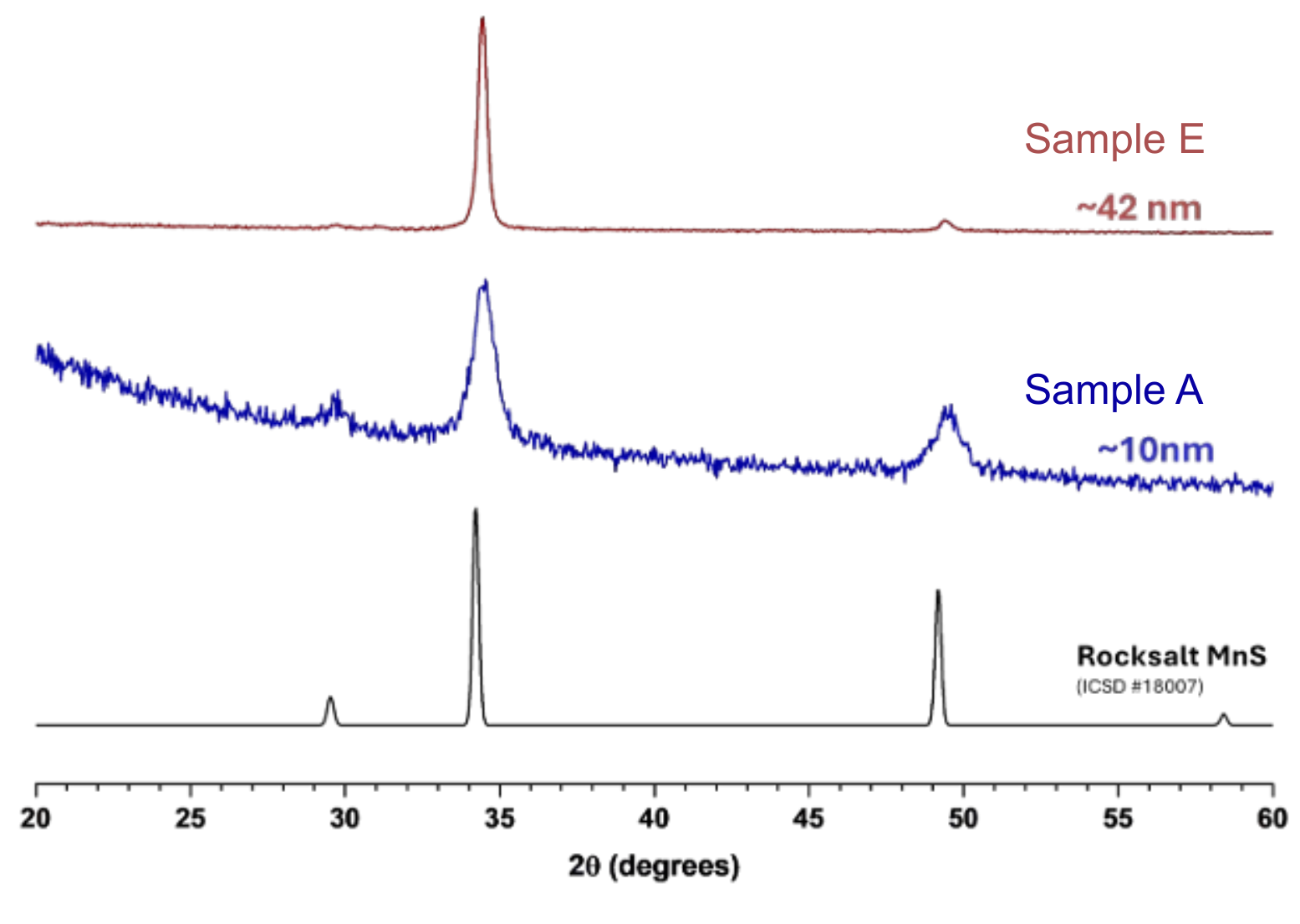}
    \caption{
Powder X-ray diffraction patterns of synthesized rock salt (RS) MnS nanocrystal samples A ($\sim 10$~nm, blue) and E ($\sim 42$~nm, red), compared with the bulk RS-MnS reference pattern.
}
    \label{fig:figs6}
\end{figure*}

Powder X-ray diffraction confirmed that all synthesized samples adopted the rock salt phase (Figure~\ref{fig:figs6}).
The apparent surface energy of NCs is calculated as the excess enthalpy of the nanosized surface relative to that of the bulk analogue, divided by the NC surface area.
Equivalently, it can be obtained from the difference in formation enthalpies between the nano and bulk species.
To minimize uncertainty, drop solution enthalpies $\Delta H_{\mathrm{ds}}$ rather than formation enthalpies $\Delta H_{\mathrm{f}}$ were used, because $\Delta H_{\mathrm{ds}}$ values are measured directly and avoid the propagation of errors from elemental reference enthalpies.
Except for sintered nanomaterials with very small total surface area, adsorbed molecules are always present on NC surfaces and stabilize the nanostructure by lowering the surface energy.
From high-temperature oxidative solution calorimetry data, two surface energies can be distinguished: the solvated surface energy, which includes the stabilizing effect of adsorbed molecules treated as bulk liquids, and the unsolvated (bare) surface energy.
The solvated surface energy is always lower because it incorporates the exothermic enthalpy of adsorption.

\begin{figure*}[htbp]
    \centering
    \includegraphics[width=0.8\linewidth]{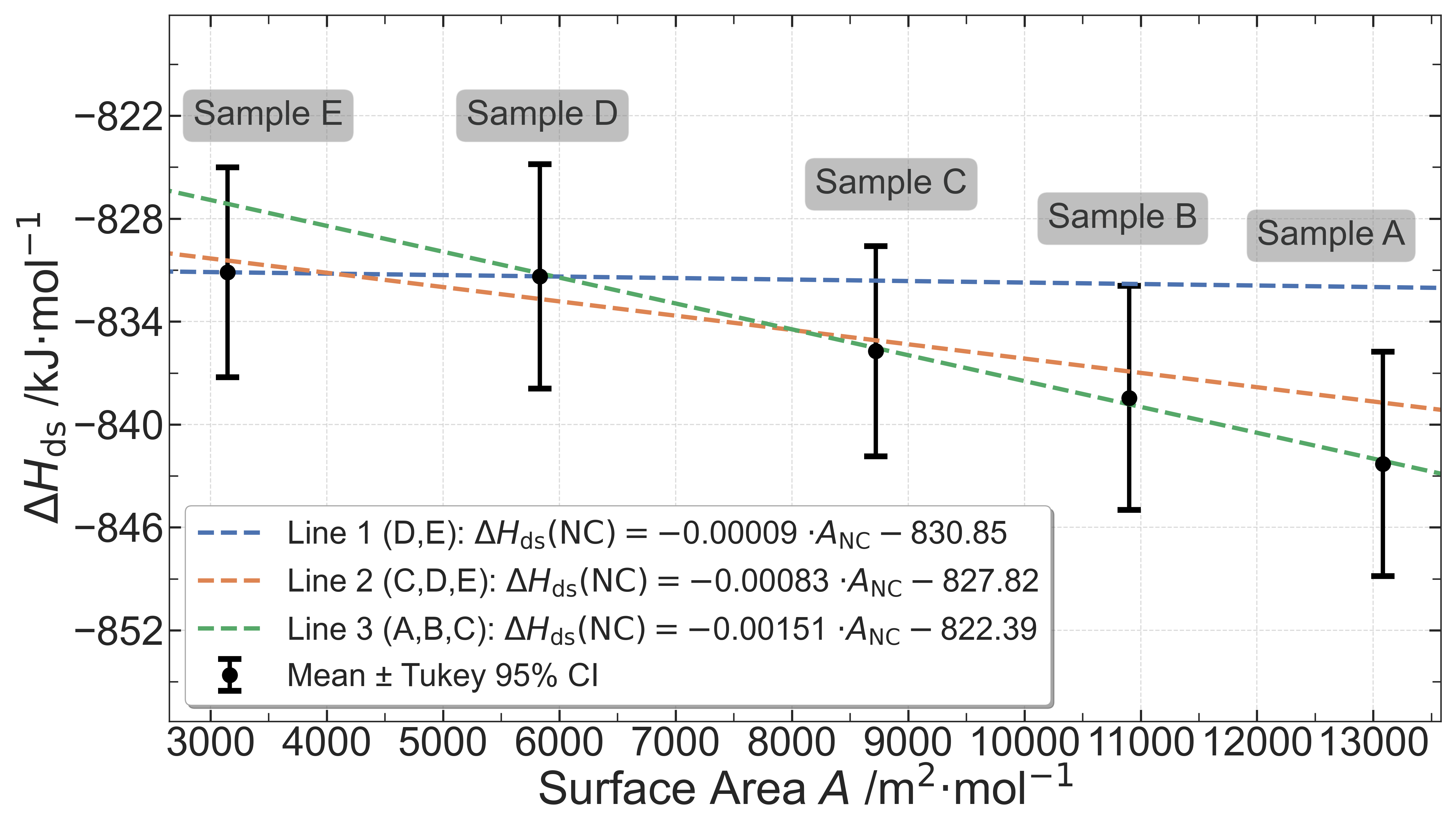}
    \caption{
Mean corrected drop solution enthalpies $\Delta H_{\mathrm{ds}}(\mathrm{NC})$ for each rock salt (RS) MnS nanocrystal sample, shown with 95\% simultaneous confidence intervals from the Tukey Honestly Significant Difference test.
Three linear fits are shown, each based on the subset of samples indicated in parentheses.
}
    \label{fig:figs7}
\end{figure*}

Table~\ref{tab:tabs12} summarizes the sample compositions and molar surface areas.
The thermochemical cycles used to correct for the oleylamine contribution and to calculate formation enthalpies are presented in Tables~\ref{tab:tabs13}~and~\ref{tab:tabs14}, respectively.
Individual corrected drop solution enthalpies and formation enthalpies for each sample are reported in Table~\ref{tab:tabs15}, and the mean values are compared with 95\% confidence intervals in Figure~\ref{fig:figs7}.
Table~\ref{tab:tabs16} compares the measured surface energy of RS-MnS NCs with literature values for other nanoscale metal oxides and sulfides.

\clearpage

\begin{table}[htbp]
\centering
\caption{
Sample labels, approximate particle sizes, calculated molar surface areas $A_{\mathrm{NC}}$, and MnS and oleylamine (OLAM) contents for the five synthesized rock salt (RS) MnS nanocrystal samples.
}
\label{tab:tabs12}
\begin{tabular}{
    l
    S[table-format=2.0]
    S[table-format=5.0]
    S[table-format=1.2]
    S[table-format=1.2]
    S[table-format=1.2]
    S[table-format=1.2]
}
\toprule
    & {Particle size}
    & {$A_{\mathrm{NC}}$}
    & \multicolumn{2}{c}{OLAM}
    & \multicolumn{2}{c}{MnS} \\
    \cmidrule(lr){4-5} \cmidrule(lr){6-7}
    {Sample}
    & {/ \unit{nm}}
    & {/ \unit{m^2.mol^{-1}}}
    & {/ wt\%} & {/ mol\%}
    & {/ wt\%} & {/ mol\%} \\
\midrule
    A & 10 & 13083 & 0.25 & 0.10 & 0.75 & 0.90 \\
    B & 12 & 10902 & 0.10 & 0.04 & 0.90 & 0.96 \\
    C & 15 &  8722 & 0.40 & 0.18 & 0.60 & 0.82 \\
    D & 23 &  5833 & 0.25 & 0.10 & 0.75 & 0.90 \\
    E & 42 &  3146 & 0.11 & 0.04 & 0.89 & 0.96 \\
\bottomrule
\end{tabular}
\begin{tablenotes}[flushleft]
\footnotesize
\item[*] Approximate particle sizes were determined from transmission electron microscopy images.
\item[*] The molar surface area of each sample was calculated as $A_{\mathrm{NC}} = (A_{\mathrm{sphere}}/V_{\mathrm{sphere}})\cdot(M_{\ce{MnS}}/\rho_{\ce{MnS}})$, where $A_{\mathrm{sphere}}$ and $V_{\mathrm{sphere}}$ are the surface area and volume of a sphere with diameter equal to the approximate particle size, $M_{\ce{MnS}}$ is the molar mass of MnS, and $\rho_{\ce{MnS}}$ is the bulk density ($3.99$~g$\cdot$cm$^{-3}$).
\end{tablenotes}
\end{table}

\begin{table}[htbp]
\centering
\caption{
Thermochemical cycle for correcting the drop solution enthalpy $\Delta H_{\mathrm{ds}}$ of rock salt (RS) MnS nanocrystal samples for the oleylamine (OLAM) contribution.
High-temperature oxidative solution calorimetry was performed at 1073~K in molten sodium molybdate (\ce{3Na2O}--\ce{4MoO3}) solvent.
}
\label{tab:tabs13}
\setlength{\tabcolsep}{4pt}
\renewcommand{\arraystretch}{1.5}
\begin{tabularx}{\linewidth}{@{}c X >{\raggedleft\arraybackslash}p{3.2cm}@{}}
\toprule
(1) &
\ce{nano\text{-}MnS$\cdot x$C18H37N \text{ (s, 298 K)} + ($27.25x+2.5$)O2 \text{ (g, 1073 K)} -> Mn^{2+} \text{ (sln, 1073 K)} + SO4^{2-} \text{ (sln, 1073 K)} + 18$x$CO2 \text{ (g, 1073 K)} + 18.5$x$H2O \text{ (g, 1073 K)} + 0.5$x$N2 \text{ (g, 1073 K)} + O^{2-} \text{ (sln, 1073 K)}} &
$\Delta H_{\mathrm{ds, uncorr}}$ \\
(2) &
\ce{C$_{18}$H$_{37}$N \text{ (l, 298 K)} + 27.25O2 \text{ (g, 1073 K)} -> 18CO2 \text{ (g, 1073 K)} + 18.5H2O \text{ (g, 1073 K)} + 0.5N2 \text{ (g, 1073 K)}} &
$\Delta H_{\mathrm{combustion}}$ \\
(3) &
\ce{nano\text{-}MnS \text{ (s, 298 K)} + 2.5O2 \text{ (g, 1073 K)} -> Mn^{2+} \text{ (sln, 1073 K)} + SO4^{2-} \text{ (sln, 1073 K)} + O^{2-} \text{ (sln, 1073 K)}} &
$\Delta H_{\mathrm{ds}}$ \\
\midrule
\multicolumn{3}{@{}c@{}}{%
$\Delta H_{\mathrm{ds}}
= \Delta H_{\mathrm{ds, uncorr}} - x\cdot\Delta H_{\mathrm{combustion}}
\qquad\text{i.e.}\qquad
(3) = (1) - x\cdot(2)$%
} \\
\bottomrule
\end{tabularx}
\end{table}

\begin{table}[htbp]
\centering
\caption{
Thermochemical cycle for calculating the formation enthalpies $\Delta H_{\mathrm{f}}$ of rock salt (RS) MnS bulk and nanocrystal samples.
High-temperature oxidative solution calorimetry was performed at 1073~K in molten sodium molybdate (\ce{3Na2O}--\ce{4MoO3}) solvent.
}
\label{tab:tabs14}
\setlength{\tabcolsep}{4pt}
\renewcommand{\arraystretch}{1.5}
\begin{tabularx}{\linewidth}{@{}c X >{\raggedleft\arraybackslash}p{3.2cm}@{}}
\toprule
(1) &
\ce{nano\text{-}MnS \text{ (s, 298 K)} + 2.5O2 \text{ (g, 1073 K)} -> Mn^{2+} \text{ (sln, 1073 K)} + SO4^{2-} \text{ (sln, 1073 K)} + O^{2-} \text{ (sln, 1073 K)}} &
$\Delta H_{\mathrm{ds}}(\text{NC})$ \\
(2) &
\ce{MnS \text{ (s, 298 K)} + 2.5O2 \text{ (g, 1073 K)} -> Mn^{2+} \text{ (sln, 1073 K)} + SO4^{2-} \text{ (sln, 1073 K)} + O^{2-} \text{ (sln, 1073 K)}} &
$\Delta H_{\mathrm{ds}}(\text{bulk})$ \\
(3) &
\ce{Mn \text{ (s, 298 K)} + 0.5O2 \text{ (g, 1073 K)} -> Mn^{2+} \text{ (sln, 1073 K)} + O^{2-} \text{ (sln, 1073 K)}} &
$\Delta H_{1}$ \\
(4) &
\ce{S \text{ (s, 298 K)} + 2O2 \text{ (g, 1073 K)} -> SO4^{2-} \text{ (sln, 1073 K)}} &
$\Delta H_{2}$ \\
\midrule
\multicolumn{3}{@{}c@{}}{%
$\Delta H_{\mathrm{f}}(\text{NC})
= -(1) + (3) + (4)
\qquad\text{and}\qquad
\Delta H_{\mathrm{f}}(\text{bulk})
= -(2) + (3) + (4)$%
} \\
\bottomrule
\end{tabularx}
\end{table}

\begin{table}[htbp]
\centering
\caption{
Individual corrected drop solution enthalpies ($\Delta H_{\mathrm{ds}}$) and formation enthalpies ($\Delta H_{\mathrm{f}}$) for each rock salt (RS) MnS nanocrystal sample.
Mean values and confidence intervals are reported in Table~2 of the main text.
}
\label{tab:tabs15}
\begin{tabular}{
    l
    S[table-format=-3.2]
    S[table-format=-3.2]
}
\toprule
    {Sample} & {$\Delta H_{\mathrm{ds}}$ / \unit{kJ.mol^{-1}}} & {$\Delta H_{\mathrm{f}}$ / \unit{kJ.mol^{-1}}} \\
\midrule
    \multirow{7}{*}{A}  & -841.72 & -197.78 \\
                        & -834.89 & -204.61 \\
                        & -836.32 & -203.18 \\
                        & -837.40 & -202.10 \\
                        & -849.87 & -189.63 \\
                        & -849.06 & -190.44 \\
                        & -846.87 & -192.61 \\
\midrule
    \multirow{7}{*}{B}  & -841.81 & -197.69 \\
                        & -835.02 & -204.48 \\
                        & -844.72 & -194.78 \\
                        & -844.68 & -194.82 \\
                        & -825.47 & -214.03 \\
                        & -843.37 & -196.13 \\
                        & -834.08 & -205.42 \\
\midrule
    \multirow{8}{*}{C}  & -820.57 & -218.93 \\
                        & -849.49 & -190.01 \\
                        & -838.63 & -200.87 \\
                        & -837.53 & -201.97 \\
                        & -849.22 & -190.28 \\
                        & -827.50 & -212.00 \\
                        & -831.91 & -207.59 \\
                        & -830.99 & -208.51 \\
\midrule
    \multirow{7}{*}{D}  & -839.55 & -199.95 \\
                        & -829.49 & -210.01 \\
                        & -830.29 & -209.21 \\
                        & -824.92 & -214.58 \\
                        & -832.31 & -207.19 \\
                        & -835.94 & -203.56 \\
                        & -827.04 & -212.46 \\
\midrule
    \multirow{8}{*}{E}  & -831.43 & -208.07 \\
                        & -822.67 & -216.83 \\
                        & -823.17 & -216.33 \\
                        & -823.41 & -216.09 \\
                        & -840.51 & -198.99 \\
                        & -850.37 & -189.13 \\
                        & -838.89 & -200.61 \\
                        & -818.58 & -220.92 \\
\bottomrule
\end{tabular}
\end{table}

\begin{table}[htbp]
\centering
\caption{
Measured surface energies of nanoscale metal oxides and sulfides from calorimetry, with the adsorbate present on the NC surface during measurement.
The MnS entry is from this work; all other values are from the literature as cited.
}
\label{tab:tabs16}
\begin{tabular}{
    llll
}
\toprule
    Compound & Crystal system & $\gamma$ / \unit{J.m^{-2}} & Adsorbate \\
\midrule
    ZnS (sphalerite)    & Cubic     & $1.25 \pm 0.21$\cite{subramani_surface_2023}          & Water \\
    ZnS (wurtzite)      & Hexagonal & $0.99 \pm 0.32$\cite{subramani_surface_2023}          & Ethylene glycol \\
    \ce{Fe3S4}          & Cubic     & $1.15 \pm 0.23$\cite{subramani_greigite_2020}         & Water \\
    MnS (rock salt)     & Cubic     & $1.15 \pm 0.38$ (this work)                           & OLAM \\
    ZnO (wurtzite)      & Hexagonal     & $1.31 \pm 0.07$\cite{zhang_surface_2007}         & Water \\
    \ce{Fe3O4}          & Cubic     & $0.80 \pm 0.05$\cite{lilova_energetics_2014}          & Water \\
    \ce{Mn2O3}          & Cubic     & $1.29 \pm 0.10$\cite{birkner_thermodynamics_2012}     & Water \\
    \ce{Mn3O4}          & Tetragonal     & $0.96 \pm 0.08$\cite{birkner_thermodynamics_2012}     & Water \\
\bottomrule
\end{tabular}
\end{table}

\clearpage
\subsection{Spherical average surface energy approximation}
\label{subsection:s2.10}

\begin{figure}
    \centering
    \includegraphics[width=0.8\linewidth]{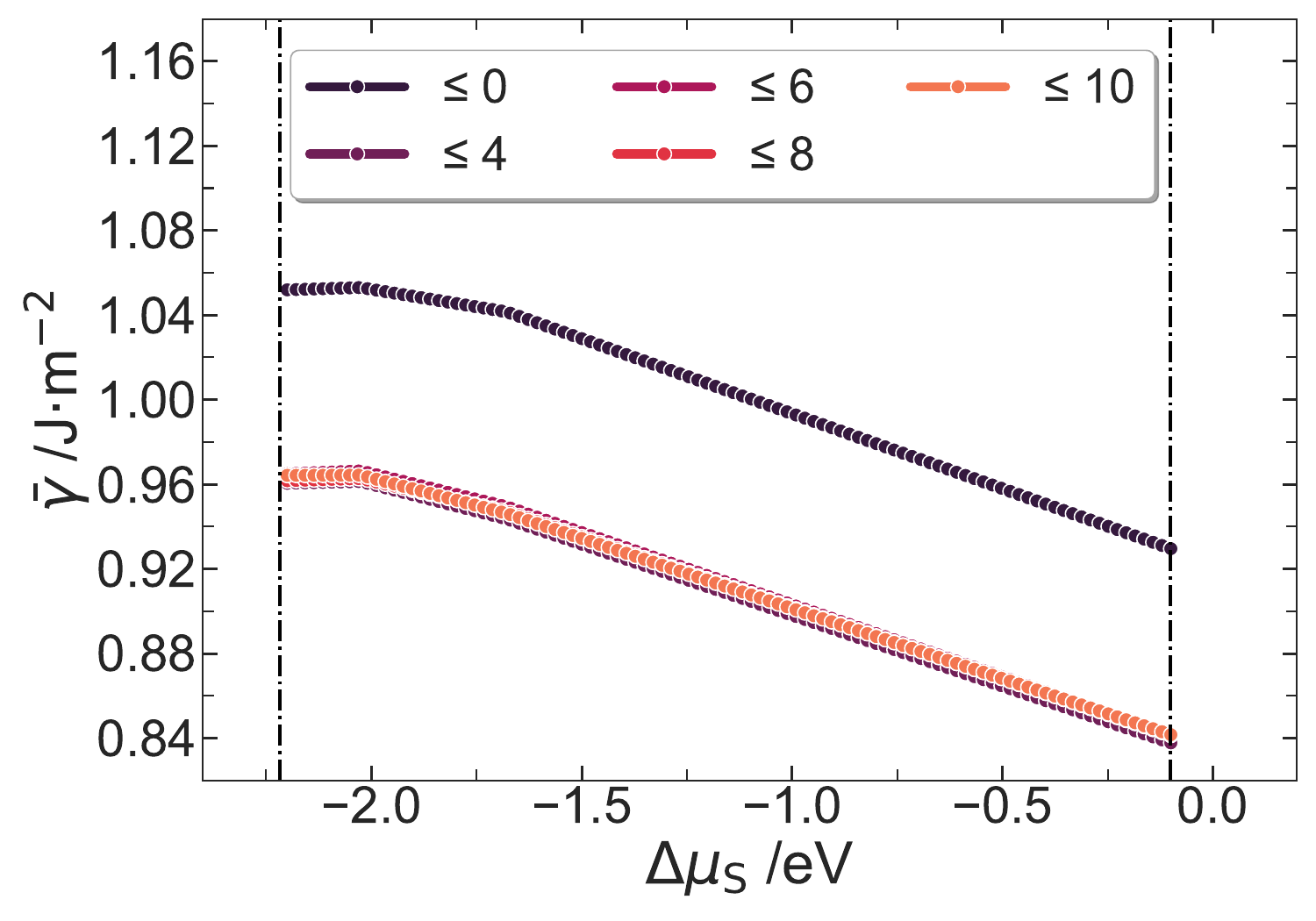}
    \caption{
Convergence of the spherical average surface energy of rock salt (RS) MnS with respect to the truncation order of the cubic-symmetric polynomial expansion.
Each curve corresponds to a maximum polynomial order, as indicated in the legend. The spherical average converges by sixth order.
    }
    \label{fig:figs8}
\end{figure}

\begin{figure}
    \centering
    \includegraphics[width=0.8\linewidth]{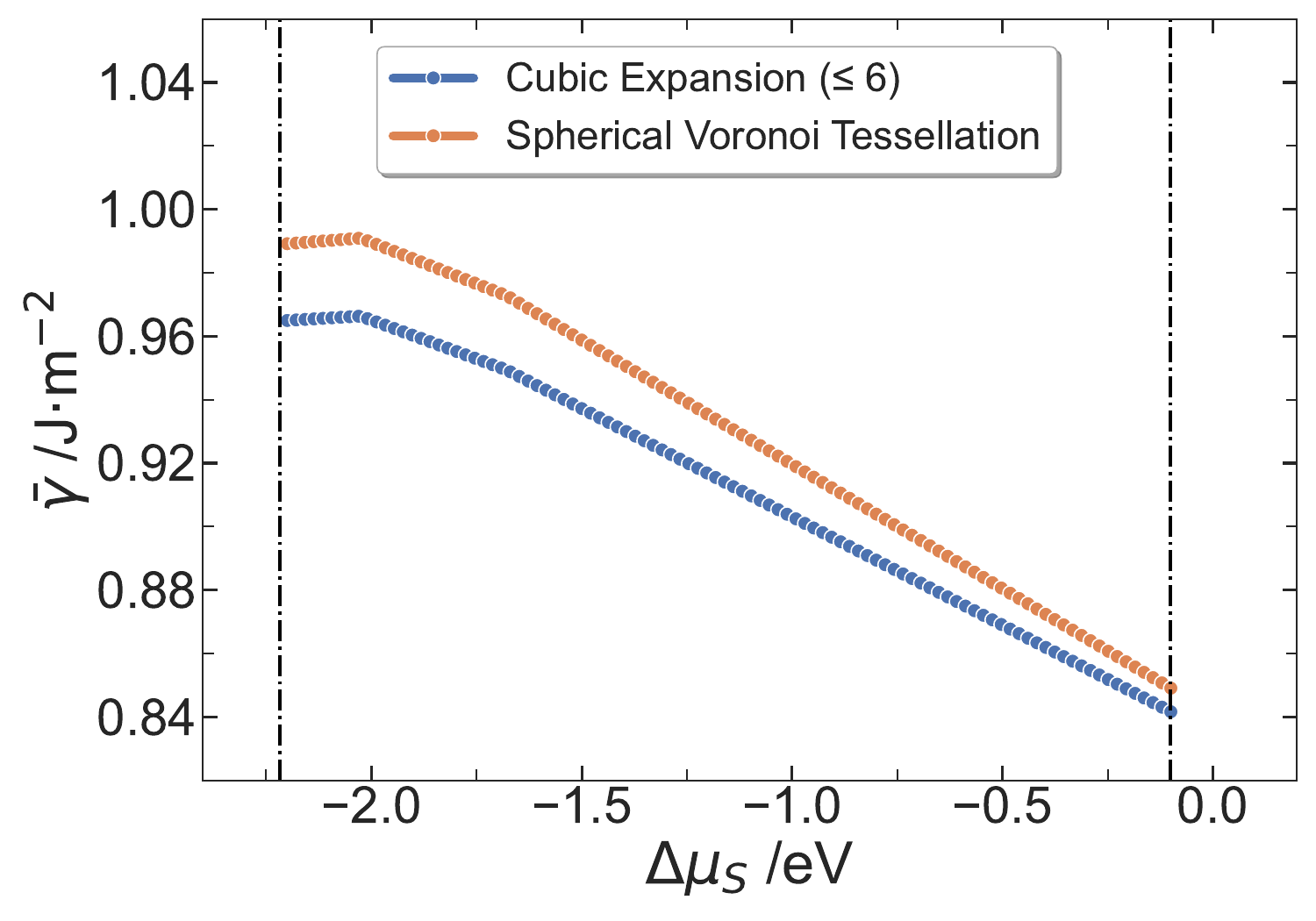}
    \caption{
Dependence of the spherical average surface energy of rock salt (RS) MnS on the relative chemical potential of sulfur, $\Delta \mu_{\ce{S}}$, obtained from the sixth-order cubic expansion and spherical Voronoi tessellation.
    }
    \label{fig:figs9}
\end{figure}

To quantify the orientation-averaged surface energy in the spherical limit, we employed a cubic-symmetric polynomial expansion and a spherical Voronoi tessellation, as described in the Methods section of the main text.
Table~\ref{tab:tabs17} lists the cubic-invariant basis functions used in the polynomial expansion.
The convergence of the cubic expansion was examined by progressively increasing the truncation order up to tenth order (Figure~\ref{fig:figs8}).
The spherical average converges by sixth order, beyond which higher-order terms introduce negligible changes.
Therefore, the sixth-order basis (including $B_0$, $B_4$, and $B_6$ terms) was adopted in the main analysis.
As shown in Figure~\ref{fig:figs9}, the sixth-order cubic expansion and the Voronoi tessellation yield consistent trends with respect to $\Delta \mu_{\ce{S}}$, indicating that the result does not depend on the choice of orientation-averaging scheme.

\begin{table}[htbp]
    \centering
    \caption{
Cubic-invariant basis functions used in the polynomial expansion of the surface energy.
The unit normal along a given Miller index direction is denoted by $\mathbf{n} = (n_x, n_y, n_z)$.
    }
    \label{tab:tabs17}
    \begin{tabular}{ll}
        \toprule
        Cubic-symmetric polynomial basis & Expression \\
        \midrule
        $B_0$              & 1 \\                                          
        $B_4$              & $n_{\mathrm{x}}^4+n_{\mathrm{y}}^4+n_{\mathrm{z}}^4$ \\
        $B_6$              & $n_{\mathrm{x}}^6+n_{\mathrm{y}}^6+n_{\mathrm{z}}^6$ \\
        $B_{8\mathrm{a}}$  & $n_{\mathrm{x}}^8+n_{\mathrm{y}}^8+n_{\mathrm{z}}^8$ \\
        $B_{8\mathrm{b}}$  & $n_{\mathrm{x}}^4 \cdot n_{\mathrm{y}}^4+n_{\mathrm{y}}^4 \cdot n_{\mathrm{z}}^4+n_{\mathrm{z}}^4 \cdot n_{\mathrm{x}}^4$ \\
        $B_{\mathrm{10}}$  & $n_{\mathrm{x}}^{10}+n_{\mathrm{y}}^{10}+n_{\mathrm{z}}^{10}$ \\

        \bottomrule
    \end{tabular}
\end{table}

\clearpage


\section{Bulk and surface models}

This section provides structural details, computational methodologies, and benchmark results for the bulk (\ref{subsection:s3.1}), surface slab (\ref{subsection:s3.2}), and surface wedge (\ref{subsection:s3.3}) models used in this study.
All corresponding structural files and Vienna Ab initio Simulation Package input/output files are available in the NOMAD repository.\cite{nomad2025dataset}

\subsection{Bulk crystals}
\label{subsection:s3.1}

All bulk structures were retrieved from the Materials Project\cite{10.1063/1.4812323, horton_accelerated_2025} and are summarized in Table~\ref{tab:tabs18}.
Spin-polarized DFT calculations were performed to relax the lattice constants, angles, and ionic positions.
The valence electron configurations and projector augmented wave data set details for each element are listed in Table~\ref{tab:tabs19}.
Although different magnetic configurations of $\alpha$-Mn yield energy differences of up to $\sim 300$~meV$\cdot$atom$^{-1}$ with coarse $k$-point grids (Table~\ref{tab:tabs20}), the reaction energy evaluation requires only the per-atom energy of the ground-state configuration.
This variability therefore does not affect our results.
A low-spin ferromagnetic collinear configuration was therefore adopted for $\alpha$-Mn, reducing computational cost while maintaining accuracy.
Owing to the high computational cost of HSE06 and its dispersion-corrected variant HSE06-D3, single-point SCF calculations with these functionals were performed on r$^2$SCAN-optimized structures.
The Hubbard $U$ correction was not applied to any bulk structure optimization or energy evaluation, as r$^2$SCAN predictions for reaction energies and lattice constants of MnS polymorphs already show good agreement with experiment (see Sections~\ref{subsection:s1.1}~and~\ref{subsection:s1.2}).

\begin{table}[htbp]
\centering
\caption{
Summary of bulk crystal models used in this study, with crystal phase, space group, and Materials Project identifier.
}
\label{tab:tabs18}
\begin{tabular}{
    llll
}
\toprule
    Crystal & Phase & Space group & Materials Project ID \\
\midrule
    Manganese           & Alpha              & Cubic $I\overline{4}3m$        & mp-35\cite{noauthor_materials_2020_Mn} \\
    Sulfur              & Alpha              & Orthorhombic $Fddd$            & mp-77\cite{noauthor_materials_2020_S} \\
    Iodine              & ---                & Orthorhombic $Cmce$            & mp-23153\cite{noauthor_materials_2020_I} \\
    Manganese sulfide   & Rock salt          & Cubic $Fm\overline{3}m$        & mp-2065\cite{noauthor_materials_2020_RS-MnS} \\
    Manganese sulfide   & Wurtzite           & Hexagonal $P6_3mc$             & mp-2562\cite{noauthor_materials_2020_WZ-MnS} \\
    Manganese sulfide   & Zinc blende        & Cubic $F\overline{4}3m$        & mp-1783\cite{noauthor_materials_2020_ZB-MnS} \\
    Manganese disulfide & Pyrite             & Cubic $Pa\overline{3}$         & mp-1455\cite{noauthor_materials_2020_MnS2} \\
    Manganese chloride  & Trigonal omega-like & Rhombohedral $R\overline{3}m$ & mp-28233\cite{noauthor_materials_2020_MnCl2} \\
    Manganese iodide    & Trigonal omega     & Hexagonal $P\overline{3}m1$    & mp-28013\cite{noauthor_materials_2020_MnI2} \\
\bottomrule
\end{tabular}
\end{table}

\begin{table}[htbp]
\centering
\caption{
Projector augmented wave (PAW) data set details for elements used in the DFT calculations, including the Vienna Ab initio Simulation Package PAW label, valence electron configuration, and dataset release date.
}
\label{tab:tabs19}
\begin{tabular}{
    llll
}
\toprule
    Element & PAW label & Valence electron state & Release date \\
\midrule
    S  & PAW\_PBE S      & $3s^2 3p^4$       & 06/09/2000 \\
    Cl & PAW\_PBE Cl     & $3s^2 3p^5$       & 06/09/2000 \\
    Mn & PAW\_PBE Mn\_pv & $3p^6 4s^2 3d^5$  & 02/08/2007 \\
    I  & PAW\_PBE I      & $5s^2 5p^5$       & 08/04/2002 \\
\bottomrule
\end{tabular}
\end{table}

\begin{table}[htbp]
\centering
\caption{
Comparison of bulk energies of alpha manganese ($\alpha$-Mn) across different magnetic configurations and sources.
Energies are reported relative to the lowest-energy configuration.
The reciprocal spacing between $k$-points was set to no larger than $0.50$~\AA$^{-1}$ to reduce computational cost.
}
\label{tab:tabs20}
\begin{tabular}{
    l
    l
    l
    l
    S[table-format=3.2]
}
\toprule
    Crystal & Source & Magnetic order & Type & {$\Delta E_{\mathrm{bulk}}$ / \unit{meV\per atom}} \\
\midrule
    $\alpha$-Mn & MAGNDATA (\#1.85)
        & Ferrimagnetic (BCS)
        & Noncollinear & 0.00 \\
    $\alpha$-Mn & MP (mp-35)\cite{noauthor_materials_2020_Mn}
        & Ferrimagnetic (BCS)
        & Noncollinear & 1.45 \\
    $\alpha$-Mn & MP (mp-35)\cite{noauthor_materials_2020_Mn}
        & $(1,0,0)$ for all Mn
        & Noncollinear & 81.95 \\
    $\alpha$-Mn & MP (mp-35)\cite{noauthor_materials_2020_Mn}
        & Ferrimagnetic (MP)
        & Collinear & 81.15 \\
    $\alpha$-Mn & MP (mp-35)\cite{noauthor_materials_2020_Mn}
        & Low-spin ferromagnetic
        & Collinear & 64.16 \\
    $\alpha$-Mn & MP (mp-35)\cite{noauthor_materials_2020_Mn}
        & High-spin ferromagnetic
        & Collinear & 300.82 \\
\bottomrule
\addlinespace[2pt]
\multicolumn{5}{l}{\footnotesize BCS: Bilbao Crystallographic Server; \quad MP: Materials Project}
\end{tabular}
\end{table}

\clearpage

\subsection{Slab models}
\label{subsection:s3.2}

Two-dimensional periodic surface slab models were constructed from r$^2$SCAN-relaxed MnS polymorphs using custom Python codes built on the \verb|surface| module of Pymatgen.\cite{sun_efficient_2013, tran_surface_2016}
Initial magnetic moments were inherited from the corresponding bulk structures and allowed to reorder during geometry optimizations.
High-index facets were considered for RS-MnS (Miller indices up to 3; Table~\ref{tab:tabs3}) and ZB-MnS (up to 2; Table~\ref{tab:tabs6}); only low-index facets were studied for WZ-MnS (Table~\ref{tab:tabs9}).
Slabs possessing appropriate symmetry operations were constructed symmetrically with identical terminations on both sides;\cite{sun_efficient_2013} where such symmetry was absent, asymmetric slabs were generated by enumerating all possible combinations of surface terminations.
Surfaces are labeled using the notation ``Miller indices--termination'': for example, (111)--S denotes the sulfur-terminated (111) facet.
When multiple terminations of the same type exist for a given orientation, they are distinguished by the coordination number of the outermost surface atoms; for example, ($10\overline{1}1$)--S$^1$ and ($10\overline{1}1$)--S$^2$ correspond to the ($10\overline{1}1$) facets with singly and doubly coordinated surface sulfur atoms, respectively.

\begin{figure*}[htbp]
    \centering
    \includegraphics[width=1.0\linewidth]{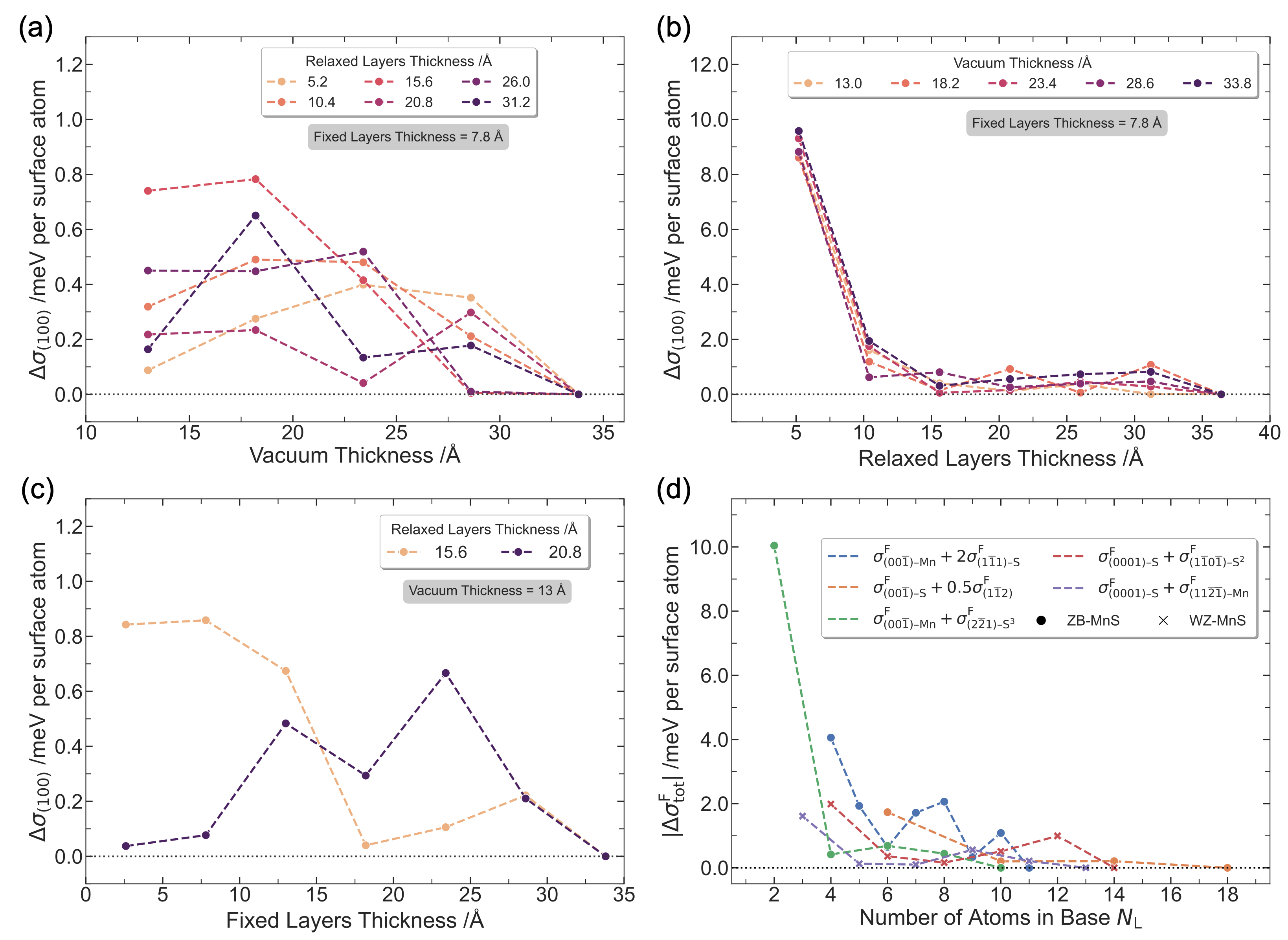}
    \caption{
Convergence of the total surface energy with respect to (a) vacuum thickness, (b) relaxed region thickness, and (c) fixed region thickness for the (100) slab in rock salt (RS) MnS.
(d) Convergence of the total surface energy with respect to wedge size $N_{\mathrm{L}}$ for zinc blende (ZB) and wurtzite (WZ) MnS wedges.
All results are from r$^2$SCAN calculations.
}
    \label{fig:figs10}
\end{figure*}

To determine the minimum thicknesses of the vacuum layer, central fixed region, and relaxed region, convergence tests were performed on the (100) slab of RS-MnS, aiming to minimize interactions between periodic images and between opposing terminations.
As shown in Figures~\ref{fig:figs10}a--c, a total thickness of 31.2~\AA~(comprising a 13~\AA~vacuum layer, a 15.6~\AA~relaxed region, and a 2.6~\AA~fixed region) was sufficient, as the difference in total surface energy was no larger than 1~meV~per~surface~atom.
For each surface, two slab models of different thicknesses were then constructed using these parameters as the minimum baseline; the surface energy was deemed converged when the difference between the two evaluations was smaller than 0.01~J$\cdot$m$^{-2}$.
Fully fixed and half-fixed slabs were also generated for calculating the individual polar surface energies of ZB-MnS and the relative surface energies of WZ-MnS (see Figures~1c~and~1d of the main text).
Although slab relaxations were carried out using r$^2$SCAN, this functional yielded incorrect surface energies for certain S-terminated polar facets, including the (111)--S, (131)--S, and (311)--S surfaces of RS-MnS.
To correct this deficiency, r$^2$SCAN+$U$ single-point SCF calculations ($U = 2.7$~eV on Mn 3d) were employed to obtain corrected surface energies for all three MnS polymorphs, as described in the Computational Framework Validation section of the main text.

\clearpage

\subsection{Wedge models}
\label{subsection:s3.3}

\begin{figure*}[htbp]
    \centering
    \includegraphics[width=1.0\linewidth]{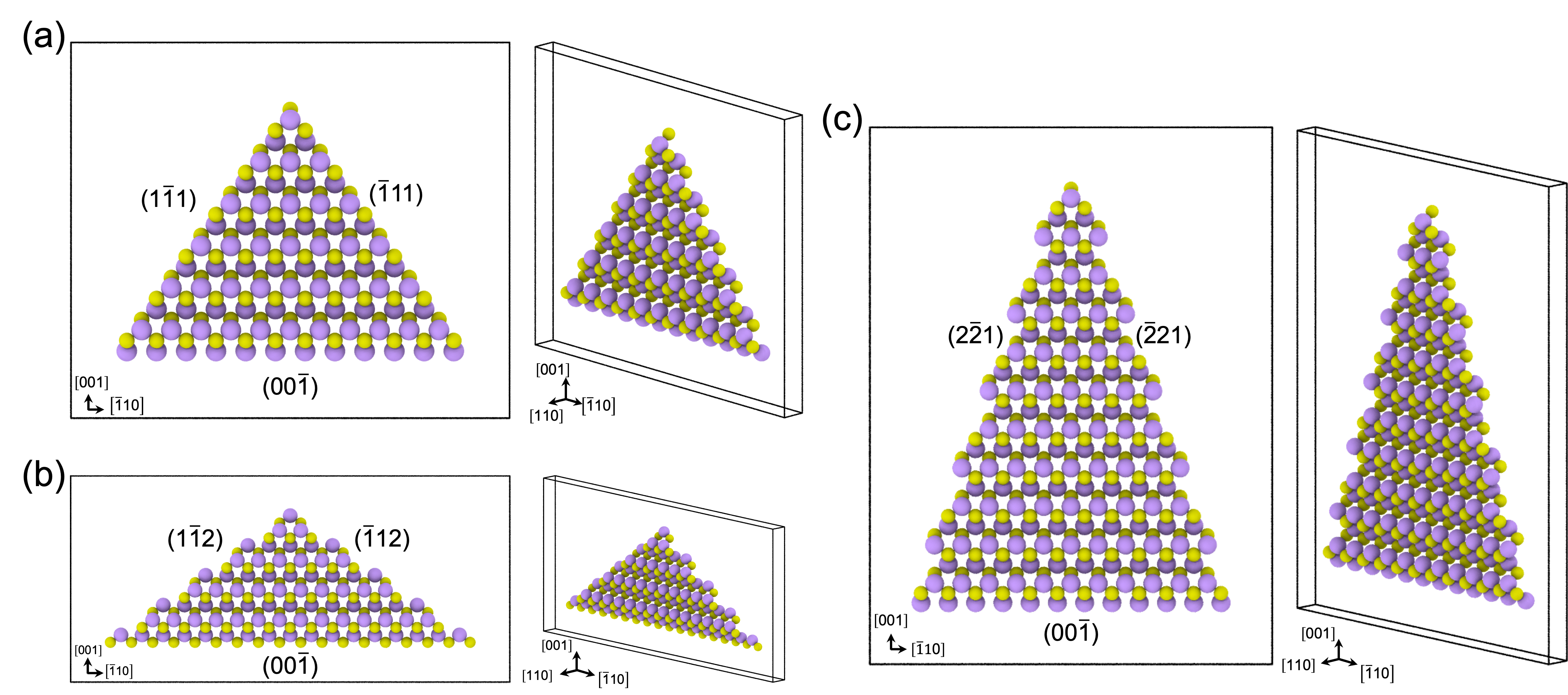}
    \caption{
Front (left) and perspective (right) views of zinc blende (ZB) MnS wedge models with a ($00\overline{1}$) base facet and two crystallographically equivalent side facets: (a) ($1\overline{1}1$) and ($\overline{1}11$), (b) ($1\overline{1}2$) and ($\overline{1}12$), and (c) ($2\overline{2}1$) and ($\overline{2}21$).
}
    \label{fig:figs11}
\end{figure*}

\begin{figure*}[htb]
    \centering
    \includegraphics[width=1.0\linewidth]{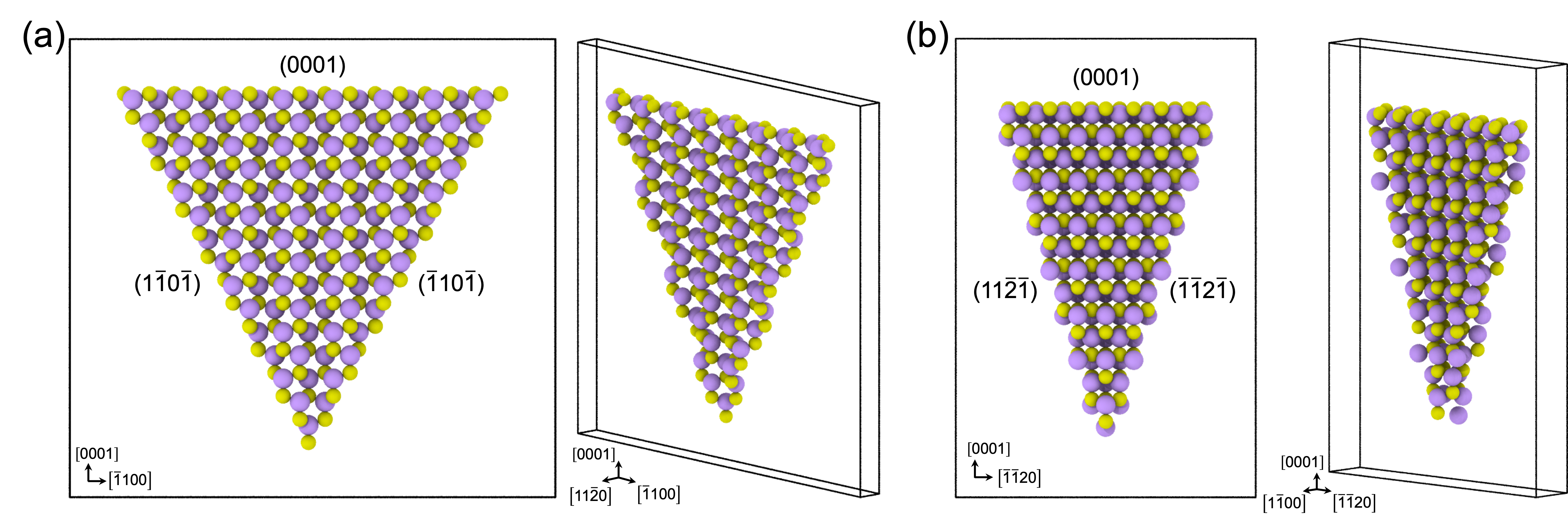}
    \caption{
Front (left) and perspective (right) views of wurtzite (WZ) MnS wedge models with a (0001) base facet and two crystallographically equivalent side facets: (a) ($1\overline{1}0\overline{1}$) and ($\overline{1}10\overline{1}$), and (b) ($11\overline{21}$) and ($\overline{11}2\overline{1}$).
}
    \label{fig:figs12}
\end{figure*}

One-dimensional periodic wedge models were generated from r$^2$SCAN-relaxed ZB- and WZ-MnS crystals using custom Python codes built on the \verb|structure| and \verb|lattice| modules of Pymatgen.\cite{ONG2013314}
Initial magnetic moments were inherited from the corresponding bulk structures and allowed to reorder during geometry optimizations.
As illustrated in Figures~\ref{fig:figs11}~and~\ref{fig:figs12}, each wedge has a triangular cross-section consisting of a base facet and two crystallographically equivalent side facets.
The wedge size is denoted by the number of atoms along the base of the triangle, $N_{\mathrm{L}}$.
Wedges were built using the smallest periodic unit along the lattice vector parallel to the infinite extension, with at least 15~\AA~of vacuum added along the other two directions to eliminate interactions between periodic images.
Convergence tests (Figure~\ref{fig:figs10}d) determined the minimum $N_{\mathrm{L}}$ required for reliable surface energies.
All wedge structures were unrelaxed and evaluated using r$^2$SCAN+$U$ single-point SCF calculations ($U = 2.7$~eV on Mn 3d), consistent with the approach described in the main text (see Methods).
Hydrogen passivation was not applied, as convergence tests on clean-surface wedges demonstrated sufficient convergence.

\clearpage

\bibliography{si-references}